\documentclass[fleqn]{svmult}
\usepackage{ae}
\usepackage{german}
\usepackage{aecompl}
\usepackage{amsmath}
\usepackage{amssymb}
\usepackage{epsfig}
\usepackage{epsf}
\usepackage{svaips}      
\usepackage{makeidx}     
\usepackage{graphicx}    
\usepackage{multicol}    
\usepackage{times}

\newcommand{\poldens}{{\varrho}}
\newcommand{\red}{{R_e^d}}
\newcommand{\dr}{{{\rm d}^d{\bf r}}}
\newcommand{\etal}{{et~al.}}
\newcommand{\N}{{\bar{\cal N}}}  
\newcommand{\G}{{\cal G}}

\newcommand{\V}{{\cal V}}
\newcommand{\EP}{{EP}}
\newcommand{\HH}{{\cal H}}
\newcommand{\DD}{{\cal D}}

\newcommand{\xx}{{\bf x}}
\newcommand{\rr}{{\bf r}}
\newcommand{\qq}{{\bf q}}
\newcommand{\kk}{{\bf k}}
\renewcommand{\ggg}{{\bf g}}
\newcommand{\dd}{{\bf d}}
\newcommand{\GG}{{\bf G}}
\newcommand{\one}{{\bf 1}}
\newcommand{\ie}{{\em i.~e.}}
\newcommand{\eg}{{\em e.~g.}}
\newcommand{\Sec}{{Sec.}}

\begin{document}
\title*{Incorporating fluctuations and dynamics in self-consistent field theories for polymer blends}
\titlerunning{Fluctuations and dynamics in self-consistent field theories}
\author{Marcus M\"uller\inst{1}\and
Friederike Schmid\inst{2}}
\institute{Institut f{\"u}r Physik, WA331, Johannes Gutenberg Universit{\"a}t, D-55099 Mainz, Germany \textit{marcus.mueller@uni-mainz.de}
\and Fakult{\"a}t f{\"u}r Physik, Universit{\"a}t Bielefeld, D-33615 Bielefeld, Germany \textit{schmid@physik.uni-bielefeld.de}}

\maketitle
\begin{abstract}
We review various methods to investigate the statics and the dynamics of collective 
composition fluctuations in dense polymer mixtures within fluctuating-field approaches.
The central idea of fluctuating-field theories is to rewrite the partition function of the 
interacting multi-chain systems in terms of integrals over auxiliary, often complex, fields, 
which are introduced by means of appropriate Hubbard-Stratonovich transformations.
Thermodynamic averages like the average composition and the structure factor can be 
expressed exactly as averages of these fields. 
We discuss different analytical and numerical approaches to studying such a theory:
The self-consistent field approach solves the integrals over the fluctuating fields in
saddle-point approximation. Generalized random phase approximations allow to incorporate
Gaussian fluctuations around the saddle point. Field theoretical polymer simulations 
are used to study the statistical mechanics of the full system with Complex Langevin or 
Monte Carlo methods. Unfortunately, they are hampered by the presence of a sign
problem. In dense system, the latter can be avoided without losing essential
physics by invoking a saddle point approximation for the complex field that couples
to the total density. This leads to the external potential theory. We investigate 
the conditions under which this approximation is accurate. Finally, we discuss
recent approaches to formulate realistic dynamical time evolution equations
for such models.
The methods are illustrated by two examples: A study of the fluctuation-induced formation 
of a polymeric microemulsion in a polymer-copolymer mixture, and a study of
early-stage spinodal decomposition in a binary blend.

\end{abstract}
\dominitoc
\begin{abbrsymblist}
\item[$\alpha$] Index for monomer species ($A$ or $B$)
\item[$J$] Index for polymer species (homopolymer $A$, homopolymer $B$, copolymer)
\item[$N$]    Number of segments of a chain
\item[$\chi$] Flory-Huggins parameter
\item[$\V$]   Volume
\item[$\poldens$] Polymer number density
\item[$R_e$]  End-to-end distance of a polymer
\item[$\N$]   Invariant degree of polymerization, $\N = \poldens R_e^d$.
\item[$\bar \phi_J$] Volume fraction of polymers of type $J$
\item[$\gamma_{J\alpha}$] Distributions of segments of species {\protect{$\alpha$}} along a polymer of type {\protect{$J$}}
\item[$\hat \phi_\alpha({\bf r})$] Microscopic segment density of species \protect$\alpha$ (Eq.~(3))
\item[$\G$] Complex field-theoretical Hamiltonian as a function of fluctuating fields (Eqs.~(10),(11))
\item[$Q_J$] Single chain partition function of type $J$ (Eqs.~(12),(13))
\item[$q_J({\bf r},\tau)$] End-segment distribution (Eqs.~(17),(18))
\item[$\HH$]  Hamiltonian of the external potential theory (Eqs.~(41),(42))
\item[${\cal F}^*$] Hamiltonian of the (dynamic) SCF theory (Eq.~(53))
\item[$\Lambda$] Onsager coefficient (Eq.~(110))
\item[$D$] single chain self-diffusion constant
\end{abbrsymblist}

%
%
\section{Introduction}
Polymeric materials in daily life are generally multicomponent systems.
Chemically different polymers are ``alloyed'' as to design a material which
combines the favorable characteristics of the individual components~\cite{APPLICATION}. 
Understanding the miscibility in the molten state is crucial for understanding 
and tailoring properties that are relevant for
practical applications. Miscibility on a microscopic length scale is desirable,
for instance, to increase the tensile strength of the material. Unfortunately,
different polymers are typically not microscopically miscible, because the entropy 
of mixing is much smaller for polymers than for small molecules. 
The mixture separates into domains in which one of the components is enriched.
The domains are separated by interfaces. Tuning the interface tension between 
the coexisting phases, and the morphology of the material on a mesoscopic length 
scale, is a key to tailoring material properties and has attracted abiding interest.

One strategy to improve the miscibility and the interfacial properties 
has been to synthesize copolymers by chemically linking polymers of different
type to each other. Added in small amounts to a homopolymer mixture, the copolymers
adsorb at the interfaces, change their local structure, and reduce the
interfacial tension~\cite{RUSSELL,KRAMER,BUD,HASHI}. Mixtures containing substantial amounts 
of copolymers form new types of phases, where different monomer types aggregate into mesoscopic 
``phase separated'' domains (microphase separation). By varying the composition of
the mixture and the architecture of the individual (co)polymers, one can create
a wide range of different morphologies, corresponding to different materials,
each with unique material properties.

Unfortunately, copolymers are often rather expensive components.
Another, cheaper, strategy for tuning the domain structure of blends 
takes advantage of the fact that in many practical applications, polymeric blends 
never reach thermal equilibrium on larger length scales.
The morphology of the blend strongly depends on the kinetics of phase separation.
Finer dispersed morphologies can be obtained by optimizing the processing 
conditions~\cite{jansen,tucker}.

One of the most powerful methods to assess such phenomena theoretically is the 
self-consistent field (SCF) theory. Originally introduced by Edwards~\cite{EDWARDS} 
and later Helfand \etal~\cite{helfand}, it has evolved into a versatile tool to describe 
the structure and thermodynamics of spatially inhomogeneous, dense polymer 
mixtures~\cite{hong,scheutjens,matsen,review}.
The SCF theory models a dense multi-component polymer 
mixture by an incompressible system of Gaussian chains with short-ranged binary 
interactions, and solves the statistical mechanics within the mean-field 
approximation.

Much of the success of the SCF theory can be traced back to the extended size of 
the polymer molecules: 

First,the large size of the chain molecules imparts a rather universal behavior
onto dense polymer mixtures, which can be characterized by only a small number
of coarse-grained parameters, \eg, the end-to-end distance $R_e$ of the
molecules, and the incompatibility per chain $\chi N$, where the Flory-Huggins
parameter $\chi$~\cite{FH} describes the repulsion between segments belonging to
different components, and $N$ is the number of segments per molecule.  
$R_e$ sets the characteristic length
scale of spatial inhomogeneities. 
Since this length scale is much larger than
the size of the repeat units along the backbone of the polymer chain, these
systems can be described successfully by coarse-grained chain models.  
The parameters, $R_e$ and $\chi N$, encode the chemical structure of the polymer 
on microscopic length scales. Indeed, comparisons between the predictions of the SCF theory
and experiments or simulations have shown that the properties of many blends
on large length scales depend on the atomistic structure only via the coarse-grained 
parameters $R_e$ and $\chi N$. The comparison works best if the parameters $R_e$  and 
$\chi$ are considered as ``black box'' input parameters. Their determination 
from first principles or even from a well-defined microscopic model (\eg, atomistic
force fields), is a formidable theoretical challenge: Even in the simplest case,
the Flory-Huggins parameter $\chi$ stems from small differences of dispersion 
forces between the different chemical segments. The value of the end-to-end distance $R_e$
of a polymer in a homogeneous melt results of a subtle screening of excluded volume
interactions along the chain by the surrounding molecules, and $R_e$ depends 
on the density and temperature. Thus $\chi$ and $R_e$ cannot be calculated rigorously.
If these coarse-grained parameters are determined independently (\eg, by experiments) 
and used as an input, the SCF theory is successful in making quantitative predictions.

These arguments rationalize the success of a coarse-grained approach. The second reason 
for the success of a the SCF theory, which is a mean-field theory, is the fact that 
due to their extended shape, the molecules have many interaction partners.
Let $\poldens$ denote the polymer number density. The
number of molecules inside the volume of a reference chain 
is then given by
$\sqrt{\N} = \poldens \red $ (where $d$ denotes the spatial dimension). 
This quantity measures the degree of interdigitation of the
molecules. In a dense three dimensional melt, ${\N}$ is proportional to the number of
segments $N$ per chain, and it is refered to as the invariant degree of polymerization~\cite{COMMENT}.
For systems which differ in ${\N}$ but are characterized 
by the same $\chi N$ and $R_e$ the SCF theory will make identical predictions. The
quantity ${\N}$ plays an important role as it controls the strength of fluctuations.

In this contribution, we shall discuss improvements which incorporate fluctuations into the SCF theory.
Those fluctuations are particularly important (i) in the vicinity of phase transitions 
(\eg, the unmixing transition in a binary blend or the onset of microphase separation 
in block copolymers) or (ii) at interfaces, where the translational symmetry is broken 
and the local position of the interface can fluctuate. Both type of fluctuations can 
qualitatively change the behavior compared to the prediction of mean-field theory.
The outline of the manuscript is as follows:
First we introduce the Edwards model, which is employed as a coarse-grained description of
a dense polymer melt throughout the article. Then, we briefly summarize the SCF theory for 
equilibrium properties and discuss various numerical strategies. Subsequently, we proceed to discuss
possibilities to go beyond the mean-field approximation and obtain a description of the
dynamics. Finally, we discuss some selected applications and close with a brief outlook
on interesting problems.

\section{The Edwards model for polymer blends}
\subsection{Gaussian model}
In dense binary blends, many of the interesting phenomena occur on length scales
much larger than the size of a monomeric unit. Hence, a theoretical description
can use a coarse-grained model, which only incorporates the relevant
ingredients that are necessary to bring about the observed universal behavior:
chain connectivity and thermal interaction between unlike monomeric units. The
universal properties on length scales much larger than the size of a monomeric
unit do not depend on the details of these interactions. One can therefore
choose a convenient mathematical model, the Edwards Hamiltonian~\cite{EDWARDS}, in which the
local microscopic structure enters only via three phenomenological parameters,
$R_e$, $\chi N$, and $\N$.

In a dense melt, the excluded volume of the monomeric units is screened and
chains adopt Gaussian conformations on large length scales. In the following, we
shall describe the conformations of a polymer as space curves ${\bf r}(\tau)$, where
the contour parameter $\tau$ runs from 0 to 1. The probability distribution ${\mathcal{P}}[{\mathbf{r}}]$ of such
a path ${\bf r}(\tau)$ is given by the Wiener measure
\begin{equation}
{\mathcal{P}}[{\mathbf{r}}(\tau)]\sim\exp\left[-\frac{3}{2R_e^2}
\int_{0}^{1}d\!\tau\left(\frac{d {\mathbf{r}}}{d\tau}\right)^2\right].
\label{eq:Wien_mes}
\end{equation}
This describes the Gaussian chain conformations of a non-interacting chain.  It
is characterized by a single length scale, $R_e$ the end-to-end distance of the
polymer chain. The structure is Gaussian on {\em all} length scales, \ie, the
model ignores the structure on short length scales~\cite{deGennes}: Self-avoiding walk
statistics inside the excluded volume blob, rod-like behavior on the length
scale of the persistence length, and the details of the monomeric
building units at even shorter length scales. All this structure on the
microscopic length scale enters the Gaussian chain model only through the
single parameter $R_e$. This parameter (and its dependence on the thermodynamic
state, \ie, temperature, density or pressure) cannot be predicted by the
theory but is used as input. The Gaussian model provides an appropriate
description if the smallest length scale $l_{\rm min}$ of interest in the calculations is
much larger than microscopic length scales. The coarse-graining length scale $l_{\rm min}$
should be at least on the order of a few persistence lengths. If this condition 
is violated, the Gaussian chain model can produce qualitatively incorrect 
predictions (\eg, at low temperatures where both the persistence length
length becomes large and the interface width becomes small). In this case,
other chain models have to be employed that take due account of the local
structure (\eg, the worm-like chain model~\cite{WORM}, which captures the crossover from
Gaussian conformations at large length scales to rod-like behavior on the
length scale of the persistence length, or enumeration schemes~\cite{SZLEIFER,MSROD,LV,MCONF} which
can deal with arbitrary chain architecture).

%
In the following we consider a polymer blend of homopolymers of species $A$ and
$B$ and/or copolymers containing both types of monomers. For simplicity,
we assume that segments of type $A$ and $B$ are perfectly symmetrical
and that all polymers have the same chain characteristics, in particular,
the same chain length. The generalization to more asymmetrical situations
is straightforward. We characterize the distribution of segments $A$ and $B$
along a chain of given type $J$ (corresponding to a $A$-homopolymer, a $B$-homopolymer or 
an $A:B$ block copolymer)
by the segment occupation functions $\gamma_{JA}(\tau)$, $\gamma_{JB}(\tau)$~\cite{MatsenRev},
which take the value 0 or 1 and
fulfill
\begin{equation}
\gamma_{JA}(\tau) + \gamma_{JB}(\tau) = 1
\qquad
\mbox{for all $J$ and $\tau$}.
\end{equation}
Combining these with the chain conformations $\rr (\tau)$, we can define
a microscopic density $\hat \phi_{\alpha}(\rr)$ for segments of type
$\alpha = A$ or $B$
\begin{equation} 
\hat{\phi}_{\alpha}=\frac{1}{\poldens}
\sum_J \sum^{n_J}_{i_J=1}\int_{0}^{1} {\rm d}\tau \;
\delta({\mathbf{r}}-{\mathbf{r}}_{i_J}(\tau)) \gamma_{J \alpha} (\tau)
\label{eq:mikro_dens}.
\end{equation}
Here the sum $i_J$ runs over all $n_J$ polymers of type $J$, and
${\bf r}_{i_J}(\tau)$ denotes the conformation of the $i_J^{\rm th}$ polymer.
The segment density is normalized by the polymer number density,
$\poldens \equiv \sum_J n_J/\V$. 

In addition to the chain connectivity, the coarse-grained model has to capture
the interactions between segments. In general, a compressible binary polymer
mixture~\cite{CMIX} exhibits both liquid-liquid immiscibility as well as liquid-vapor type
phase separation (cf.\ {\em Polymer+solvent systems: Phase diagrams,
interface free energies, and nucleation} in this issue). In the following, however,
we regard only liquid-liquid unmixing into 
$A$-rich and $B$-rich phases or microphase separation into domains comparable in size to the molecular extension.

To describe the interaction in the polymer liquid, the interaction potential 
between segments can be decomposed in a short-ranged repulsive part and a 
longer-ranged attractive part (\eg, using the Barker-Henderson scheme~\cite{BH} or the
Weeks-Chandler-Anderson decomposition~\cite{HMC}). The short-ranged repulsive contribution determines the 
packing and structure of the fluid. The length scale of fluctuations in the
total density is set by the excluded volume correlation length $\xi_{\rm ev}$, 
which also sets the length scale below which the chain conformations do not
obey Gaussian but rather self-avoiding walk statistics. In a dense melt,
$\xi_{\rm ev }$ is a microscopic length scale. On length scales larger 
than $\xi_{\rm ev}$ density fluctuations have already decayed, \ie, the
fluid is incompressible. For $\xi_{\rm ev} \ll l_{\rm min}$, we can represent 
the effect of the short-ranged repulsive interactions by an incompressibility 
constraint:
\begin{equation}
\hat \phi_A({\bf r}) + \hat \phi_B({\bf r}) = 1
\end{equation}
where we have assumed that both segments occupy identical volumes.
Of course, this incompressibility constraint can only be enforced after the 
coarse-graining procedure, \ie, on length scales larger than $\xi_{\rm ev}$.
The short-ranged repulsive interactions, hence, do not enter the Edwards Hamiltonian
explicitely, but only via the number density of polymers $\poldens$. The dependence
of $\poldens$ on the thermodynamic state (\ie, the temperature, pressure, and composition of
the mixture) -- the equation of state -- is not predicted by this computational scheme.

The longer-ranged attractive interactions between the segments do not strongly influence
the local fluid structure, but drive the phase separation. Again we assume that the
range of the attractive interactions is small compared to smallest length scale $l_{\rm min}$
of interest in our calculations. Then, the detailed shape of the longer-ranged potential 
does not matter either, and we can represent the interactions as zero-ranged. By virtue of the 
incompressibility constraint, there is only one independent bilinear form of densities
and so we use as interaction free energy the expression
\begin{equation}
k_BT \poldens \chi N \int \dr \; \hat \phi_A({\bf r})\hat \phi_B({\bf r})
\end{equation}
The combination of Flory-Huggins parameter and chain length, $\chi N$, parameterizes all the
subtle interplay between liquid-like structure of the polymeric liquid and the small differences
in the attractive interactions between segments of different types. 
It is worth noting that the Flory-Huggins parameter
for neutral polymers without specific interactions is on the order $10^{-4}-10^{-2}$~\cite{PH}, while the
strength of the attractive interactions in the fluid are of order 1 $k_B T$. Hence the $\chi$ parameter,
which describes the difference in attractive interactions between the species, arises from a
strong cancellation. Small changes in the thermodynamic state might perturb this balance. 
As a result, the $\chi$-parameter depends on the temperature, pressure and composition, 
as is often observed in experiments.

Within the coarse-grained model the canonical partition function of a polymer mixture with $n_J$ chains
of type $J$ takes the form:
\begin{eqnarray}
{\cal Z} &\sim & 
\frac{1}{\prod_J n_J!} \int \prod_{i=1}^n {\cal{D}}[{\mathbf{r}}_{i}] {\cal{P}}[{\mathbf{r}}_{i}] 
\exp\left[-\poldens \int \dr \; 
\chi N\hat{\phi}_A\hat{\phi}_B\right] \delta(\hat{\phi}_A+\hat{\phi}_B-1) \nonumber \\
	& \sim & 
\frac{1}{\prod_J n_J!} \int \prod_J \prod_{i_J=1}^{n_J} {\cal{D}}[{\mathbf{r}}_{i_J}] {\cal{P}}[{\mathbf{r}}_{i_J}] 
\exp\left[+\frac{\sqrt{\N} \chi N}{4 \red}\int \dr \; (\hat{\phi}_A-\hat{\phi}_B)^2 \right]
		      \nonumber \\
	&& \hspace*{1.5cm}      
		      \exp\left[-\frac{\sqrt{\N} \chi N}{4 \red}\int \dr \; (\hat{\phi}_A+\hat{\phi}_B)^2 \right] 
                      \delta(\hat{\phi}_A+\hat{\phi}_B-1) \
\label{eq:part_bulk}
\end{eqnarray}
The product $\prod_i$ runs over all $n = \sum n_J$ chains, and
the functional integral ${\mathcal{D}}[{\mathbf{r}}_i]$ sums over all possible conformations of the chain $i$. 
Within the Edwards Hamiltonian, the system is thus described by the following parameters: 
$\chi N$ describes the incompatibility of the different polymer species, 
$R_e$ sets the size of the molecules, 
$\N$ is the invariant chain length, which describes the degree of interdigitation of the molecules,
$\bar \phi_J \equiv n_J/\poldens \V$ is the average density of type $J$,
and $\gamma_{J\alpha}(\tau)$ the distribution of segments along the molecules of type $J$.
The quantity $\N$ sets the scale of the free energy, but does not influence the SCF solution
(\ie, the location of the extrema of the free energy, see \Sec~\ref{sec:ep}).

\subsection{Hubbard Stratonovich transformation and fluctuating fields}
The partition function (\ref{eq:part_bulk}) describes a system of mutually interacting chains.
Introducing auxiliary fields $U$ and $W$ via a Hubbard-Stratonovich transformation, one can 
decouple the interaction between the chains and rewrite the Hamiltonian in terms of independent 
chains in fluctuating fields. Then, one can integrate over the chain conformations and obtain 
a Hamiltonian which only depends on the auxiliary fields. Thermodynamic averages like density or
structure factors can be expressed as averages over the field $U$ and $W$ without approximation.

In the following we introduce a minimal transformation to decouple the interactions between
polymer chains. Two different schemes have to be employed for the thermal interactions between
the monomers and the incompressibility constraint. For the thermal interactions which give rise
to the term $(\hat \phi_A - \hat \phi_B)^2$ in the Hamiltonian, we use the Hubbard-Stratonovich
formula 
\begin{equation}
\exp[ \alpha x^2] = \frac{1}{2\sqrt{\pi \alpha}} \int_{-\infty}^{+\infty} {\rm d}y \; \exp \left[ - \left( \frac{y^2}{4\alpha} + x y \right) \right] 
\end{equation}
at each point in space and identify $x=\hat \phi_A - \hat \phi_B$, $y=\poldens W/2$, and 
$\alpha={\sqrt{\N} \chi N}/{4 \red}$. This introduces a real auxiliary field $W$.

To rewrite the incompressibility constraint, we use the Fourier representation of the $\delta$-function:
\begin{equation}
\delta(x-1) = \frac{1}{2\pi} \int_{-\infty}^{+\infty} {\rm d}y \; \exp \left[ - i y (x-1) \right]
\end{equation}
using $x=\hat \phi_A + \hat \phi_B$ and $y=\poldens U/2$. This introduces a real auxiliary field $U$.

The field $W$, which couples to the composition of the mixture, gives rise to a real contribution in the
exponent, while the field $U$, which enforces incompressibility, yields a complex contribution. 
Replacing the incompressibility constraint by a finite compressibility
\begin{equation}
\poldens N \frac{\kappa}{2} \int \dr \; (\hat \phi_A + \hat \phi_A -1)^2
\end{equation}
where $\kappa$ denotes the isothermal compressibility of the polymer liquid,
would also introduce a complex term in the exponential via the Hubbard-Stratonovich formula (cf.~\Sec~\ref{sec:FTMC}).

Using these two expressions we can exactly rewrite the partition function as
\begin{eqnarray}
{\cal Z} & \sim & 
\frac{1}{\prod_J n_J!}\int \prod_i {\cal{D}}[{\mathbf{r}}_{i}] {\cal{P}}[{\mathbf{r}}_{i}]
                           {\cal D}U[{\mathbf{r}}] {\cal D}W[{\mathbf{r}}]
\exp \left[ - \frac{\poldens \V \chi N }{4} \right] \nonumber \\
	 && \hspace*{.5cm} \exp \left[ - \poldens \int \dr  \; 
\left\{  \frac{W^2}{4 \chi N} + \frac{W}{2}(\hat \phi_A - \hat \phi_B) 
	 + \frac{i U}{2} \left( \hat \phi_A + \hat \phi_B - 1\right) \right\} \right] \nonumber \\
         &  \sim & \int {\cal{D}} U {\cal{D}} W \exp \left[ - \frac{\G[U,W]}{k_BT} \right],
\end{eqnarray}
which defines a Hamiltonian $\G[U,W]$~\cite{ELLEN_DIS,ELLEN1} for the fields $U$ and $W$,
\begin{equation}
\frac{\G[U,W]}{\sqrt{\N}k_BT (\V/\red)} = 
- \sum_J \bar \phi_J \ln \frac{\V Q[W_A,W_B]}{n_J}
  + \frac{1}{\V} \int \dr \; \frac{W^2}{4\chi N} .
		    \label{eqn:Gt1}
\end{equation}
Here $Q_J[W_A,W_B]$ denotes the partition function of a single noninteracting Gaussian chain of type $J$ 
\begin{eqnarray}
Q_J[W_A,W_B] &=& \frac{1}{\V} \int{\mathcal{D}}[{\mathbf{r}}]{\mathcal{P}}[{\mathbf{r}}]
\exp\left[-\int_0^{1}d\tau \; \sum_{\alpha=A,B} W_\alpha({\mathbf{r}}(\tau))\gamma_{J\alpha}(\tau) \right]
\label{eq:si_ch_part} \\
             &=& \frac{1}{\V} \int {\cal D}[{\bf r}]{\cal P}[{\bf r}] \; 
	     \exp \left[- \frac{1}{\V} \int \dr \; \sum_{\alpha} W_\alpha({\bf r}) \hat \phi^{sc}_{J\alpha}({\bf r}) \right]
\end{eqnarray}
in the fields
\begin{equation}
\label{eqn:wawb}
W_A = \frac{iU + W}{2} \qquad W_B = \frac{iU - W}{2}.
\end{equation}
Here we have normalized $Q_J$ by the volume $\V$ in order to make the trivial 
contribution of the translational entropy explicit. 
The dimensionless, microscopic single chain density is defined as
\begin{equation}
\hat \phi^{sc}_{J\alpha}({\bf r}) = \V \int {\rm d}\tau \; \delta({\bf r}-{\bf r}(\tau))
\label{eqn:scdens}
\end{equation}
where $\alpha=A$ or $B$ describes the segment species and $J$ denotes the type of polymer.
In Eq. (\ref{eqn:Gt1}), we have omitted a term 
$
- ({1}/{2\V}) \: \int \dr \; \left( iU - {\chi N}/{2}\right).
$
This is legitimated by the fact that $\G[U,W]$ is invariant under adding a spatially constant field 
$\Delta \bar U$ to $U$. Therefore, we can fix 
\begin{equation}
\frac{1}{\V}\int \dr \; iU({\bf r}) = \frac{\chi N}{2}.
\label{eqn:convention}
\end{equation}

To calculate the single chain partition function $Q_J$, it is useful to define the end segment 
distribution $q_J({\mathbf{r}},t)$, which describes the probability of finding the end of a chain 
fragment containing the segments $[0:t]$ and exposed to the fields $W_{\alpha} (\alpha = A,B)$
at position ${\mathbf{r}}$~\cite{helfand}: 
\begin{equation}
\label{eq:end_seg_dis}
q_J({\mathbf{r}},t)=\int{\mathcal{D}}[{\mathbf{r}}(t)]{\mathcal{P}}[{\mathbf{r}}(t)]
\delta({\mathbf{r}}(t)-{\mathbf{r}}) \: e^{-\int^t_0 d\tau 
\sum_{\alpha} W_{\alpha}({\mathbf{r}}(\tau)) \gamma_{J \alpha}(\tau)}
\end{equation} 
Similarly, we define
\begin{equation}
q_J^{\dagger}({\mathbf{r}},t)=\int{\mathcal{D}}[{\mathbf{r}}(t)]{\mathcal{P}}[{\mathbf{r}}(t)]
\delta({\mathbf{r}}(t)-{\mathbf{r}}) \: e^{-\int_t^1 d\tau 
\sum_{\alpha} W_{\alpha}({\mathbf{r}}(\tau)) \gamma_{J \alpha}(\tau)}.
\label{eq:end_seg_dis_d}
\end{equation} 
The end segment distributions obey the following diffusion equation~\cite{helfand}
\begin{eqnarray}
\label{eq:diff_eq_end}
\frac{\partial q_J({\mathbf{r}},t)}{\partial t} 
 &= &
\frac{R_e^2}{6}\nabla^2q_J({\mathbf{r}},t) 
- \sum_{\alpha=A,B} W_{\alpha} \gamma_{J \alpha}(t) q_J({\mathbf{r}},t) \\
\label{eq:diff_eqd_end}
-\frac{\partial q_J^{\dagger}({\mathbf{r}},t)}{\partial t}
& = &
\frac{R_e^2}{6}\nabla^2 q_J^{\dagger}({\mathbf{r}},t) 
- \sum_{\alpha=A,B} W_{\alpha} \gamma_{J \alpha}(t) q_J^\dagger({\mathbf{r}},t) 
\end{eqnarray}
with the boundary condition $q_J({\mathbf{r}},0)=1$ and $q_J^{\dagger}({\mathbf{r}},1) = 1$,
\ie, the beginning of the chain fraction is uniformly distributed. For homopolymers the
two propagators are related via $q_J({\bf r},t) = q_J^\dagger({\bf r},1-t)$.
The solution of these equations yields the single chain partition function:
\begin{equation}
Q_J 
= \frac{1}{\V} \int \dr \;q_J({\mathbf{r}},1) 
= \frac{1}{\V} \int \dr \;q_J^{\dagger}({\mathbf{r}},0) 
= \frac{1}{\V} \int \dr \;q_J({\mathbf{r}},t)q_J^{\dagger}({\mathbf{r}},t) \quad \forall t
\label{eq:si_ch_part_calc}
\end{equation}

We note that the fluctuations in the spatially homogeneous component 
$\bar W \equiv (1/\V) \int \dr \; W$ of the field $W$ are strictly Gaussian. 
To show this, we decompose $W = \bar W + W'$ and obtain:
\begin{eqnarray}
\frac{\G[U,W]}{\sqrt{\N}k_BT (\V/\red)} &=& \frac{\G[U,W']}{\sqrt{\N}k_BT (\V/\red)} \nonumber \\
&& + \frac{1}{4\chi N}\left[\bar W 
+ \chi N (\bar \phi_A- \bar \phi_B)\right]^2 - \frac{\chi N}{4}(\bar \phi_A- \bar \phi_B)^2
\end{eqnarray}

The only effect of a spatially homogeneous field acting on $A$-segments or $B$-segments 
is to introduce additional contributions to the chemical potentials $\mu_J$ --
quantities which are immaterial in the canonical ensemble.

By the Hubbard-Stratonovich transformation we have
rewritten the partition function of the interacting multi-chain systems in terms of non-interacting
chains in complex fluctuating fields $iU+W$ and $iU-W$. In field theoretical polymer simulations,
one samples the fields $U$ and $W$ via computer simulation using the above Hamiltonian (cf.~\Sec~\ref{sec:FTPS}).

To calculate thermal averages of $\hat \phi_A({\bf r}) - \hat \phi_B({\bf r})$ we introduce a local exchange
potential $\Delta \mu$ that couples to the local composition $\hat \phi_A - \hat \phi_B$~\cite{ELLEN1}:
\begin{eqnarray}
\tilde {\cal Z}[\Delta \mu] & \sim & \frac{1}{\prod_J n_J!}
\int \prod_i {\cal{D}}[{\mathbf{r}}_{i}] {\cal{P}}[{\mathbf{r}}_{i}] 
{\cal D}U[{\mathbf{r}}] {\cal D}W[{\mathbf{r}}] \nonumber \\
	 && \hspace*{.5cm} \exp \left[ - \poldens \int \dr \; \left\{  \frac{W^2}{4 \chi N} + \frac{W+\Delta \mu}{2}(\hat \phi_A - \hat \phi_B) 
					 + \frac{i U}{2} \left( \hat \phi_A + \hat \phi_B \right) \right\} \right] \nonumber \\
         &  \sim & \int {\cal{D}} U {\cal{D}} W \; \exp \left[ - \frac{\tilde \G[U,W,\Delta \mu]}{k_BT} \right] 
\end{eqnarray}
The free energy functional $\tilde \G[U,W,\Delta \mu]$ takes the form
\begin{eqnarray}
\frac{\tilde \G[U,W,\Delta \mu]}{\sqrt{\N} k_BT (\V/red)} 
 &=& - \sum_J \bar \phi_J\: \ln \frac{\V Q_J[\frac{1}{2}(iU+W+\Delta \mu),\frac{1}{2}(iU-W-\Delta \mu)]}{n_J} 
  \nonumber \\ && \qquad \qquad
     + \: \frac{1}{\V} \int \dr \; \frac{W^2}{4\chi N} \nonumber\\
		    \label{eqn:tG1} \nonumber \\
    &=& \frac{\G[U,W]}{\sqrt{\N} k_BT (\V/\red)} 
    + \frac{1}{4 \chi N \V} \int \dr \; (\Delta \mu^2 - 2 \Delta \mu W) 
		    \label{eqn:tG2}
\end{eqnarray}
where we have changed the dummy field $W$ to $W+\Delta \mu$ in the last step. Differentiating Eq.(\ref{eqn:tG2}) with respect to $\Delta \mu({\bf r})$ 
we obtain:
\begin{equation}
\langle \hat \phi_A({\bf r}) - \hat \phi_B({\bf r}) \rangle =
      - \frac{2}{\poldens} \left. \frac{1}{\tilde {\cal Z}[\Delta \mu]} 
         \frac{{\delta} \tilde {\cal Z}[\Delta \mu] }{{\delta} \Delta \mu ({\bf r}) }\right|_{\Delta \mu=0}  
      = - \frac{1}{\chi N} \langle W \rangle  
\label{eqn:Wav}  
\end{equation}
This expression relates the thermodynamic average of the composition difference to the average of
the real field $W$. Similarly, one can generate higher moments of the composition:
\begin{eqnarray}
\label{eqn:Wfluc}
\lefteqn{
\langle [\hat \phi_A({\bf r}) - \hat \phi_B({\bf r})][\hat \phi_A({\bf r'}) - \hat \phi_B({\bf r'})]\rangle}
\qquad &&\\
 &=& \frac{4}{\poldens^2} \frac{1}{\tilde {\cal Z}[\Delta \mu]} \left. \frac{{\delta}^2  
\tilde {\cal Z}[\Delta \mu] } {{\delta} \Delta \mu ({\bf r}) {\delta} \Delta \mu ({\bf r'})}\right|_{\Delta \mu=0}  
= - \frac{2 \red \delta ({\bf r}-{\bf r'})}{\sqrt{\N} \chi N}  
+ \frac{\langle W({\bf r})W({\bf r'})\rangle}{(\chi N)^2}  \nonumber
\end{eqnarray}
The two expressions (\ref{eqn:Wav}) and (\ref{eqn:Wfluc}) allow us to calculate the physically important 
thermodynamic average of the microscopic 
densities and their fluctuations from the thermodynamics average and fluctuations of the field $W$. 
Although the Hamiltonian $\G[U,W]$ is complex,
the thermodynamic average of the microscopic densities and their fluctuations can be expressed in terms of the
real field $W$~\cite{ELLEN1}.

Alternatively, one can calculate the average of the microscopic densities from Eq.(\ref{eqn:tG1}) and obtains
\begin{equation}
\langle \hat \phi_A({\bf r}) - \hat \phi_B({\bf r}) \rangle 
= \langle \phi^*_A[W_A,W_B] - \phi^*_B[W_A,W_B] \rangle,
\label{eqn:Pav}
\end{equation}
where the $\phi^*_\alpha$ ($\alpha = A,B$) are functionals of $W_A=(iU + W)/2$ and $W_B=(iU-W)/2$ 
(cf. Eq. (\ref{eqn:wawb})) given by
\begin{eqnarray}
\phi^*_{\alpha}({\bf r}) & = & - \frac{1}{\poldens} \sum_J n_J 
         \frac{\delta \ln \V Q_J[W_A,W_B]} {\delta W_{\alpha}(\rr)} 
\label{eqn:phistar}
\\
                  & = & 
\sum_J  \bar \phi_J \frac{\int{\mathcal{D}}[{\mathbf{r}}]{\mathcal{P}}[{\mathbf{r}}] 
      e^{-\frac{1}{\V} \int \dr \; \sum_{\alpha} W_\alpha({\bf r}) \hat \phi^{sc}_{J\alpha}({\bf r})}
      \V \int{\rm d}\tau\; \delta({\bf r}-{\bf r}(\tau)) \gamma_{J\alpha}(\tau))}
      {\int{\mathcal{D}}[{\mathbf{r}}]{\mathcal{P}}[{\mathbf{r}}] 
      e^{-\frac{1}{\V} \int \dr \; \sum_{\alpha} W_\alpha({\bf r}) \hat \phi^{sc}_{J\alpha}({\bf r})}} \nonumber \\
      &\equiv &  \sum_J \bar \phi_J  \langle \hat \phi_{J\alpha}^{sc} ({\bf r}) \rangle.
\end{eqnarray}
This equation identifies $\phi^*_{\alpha}$ as the density of $\alpha$-segments created
by a single chain in the external fields $W_A$ and $W_B$, averaged over all polymer types $J$.
(We note that the average normalized microscopic density of a single chain in
$\langle \hat \phi^{sc} \rangle$ in a volume $\V$ is independent from $\V$, cf.\ Eq.~(\ref{eqn:scdens}).)
The thermodynamic average of the composition is simply the Boltzmann average of the corresponding 
single chain properties averaged over the fluctuating fields. 
The functional derivatives $\phi^*_A({\bf r})$ and $\phi^*_B(\rr)$ can be calculated using 
the end-segment distribution functions, $q_J({\bf r},t)$ and $q_J^\dagger(\rr, t)$, according to:
\begin{equation}
\phi_{\alpha}^*({\mathbf{r}})=
\sum_J \bar \phi_J \frac{1}{Q_J}\int_0^1 {\rm d}t\; q_J({\mathbf{r}},t) \: 
q_J^{\dagger}({\mathbf{r}},t) \: \gamma_{J\alpha}(t)
\label{eq:dens_calc}
\end{equation}
While the thermodynamic average of the microscopic composition $\hat \phi_{\alpha}$ 
equals the average of the functional $\phi^*_{\alpha}$ over the fluctuating fields $W_A$ and $W_B$,
or, alternatively, $U$ and $W$, such a simple relation does not hold true for the fluctuations. 
Taking the second derivative with respect to $\Delta \mu$, we obtain:
\begin{eqnarray}
\lefteqn{
\langle [\hat \phi_A(\rr) - \hat \phi_B(\rr)][\hat \phi_A(\rr') - \hat \phi_B(\rr')]\rangle 
      = \langle [ \phi^*_A(\rr) - \phi^*_B(\rr)][\phi^*_A(\rr') - \phi^*_B(\rr')]\rangle 
} \qquad \qquad \qquad && \nonumber \\
&&
- \:\frac{\red}{\sqrt{\N}} \left\langle   
   \frac{{\delta}\phi^*_A({\bf r})}{{\delta}W_A({\bf r'})} 
+  \frac{{\delta}\phi^*_B({\bf r})}{{\delta}W_B({\bf r'})}   
-  \frac{{\delta}\phi^*_A({\bf r})}{{\delta}W_B({\bf r'})} 
-  \frac{{\delta}\phi^*_B({\bf r})}{{\delta}W_A({\bf r'})}   
\right\rangle  
\label{eqn:d1}
\end{eqnarray}
The fluctuations of the physically relevant microscopic monomer densities 
$ \hat \phi_A$ and $\hat \phi_B$ and the fluctuations of the Boltzmann
averaged single chain properties $\phi^*_A$ and $\phi^*_B$ due to fluctuations of 
the fields $U$ and $W$ are not identical.
The additional term accounts for the single chain correlations~\cite{COMMENTW}:
\begin{equation}
\red \frac{{\delta}\phi^*_{\alpha}({\bf r})}{{\delta}W_{\beta}({\bf r'})} = 
\sum_J \frac{\bar \phi_J \red}{\V}
\left(  \langle \hat \phi_{J \alpha}^{sc} ({\bf r})\hat \phi_{J \beta}^{sc} ({\bf r'}) \rangle
  - \langle \hat \phi_{J \alpha}^{sc} ({\bf r}) \rangle  \langle \hat \phi_{J \beta}^{sc} ({\bf r'}) \rangle   
\right).
\label{eqn:commentw}
\end{equation}
We emphasize that the additional term due to single chain correlations is of the same order of magnitude 
than the fluctuations of the physically relevant microscopic monomer densities itself. We shall
demonstrate this explicitly in the framework of the Random-Phase Approximation (RPA) in \Sec~\ref{sec:RPA}.

To calculate thermal averages of $\hat \phi_A({\bf r}) + \hat \phi_B({\bf r})$ we introduce a spatially varying total chemical potential $\delta \mu$
\begin{eqnarray}
\tilde {\cal Z}[\delta \mu] & \sim & \frac{1}{\prod_J n_J!}
\int \prod_i {\cal{D}}[{\mathbf{r}}_{i}] {\cal{P}}[{\mathbf{r}}_{i}] 
             {\cal D}U[{\mathbf{r}}] {\cal D}W[{\mathbf{r}}]
\nonumber \\
	 && \hspace*{.5cm} \exp \left[ - \poldens \int \dr \; \left\{  \frac{W^2}{4 \chi N} + \frac{W}{2}(\hat \phi_A - \hat \phi_B) 
					 + \frac{i U + \delta \mu}{2} \left( \hat \phi_A + \hat \phi_B\right) \right\} \right] \nonumber \\
         &  \sim & \int {\cal{D}} U {\cal{D}} W \; \exp \left[ - \frac{\tilde \G[U,W,\delta \mu]}{k_BT} \right] 
\end{eqnarray}
Similar to Eqs.(\ref{eqn:tG1}) and (\ref{eqn:tG2}) we can rewrite the free energy functional $\tilde \G[U,W,\delta \mu]$ as:
\begin{eqnarray}
\frac{\tilde \G[U,W,\delta \mu]}{\sqrt{\N} k_BT (\V/\red)} &=& 
\frac{\G[U,W]}{\sqrt{\N} k_BT (\V/\red)} + \frac{1}{2 \V} \int \dr \; \delta \mu  \\
 &=& - \sum_J \bar \phi_J \ln \frac{\V Q_J}{n_J} + \frac{1}{\V} \int \dr \;  \frac{W^2}{4\chi N}
\end{eqnarray}
The first derivative wrt $\delta \mu$ yields the average of the local monomer density
\begin{eqnarray}
\langle \hat \phi_A({\bf r}) + \hat \phi_B({\bf r}) \rangle 
  &=&  - \frac{2}{\poldens} \left. \frac{1}{\tilde {\cal Z}[\delta \mu]} 
  \frac{{\delta} \tilde {\cal Z}[\delta \mu] }{{\delta} \delta \mu ({\bf r}) }\right|_{\delta \mu=0}  
   = \langle \phi^*_A + \phi^*_B \rangle \\
    &=& 1 
\end{eqnarray}
The second derivative yields the density fluctuations:
\begin{eqnarray}
\lefteqn{
\langle [\hat \phi_A(\rr) + \hat \phi_B(\rr)][\hat \phi_A(\rr') + \hat \phi_B(\rr')]\rangle 
  = \frac{4}{\poldens^2} \frac{1}{\tilde {\cal Z}[\delta \mu]} \left. 
  \frac{{\delta}^2  \tilde {\cal Z}[\delta \mu] } 
       {{\delta} \delta \mu ({\bf r}) {\delta} \delta \mu ({\bf r'})}\right|_{\delta \mu=0}  
} 
\qquad && \nonumber\\
   & = &
\langle [ \phi^*_A({\bf r}) + \phi^*_B({\bf r})][\phi^*_A({\bf r'}) + \phi^*_B({\bf r'})]\rangle
 \nonumber \\
&& - \: \frac{\red}{\sqrt{\N}} \left\langle   
  \frac{{\delta}\phi^*_A({\bf r})}{{\delta}W_A({\bf r'})} 
+ \frac{{\delta}\phi^*_B({\bf r})}{{\delta}W_B({\bf r'})}   
+ \frac{{\delta}\phi^*_A({\bf r})}{{\delta}W_B({\bf r'})} 
+ \frac{{\delta}\phi^*_B({\bf r})}{{\delta}W_A({\bf r'})}   
\right\rangle  
\label{eqn:d2}\\
     &=& 1 
\end{eqnarray}
Note that the last contribution in the equation above is similar to the single chain correlations in
Eq.~(\ref{eqn:d1}). The incompressibility constraint is enforced on the
microscopic density $\hat \phi_A + \hat \phi_B$. At this stage, $\phi^*_A$ and
$\phi^*_B$ are only auxiliary functionals of the fields $U,W$ proportional to
the density distribution of a single chain in the external fields.

\section{External potential (\EP ) theory and self-consistent field (SCF) theory}
\label{sec:ep}
The reformulation of the partition function in terms of single chains in the
fluctuating, complex fields $W_A = (iU+W)/2$ and $W_B = (iU-W)/2$ is exact.  The 
numerical evaluation of the functional integral over the fields $U$ and $W$ is, however,
difficult (cf.~\Sec~\ref{sec:FTPS}), and various approximations have been devised. 

The functional integral over the fluctuating field $U$, conjugated to the
total density, can be approximated by the ``saddle point'': The integrand evaluated at 
that function $U^*[W]$ which minimizes the free energy function $\G[U,W]$. 
Carrying out this saddle point integration in the field $U$, we
obtain a free energy functional ${\cal H}[W]$. In the following we denote this
external potential (\EP ) theory , following Maurits and Fraaije who derived the
saddle point equations heuristically~\cite{maurits}. This scheme still retains the important
fluctuations in the field $W$, conjugated to the composition. Using an additional
saddle point approximation for $W$, we neglect fluctuations in the composition
and we arrive at the self-consistent field theory.

\subsection{External potential (\EP ) theory: saddle point integration in $U$}
\label{sec:EPT}
The fluctuations described by the two fields $U$ and $W$ are qualitatively
different. This is already apparent from the fact that one, $U$, gives rise to a
complex contribution to the field that acts on a chain, while the other, $W$, corresponds
to a real one. The field $U$ couples to the total density $\hat \phi_A+\hat \phi_B$ and
has been introduced to decouple the incompressibility constraint. Qualitatively,
it controls the fluctuations of the total density, and it does not directly influence
the expectation value or the fluctuations of the composition $\hat \phi_A-\hat \phi_B$.

The saddle point value $U^*$ is given by the condition:
\begin{equation}
\frac{{\delta}G[U^*,W]}{{\delta}U}=0 \quad \Rightarrow \quad \phi_A^*+\phi_B^*=1.
\label{eqn:Ustar}
\end{equation}
Instead of enforcing the incompressibility constraint on each microscopic conformations,
we thus only require that the single chain averages in the external field obey the constraint. 
We recall that the $\phi_{\alpha}^*$ are functionals of $W_A = (i U + W)/2$ and 
$W_B = (i U - W)/2$ (cf.\ Eq.~(\ref{eqn:phistar})), thus Eq.~(\ref{eqn:Ustar}) implicitly 
defines a functional $U^*[W]$. 
Substituting the saddle point value into the free energy functional (\ref{eqn:Gt1}), 
we obtain an approximate partition function
\begin{equation}
{\cal Z}_{\rm \EP} \sim \int {\cal D}W \; \exp 
\left[ - \frac{{\cal H}[W]}{k_BT}\right] 
\end{equation}
with the free energy functional
\begin{eqnarray}
\frac{{\cal H}[W]}{\sqrt{\N}k_BT (\V/\red)} &=& \frac{\G[U^*,W]}{\sqrt{\N}k_BT (\V/\red)} \nonumber \\
&=& - \sum_J \bar \phi_J\ln \frac{\V Q_J}{n_J} + \frac{1}{\V} \int \dr \; \frac{W^2}{4\chi N}
		    \label{eqn:H}
\end{eqnarray}
At this point, one has two possible choices for calculating the thermodynamic averages of the microscopic composition~\cite{ELLEN1}:
\begin{enumerate}
\item Although one performs a saddle point approximation in $U$, one can use the expressions (\ref{eqn:Wav}) and (\ref{eqn:Wfluc}), 
which relate the average of the field $W$ and its fluctuations to the average and the fluctuations of
the microscopic composition. The crucial difference is, however,
that the fluctuations of the field $W$ are now governed by the Hamiltonian ${\cal H}[W]$ of the \EP~theory 
and not by the exact Hamiltonian $\G[U,W]$. If the saddle point approximation is reliable, $\G[U,W]$ will be well described by
a parabola in $U-U^*[W]$, the fluctuations in $U$ will be nearly Gaussian and will hardly influence the fluctuations of $W$. 
Therefore, the fluctuations of $W$ with respect to the Hamiltonian 
${\cal H}[W]$ of the \EP~theory closely mimic the fluctuations of the field $W$ with respect
to the exact Hamiltonian 
$\G[U,W]$, \ie, $\langle W({\bf r})W({\bf r}')\rangle_{\rm EP} \approx \langle W({\bf r})W({\bf r}')\rangle$. By the same token, using 
Eq. (\ref{eqn:Wfluc}), we ensure that the composition fluctuations in the \EP~theory closely resemble
the exact thermodynamic average of fluctuations of the microscopic composition.\\
\item One can introduce a spatially varying exchange potential $\delta \mu$ and calculate the thermodynamic averages 
with respect to the free energy functional ${\cal H}[W]$ of the \EP~theory. This leads to expressions similar to (\ref{eqn:Pav}) 
and (\ref{eqn:d1}):
\begin{equation}
\langle \hat \phi_A({\bf r}) - \hat \phi_B({\bf r}) \rangle_{\rm \tiny \EP} 
      = \langle \phi^*_A - \phi^*_B \rangle_{\rm EP} \label{eqn:Pav_EPD} 
\end{equation}
\begin{eqnarray}
\lefteqn{
\langle [\hat \phi_A(\rr) - \hat \phi_B(\rr)][\hat \phi_A(\rr') - \hat \phi_B(\rr')]\rangle_{\rm \tiny \EP}
}\qquad \qquad \label{eqn:d1_EPD} \\ 
 & =& \langle \phi^*_A(\rr) - \phi^*_B(\rr)][\phi^*_A(\rr') - \phi^*_B(\rr')]\rangle_{\rm EP}  \nonumber\\
&&
      - \: \frac{\red}{\sqrt{\N}} \left\langle   
        \frac{{\delta}\phi^*_A(\rr)}{{\delta}W_A(\rr')} 
     +  \frac{{\delta}\phi^*_B(\rr)}{{\delta}W_B(\rr')}   
     -  \frac{{\delta}\phi^*_A(\rr)}{{\delta}W_B(\rr')}   
     -  \frac{{\delta}\phi^*_B(\rr)}{{\delta}W_A(\rr')}   
\right\rangle_{\rm EP}  \nonumber
\end{eqnarray}
We refer to the last expression as the literal composition fluctuations in the \EP~theory. Likewise, one can use the functional
derivative with respect to a spatially varying chemical potential $\Delta \mu$, which couples to $\hat \phi_A+\hat \phi_B$,
to calculate the literal average of the total density.
\begin{equation}
\langle \hat \phi_A({\bf r}) + \hat \phi_B({\bf r}) \rangle_{\rm \tiny \EP} = 
       \langle \phi^*_A + \phi^*_B \rangle_{\rm EP} = 1 
\end{equation}
For the literal fluctuations of the total density in the \EP~theory (cf.\ Eq.(\ref{eqn:d2})) one obtains:
\begin{eqnarray}
\lefteqn{
\langle [\hat \phi_A(\rr) + \hat \phi_B(\rr)][\hat \phi_A(\rr') + \hat \phi_B(\rr')]\rangle_{\rm \tiny \EP } 
} \qquad \qquad \\
\label{eqn:d2_EPD}
 &=& \langle \underbrace{[ \phi^*_A({\bf r}) + \phi^*_B({\bf r})]}_{=1}
                 \underbrace{[\phi^*_A({\bf r'}) + \phi^*_B({\bf r'})]}_{=1}\rangle_{EP} \nonumber \\
&&
     - \: \frac{\red}{\sqrt{\N}} \left\langle   
        \frac{{\delta}\phi^*_A(\rr)}{{\delta}W_A(\rr')} 
     +  \frac{{\delta}\phi^*_B(\rr)}{{\delta}W_B(\rr')}   
     +  \frac{{\delta}\phi^*_A(\rr)}{{\delta}W_B(\rr')}   
     +  \frac{{\delta}\phi^*_B(\rr)}{{\delta}W_A(\rr')}   
\right\rangle_{EP}  \nonumber
\end{eqnarray}
This demonstrates that the saddle point approximation in $U$ enforces the incompressibility constraint only on average,
but the literal fluctuations of the total density in the \EP~theory do not vanish.
\end{enumerate}
Unfortunately, the two methods for calculating composition fluctuations yield different results. Of course,
after the saddle point approximation for the field $U$ is is not {\em a priory} obvious to which extent the 
\EP~theory can correctly describe composition fluctuations. In the following we shall employ Eqs. (\ref{eqn:Wav}) 
and (\ref{eqn:Wfluc}) to calculate the thermal average of the composition and its fluctuations in the \EP~theory. As
we have argued above, these expressions will converge to the exact result if the fluctuations in $U$ become Gaussian.

\subsection{Self-consistent field (SCF) theory: saddle point integration in $U$ and $W$}
\label{sec:SCF}
In the self-consistent field (SCF) theory one additionally approximates the functional 
integral over the field $W$ by its saddle point value.
Using the functionals 
\begin{displaymath}
\phi_{\alpha}^*[W_A,W_B] = \phi_{\alpha}^*\Big[\frac{1}{2}(i U^*[W]+W), \frac{1}{2}(i U^*[W] - W)\Big]
\end{displaymath}
(cf.~Eq.~(\ref{eqn:phistar})), the derivative of the  
free energy functional ${\cal H}[W]$ (Eq. (\ref{eqn:H})) can be written as
\begin{equation}
\frac{\delta {\cal H}[W]}{{\delta}W({\bf r})} =
\frac{\sqrt{\N}k_BT}{R_e^d}\frac{W+\chi N(\phi_A^* - \phi_B^*)}{2 \chi N}
\end{equation}
where we have used the convention (\ref{eqn:convention}).
Therefore the saddle point condition simply reads:
\begin{equation}
W^*({\bf r})  = - \chi N(\phi_A^*-\phi_B^*) 
\label{eqn:Wstar}
\end{equation}
where $\phi_A^*$ and $\phi_B^*$ are functionals of the fields $U$ and $W$.

The two equations (\ref{eqn:Ustar}) and (\ref{eqn:Wstar}) define a self-consistent set of equations for the two fields
$W^*$ and $U^*$. Knowing the fields $W^*$ and $U^*$, one can calculate the thermal average of the microscopic 
composition via:
\begin{equation}
\langle \hat \phi_A - \hat \phi_B \rangle_{\rm \tiny SCF} = - \frac{W^*}{\chi N} = \phi_A^*-\phi_B^*
\end{equation}

\subsection{Alternative derivation of the SCF theory}
The SCF equations are often derived in a slightly different way by additionally introducing auxiliary fields
also for the microscopic densities. For completeness and further reference, we shall present this derivation here. 

The starting point is the partition function, Eq.~(\ref{eq:part_bulk}). Then, one proceeds to introduce 
collective densities $\Phi_A$, $\Phi_B$,
fields $\Omega_A$, $\Omega_B$, and a pressure field $U$ to rewrite the partition function in the form:
\begin{eqnarray}
{\cal Z} &\sim & \frac{1}{\prod_J n_J!}\int \prod_i {\cal{D}}[{\mathbf{r}}_{i}] {\cal{P}}[{\mathbf{r}}_{i}] 
		     {\cal{D}}\Phi_A {\cal{D}}\Phi_B 
\nonumber \\
	 && \hspace*{1.5cm} \exp\left[-\poldens\int \dr \; 
              \Big\{ \chi N{\Phi}_A{\Phi}_B\right] \delta({\Phi}_A+{\Phi}_B-1) 
	                    \delta(\Phi_A-\hat{\phi}_A) \delta(\Phi_B-\hat{\phi}_B)
\nonumber \\
         &\sim & \frac{1}{\prod_J n_J!}\int \prod_i {\cal{D}}[{\mathbf{r}}_{i}] {\cal{P}}[{\mathbf{r}}_{i}]
		     {\cal{D}}\Phi_A {\cal{D}}\Phi_B {\cal{D}}\Omega_A {\cal{D}}\Omega_B {\cal{D}}U \nonumber\\
 && \hspace*{1.5cm}\exp\left[-\poldens\int \dr \; \left\{\chi N{\Phi}_A{\Phi}_B  \frac{iU}{2} \left( \Phi_A + \Phi_B - 1\right) \right\}\right]
\nonumber \\
	 && \hspace*{1.5cm} 
	                    \exp \left[  - \poldens \int \dr \; \left\{
	                      + i \Omega_A \left( \hat \phi_A - \Phi_A\right) 
	                      + i \Omega_B \left( \hat \phi_B - \Phi_B\right)   \right\}  \right]
\nonumber \\
        &\sim & \int {\cal{D}}\Phi_A {\cal{D}}\Phi_B {\cal{D}}\Omega_A {\cal{D}}\Omega_B {\cal{D}}U \; \exp \left[-\frac{{\cal F}[\Phi_A,\Phi_B,\Omega_A,\Omega_B,U]}{k_BT} \right]
\label{eq:part_new}
\end{eqnarray}
where we have used the Fourier representation of the $\delta$-functions.
The free energy functional ${\cal F}[\Phi_A,\Phi_B,\Omega_A,\Omega_B,U]$ is given by:
\begin{eqnarray}
\lefteqn{
\frac{{\cal F}[\Phi_A,\Phi_B,\Omega_A,\Omega_B,U]}{\sqrt{\N}k_BT (\V/\red)} 
    = - \sum_J \bar \phi_J \ln \frac{\V Q_J[i\Omega_A,i\Omega_B]}{n_J}
} && \nonumber \\
     && + \frac{1}{\V} \int \dr \;  
\left( \chi N \Phi_A \Phi_B + \frac{iU}{2} \left( \Phi_A + \Phi_B - 1\right)
                                   - i \Omega_A \Phi_A - i \Omega_B \Phi_B \right)
\end{eqnarray}
The saddle point approximation of the partition functions yield the SCF theory. Let us first make the saddle point approximation
in the fields. Like before, we denote the values of the fields at the saddle point by an asterisk.
\begin{eqnarray}
\frac{\delta{\cal F}}{\delta \Omega_A}=0 \qquad & \Rightarrow&  \qquad
\Phi_A = - \sum_J \bar \phi_J\: \frac{1}{Q_J} \V \frac{\delta  Q_J}{\delta i \Omega_A} = \phi_A^*\nonumber \\
\frac{\delta{\cal F}}{\delta \Omega_B}=0 \qquad & \Rightarrow& \qquad
\Phi_B = - \sum_J \bar \phi_J \: \frac{1}{Q_J}\V \frac{\delta  Q_J}{\delta i \Omega_B} = \phi_B^*\nonumber \\
\frac{\delta{\cal F}}{\delta U}=0 & \Rightarrow& \Phi_A + \Phi_B = 1
\label{eqn:omega}
\end{eqnarray}
Note that the saddle point values, $i\Omega_A^*$ and $i\Omega_B^*$ are real.
These equations correspond to Eq.~(\ref{eqn:Ustar}).
Inserting these saddle point values into the free energy functional, we obtain 
a functional that depends only on the composition $\Phi_A$: 
\begin{eqnarray}
\lefteqn{
\frac{{\cal F}^*[\Phi_A]}{\sqrt{\N}k_BT 
(\V/\red)} \equiv  \frac{{\cal F}[\Phi_A,1-\Phi_A,\Omega_A^*,\Omega_B^*,U^*]}{\sqrt{\N}k_BT (\V/\red)} 
} 
\nonumber \\
    &=& - \sum_J \bar \phi_J \ln \frac{\V Q_J[i \Omega_A^*[\Phi_A], i\Omega_B^*[\Phi_A]]}{n_J} \nonumber \\
    && + \frac{1}{\V} \int \dr \;  \left\{ \chi N \Phi_A (1-\Phi_A) - i \Omega_A^*[\Phi_A] \Phi_A - i \Omega_B^*[\Phi_A] (1-\Phi_A) \right\}
\label{eqn:dscf}
\end{eqnarray}
This functional can be used to 
investigate the dynamics of collective composition fluctuations (cf.~\Sec~\ref{sec:dynamics}). 
If we proceed to make a saddle point approximation
for the collective density $\Phi_A$, we arrive at 
\begin{equation}
\frac{{\cal F}^*[\Phi_A]}{\delta \Phi_A}=0 \Rightarrow i(\Omega_A^*-\Omega_B^*) + \chi N (\Phi_A^*-\Phi_B^*) = 0
\end{equation}
which is equivalent to Eq.~(\ref{eqn:Wstar}) when we identify $W^*=i\Omega_A^*-i\Omega_B^*$.

\subsection{Numerical techniques}

Both in the SCF theory and in the \EP~theory, 
the central numerical task is to solve a closed set of self-consistent 
equations one or many times. This is usually achieved according to the 
general scheme:

\begin{enumerate}
\item[(0)] Make an initial guess for the fields $U$ or $(W,U)$.
\item[(1)] From the fields, calculate the propagators $q$ and $q^{\dagger}$
  (Eqs. (\ref{eq:diff_eq_end} and (\ref{eq:diff_eqd_end})).
\item[(2)] From the propagators, calculate a new set of fields.
\item[(3)] Mix the old fields and the new fields according to
  some appropriate prescription.
\item[(4)] Go back to (1). Continue this until the fields
  have converged to some value within a pre-defined accuracy.
\end{enumerate}

The numerical challenge is thus twofold. First, one needs an
efficient iteration prescription. Second, one needs a 
fast method to solve the diffusion equation. 

We will begin with presenting some possible iteration prescriptions. 
Our list is far from complete. We will not derive the methods, 
nor discuss the numerical background and the numerical stability.
For more information, the reader is referred to the references.

We consider a general nonlinear set of equations of the form
\begin{equation}
\ggg (\xx) = \xx.
\end{equation}
The value of $\xx$ after the $n$th iteration step is denoted $\xx_n$. 

One of the most established methods to solve nonlinear systems
of equation, which has often been used to solve SCF equations,
is the Newton-Raphson method~\cite{numerical_recipes}.
In this scheme, one calculates not only $\ggg(\xx_n)$ from a given 
$\xx_n$, but also
the whole Jacobian matrix $ \GG_n = \frac{\delta \ggg}{\delta \xx'}$.
The $(n+1)$th iteration is then given by
\begin{equation}
\xx_{n+1} = \xx_n + [ \one - \GG_n ]^{-1} (\ggg(\xx_n) - \xx_n).
\end{equation}
If this method converges at all, \ie, if the initial
guess is not too far from the solution, it usually requires very few
iteration steps. Unfortunately, it has the drawback that the Jacobian
must be evaluated in every iteration step. If one has $m$ independent 
parameters (fields), one must solve $2 m$ diffusion equations. 
Therefore, the method becomes inefficient if the number of degrees 
of freedom $m$, \ie, the dimension of $\xx$, is very large.

To speed up the calculations, one can use the information on the
Jacobian from previous steps and just iterate it. This is the
idea of Broyden's method~\cite{numerical_recipes} and other gradient-free methods~\cite{QNEWTON}. Given
the Jacobian $\GG_n$ for one step, the Jacobian for the next
step $\GG_{n+1}$ is approximated by
\begin{equation}
\GG_{n+1} - \GG_n \approx 
\frac{[\delta \ggg_n - \GG_n \delta \xx_n]}{(\delta \xx_n)^2}
\otimes \delta \xx_n 
=
\frac{[\ggg(\xx_{n+1}) - \xx_{n+1}]}{(\delta \xx_n)^2}
\otimes \delta \xx_n 
\end{equation}
Here we have used the short notation $\delta \xx_n = \xx_{n+1}-\xx_n$ 
and $\delta \ggg_n = \ggg(\xx_{n+1})-\ggg(\xx_n)$, and $\otimes$ denotes
the tensor product. The full Jacobian has to be calculated in the 
first step, and must possibly be updated after a number of steps. 
The method is a major improvement over Newton-Raphson, 
but it still requires the evaluation of at least one Jacobian.

A very simple scheme which uses little computer time for one iteration 
is simple mixing. The $(n+1)$th guess $\xx_{n+1}$ is simply given by
\begin{equation}
\xx_{n+1} = \xx_n + \lambda (\ggg(\xx_n) - \xx_n)
\end{equation}
with some appropriate $\lambda$. Single iterations are thus cheap, 
but the method does not converge very well and many iteration steps 
are needed. The optimal values of $\lambda$ have to be established
empirically (typically of the order of 0.1 or less). 

The simple mixing method can be improved by using information from
previous steps to adjust $\lambda$. 
For example, the results from the two previous steps can be used to
determine a new value for $\lambda_n$ at every step~\cite{scft}
\begin{equation}
\lambda_n = \sqrt{\frac{(\xx_n - \xx_{n-1})^2}
{(\ggg(\xx_n)-\xx_n - \ggg(\xx_{n-1}) + \xx_{n-1})^2}}.
\end{equation}

A more systematic approach is ``Anderson mixing'', which is
another standard method to solve systems of nonlinear equations
in high dimensions~\cite{anderson,eyert,russell}. From the remaining
deviations $\dd_n = \ggg(\xx_n) - \xx_n$ after $n$ steps,
one first determines
\begin{eqnarray}
U_{ij} & = & (\dd_n - \dd_{n-i})(\dd_n - \dd_{n-j}) \\
V_{j} & = & (\dd_n - \dd_{n-j})\dd_n  
\end{eqnarray}
and then calculates
\begin{eqnarray}
\xx_n^A & = & \xx_n + \sum_{i,j} U_{ij}{}^{-1} V_j (\xx_{n-i}-\xx_n)
\\
\ggg_n^A & = & \ggg(\xx_n) + 
\sum_{i,j} U_{ij}{}^{-1} V_j (\ggg(\xx_{n-i})-\ggg(\xx_n)).
\end{eqnarray}
In the original Anderson mixing scheme, the $(n+1)$th guess 
$\xx_{n+1}$ is given by $\xx_{n+1} = \ggg_n^A$.
More generally, the method can be combined with simple mixing
\begin{equation}
\xx_{n+1} = \xx_n^A + \lambda (\ggg_n^A - \xx_n^A).
\end{equation}

This summarizes some methods which can be used to solve the 
self-consistent equations. Regardless of the particular choice,
the single most time-consuming part in every SCF or \EP~calculation 
is the repeated solution of the diffusion equation 
(Eqs. (\ref{eq:diff_eq_end}) and (\ref{eq:diff_eqd_end})). 
Therefore the next task is to find an efficient 
method to solve equations of the type
\begin{equation}
\label{eq:diff_real}
\frac{\partial q(\rr,t)}{\partial t} 
= \frac{R_e^2}{6} \Delta q(\rr,t) - W (\rr) q(\rr,t).
\end{equation}
with the initial condition $q(\rr,0) \equiv 1$.
Unfortunately, the two terms on the right hand side of this equation
are incompatible. The Laplace operator $\Delta$ is best treated in Fourier 
space.  The contribution of the field $W(\rr)$ can be handled much more 
easily in real space.

In SCF calculations of equilibrium mean-field structures,
it is often convenient to operate in Fourier space, because crystal
symmetries can be exploited. Relatively few Fourier components 
are often sufficient to characterize the structure of a phase 
satisfactorily. In Fourier space, Eq.~(\ref{eq:diff_real})
turns into a matrix equation
\begin{equation}
\label{eq:diff_fourier}
\frac{\partial q(\kk,t)}{\partial t} 
= - \sum_{\kk,\kk'} {\bf A}_{\kk,\kk'}(t) q(\kk',t)
\end{equation}
with
\begin{equation}
{\bf A}_{\kk,\kk'} = \frac{R_e^2}{6} k^2 \delta_{\kk,\kk'} + W (\kk,\kk') 
\end{equation}
and the initial condition $\qq(\kk,0) = \delta(\kk,0)$.
The formal solution of this equation is
\begin{equation}
q(\kk,t) = \sum_{\kk'} [\exp(- {\bf A} t)]_{\kk,\kk'} \: q(\kk',0).
\end{equation}
It can be evaluated by diagonalizing the matrix ${\bf A}$.
The approach has the advantage that the diffusion equation 
is solved exactly in time, \ie, one has no  errors 
due to a time discretization.
Unfortunately, the efficiency drops rapidly if the number of 
Fourier components and hence the dimension of the matrix ${\bf A}$
is very large. This hampers studies of fluctuations and dynamic phenomena, 
because one can no longer exploit crystal symmetries of the bulk structure, 
to reduce the number of variables.
It is also an issue in SCF calculations
if the symmetry of the resulting structure is not yet 
known~\cite{drolet}.

Therefore real space methods have received increased interest in
recent years~\cite{drolet}. In real space, the number of degrees
of freedom $m$ can be increased more easily (the computing
time typically scales like $m$ or $m \ln m$).
However, the time integration 
of the diffusion equation can no longer be performed exactly.

Several different integration methods are available in the literature. 
Many of them approximate the effect of the Laplace operator 
by finite differences~\cite{numerical_recipes,fdm}. 
Among the more elaborate schemes of this kind we mention the
Crank-Nicholson scheme~\cite{numerical_recipes} and the
Dufort-Frankel scheme~\cite{fdm}. 

In our own applications, we found that a different 
approach was far superior to finite difference methods: The
pseudo-spectral split-operator scheme~\cite{feit,rasmussen}.
This method combines Fourier space and real space steps,
treating the contribution of the fields $W(\rr)$ in real space, 
and the contribution of the Laplace operator in Fourier space. 
In that approach, the dynamical evolution during a single
discrete time step of length $h$ is formally described by
\begin{equation}
q(\rr,t+h) = 
e^{-W(\rr) h/2} \: e^{h \Delta} \: e^{-W(\rr) h/2}  \: q(\rr,t),
\end{equation}
where the effect of the exponential $\exp(h \Delta)$ is
evaluated in Fourier space. Taking as starting point
a given $q(\rr,t)$, the different steps of the algorithm are
\begin{enumerate}
\item[1.] Calculate
$
 q'(\rr,t) = \exp(- W(\rr) h/2) \: q(\rr,t)
$
in real space for each $\rr$.
\item[2.] 
Fourier transform
$
 q'(\rr,t)  \rightarrow q'(\kk,t)
$
\item[3.]
Calculate
$
 q''(\kk,t) = \exp(- h \Delta) \: q'(\kk,t)
 =\exp(- h (2 \pi \kk/L)^d)  \: q'(\kk,t)
$,
where $d$ is the dimension of space and $L$ the system size.
\item[4.] 
Inverse Fourier transform
$
 q''(\kk,t)  \rightarrow q''(\rr,t)
$
\item[5.] Calculate
$
 q(\rr,t+h) = \exp(- W(\rr) h/2) \: q''(\rr,t)
$
in real space for each $\rr$.
\item[5.]
Go back to 1.
\end{enumerate}
The algorithm thus constantly switches back and forth between the
real space and the Fourier space. Even if this is done with efficient
Fast Fourier Transform (FFT) routines, a single time step takes longer
in this scheme than a single time step in one of the most efficient 
finite difference schemes.
Moreover, the computing time now scales with $m \log m$ with the
system size, instead of $m$. Nevertheless, we found that this is
more than compensated by the vastly higher accuracy of the
pseudo-spectral method, compared to finite difference methods.
In order to calculate $q(\rr,t)$ with a given numerical accuracy,
the time step $h$ could be chosen an order of magnitude larger in 
the pseudo-spectral method than, \eg, in the 
Dufort-Frankel scheme~\cite{dominik_diss}.

\subsection{Extensions}
In order to make connection to experiments a variety of extension to the
Gaussian chain model have been explored. Often it is desirable to include some
of the architectural details for a better description on the length scale of
the statistical segment length.  Of course, the Gaussian chain model is only
applicable if the conformational statistics on the length scale on which the
composition varies is Gaussian. At strong segregation, $\chi N \gg 1$, or at
surfaces to substrates or to vapor/vacuum, this assumption is often not
fulfilled.

The crossover from the Gaussian chain behavior on large length scales to the
rod-like behavior on the length scale of the statistical segment length can be
described by the worm-like chain model~\cite{WORM}. In this case, the
end-segment distribution $q$ depends both on the spatial coordinate ${\bf r}$
as well as on the orientation ${\bf u}$, defined by the tangent vector of the
space curve at position ${\bf r}$~\cite{W1,W2}. 

While the worm-like chain model still results in an analytical diffusion
equation for the propagator, the structure of a polymer in experiments or
computer simulation model often is not well-described by rod-like behavior. In
order to incorporate arbitrary chain architecture and incorporate details of
the molecular structure on {\em all} length scales, the sum ($ \int {\cal
D}_1[{\bf R}] {\cal P}_1[{\bf R}]$) over the single chain conformations can be
approximated by a partial enumeration over a large number of explicit chain
conformations~\cite{SZLEIFER,LV,MCONF,MREV,WET}. Ideally they have been
extracted from a computer simulation of the liquid phase, or they are obtained
from a molecular modeling approach.  Typically $10^6-10^7$ single chain
conformations are employed in one-dimensional calculations. The longer the
chains and the stronger the changes of the chain conformations in the spatially
inhomogeneous field, the more conformations have to be considered. The sample
size should be large enough that sufficiently many chains contribute significantly
to the Boltzmann weight.  The enumeration over the chain
conformations is conveniently performed in parallel.  To this end a small
fraction of single chain conformations is assigned to each processor. Then,
each processor calculates the Boltzmann weight of its conformations, the
corresponding density profile, and the weight of its fraction of chain
conformations.  Subsequently, the total density profile is constructed by
summing the weighted results of all processors.  Typically, 64 or 128
processors are employed in parallel and a SCF calculation of a profile takes a
few minutes on a CRAY T3E. As it is apparent, the detailed description of chain
architecture comes at the expense of a large increase in computational demand.

Also, in addition to the zero-ranged repulsion between unlike
species, other interactions can be included into the theory. In order to capture some fluid-like
correlations and describe the details of packing in the vicinity of surfaces
weighted density functionals have been used successfully. 

Electrostatic interactions have been included into the theory. This is important
for tuning the orientation of self-assembled structures in block copolymers, and also
to describe biological systems.

\section{Fluctuations}
\subsection{Examples of fluctuation effects and the role of $\N$}
\label{sec:fluc_examples}
From the general expression of the free energy functional (\ref{eqn:Gt1}),
we infer that the parameter combination $k_BT \sqrt{\N}$ sets the scale 
of free energy fluctuations in 
a volume comparable to the size of the chain extension. Generally, if $\N$ is large, 
thermal fluctuations will not be important, and the self-consistent field
theory will provide an adequate description. Qualitatively, the quantity
$\N$ describes how strongly the molecules interdigitate: Fluctuation effects
decrease with increasing number of interaction partners per molecule. There are, however, 
important exceptions where fluctuations change the qualitative behavior. Some
examples of fluctuation effects in dense multicomponent mixtures shall be briefly mentioned:

\begin{description}
\item[{\bf 1) Unmixing transition in binary homopolymer blends}]~\\
      In the vicinity of the critical point of a binary mixture one observes universal behavior,
      which mirrors the divergence of the correlation length of composition fluctuations. The
      universal behavior does not depend on the details of the system, but only on the dimensionality
      of space and the type of order parameter. Therefore binary polymer blends fall into the same
      universality class as mixtures of small molecules, metallic alloys, or the three-dimensional 
      Ising model. In the 
      vicinity of the critical point, $\chi_c N = 2$ for a symmetric blend~\cite{FH}, the difference of the 
      composition of the two coexisting phases  -- the order parameter $m$ -- vanishes like 
      $m \sim (\chi N - \chi_c N)^\beta$, where the critical exponent $\beta=0.324$ is 
      characteristic of the 3D Ising universality class. This behavior is, however, only 
      observable in the close vicinity of the critical point~\cite{GINZ}
      \begin{equation}
      t \equiv \frac{|\chi N - \chi_c N|}{\chi_c N} \ll {\rm Gi} \sim \frac{1}{\N} 
      \label{eqn:Ginz}
      \end{equation}
      Fluctuations also shift the location of the critical point away from the prediction of
      the mean-field theory, $\frac{|\chi_c N - \chi_c^{\rm MF} N|}{\chi_c N} \ll \sqrt{\rm Gi} \sim \frac{1}{\sqrt{\N}}$~\cite{MON,MREV}.
      Outside the region marked by the Ginzburg criterion (\ref{eqn:Ginz}), fluctuations do not change
      the qualitative behavior. In the limit of large $\N$, the region where fluctuations dominate the behavior
      becomes small. The cross-over from Ising to mean-field behavior as a function of chain length has
      attracted much experimental~\cite{SCHWAHN} and simulational interest~\cite{HPD,M0}.
\\

\item[{\bf 2) Fluctuations at the onset of microphase separation}]~\\
      In case of diblock copolymers, the ordering is associated with composition waves of a finite periodicity in space.
      In contrast to homopolymer mixtures, not long-wavelength fluctuations are important, but those with wavevectors
      that correspond to the periodicity of the ordering.  For symmetric block copolymers one finds a fluctuation-induced first
      order transition (Brazovskii-mechanism~\cite{brazovskii}) and the shift of the transition temperature is~\cite{glenn_fluc}
      \begin{equation}
      \chi_t N = \chi_{t{\rm MF}} + 41.022 \N^{-1/3} \qquad \mbox{with} \qquad \chi_{t{\rm MF}}=10.495
      \end{equation}
\\

\item[{\bf 3) Capillary waves at the interface between two homopolymers}]~\\
      If the systems contains interfaces, the local position of the interface can fluctuate. These capillary waves increase the
      area of the interface, and the free energy costs are proportional to the interface tension $\gamma$~\cite{CAP}. Long-wavelength fluctuations
      are easily excited and influence measurements of the interfacial width $w$ in experiments or computer
      simulations~\cite{A0}. The resulting ``apparent width'' $w$
      \begin{equation}
      \frac{w^2}{R_e^2} = \frac{w_0^2}{R_e^2} + \frac{k_BT}{4\pi \gamma R_e^2} \ln \frac{L}{B_0}, 
      \end{equation}
      depends on the lateral length scale $L$ on which the interface is observed. $B_0$ denotes a 
      short-length scale cut-off on the order of the intrinsic width $w_0$ predicted by
      the self-consistent field theory~\cite{andreas}. Since $\gamma R_e^2$ is proportional to $\sqrt{\N}$, the capillary
      wave contribution to the apparent interfacial width, measured in
      units of $R_e$, scales like $1/\sqrt{\N}$~\cite{albano}.
\\

\item[{\bf 4) Formation of microemulsion-like structure in the vicinity of Lifshitz points}]~\\
\begin{figure}[!t]
\begin{center}
\epsfig{file=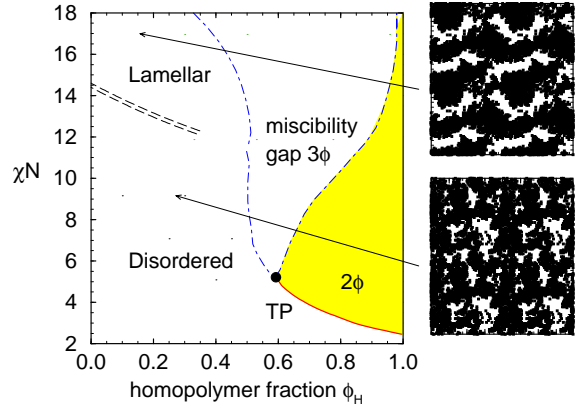,width=0.6\textwidth}
\end{center}
\caption{\label{fig:phase_mc}
Phase diagram for a ternary symmetrical
A+B+AB blend from Monte Carlo simulations of the bond fluctuation model
as a function of the incompatibility parameter
$\chi N$ and the homopolymer volume fraction $\Phi_C$.
The chain length of all three species is equal, $N=32$ corresponding to $\N = 91$.
2$\phi$ denotes a region of two-phase coexistence
between an A-rich and a B-rich phase, 3 $\phi$ one of three-phase
coexistence between an A-rich, a B-rich, and a lamellar phase.
Slice through the three-dimensional system of linear dimension $L=4.7R_e$ and  three periodic images 
are shown on the left hand side.
From Ref.~\cite{MMCOP}.
}
\end{figure}
      Capillary waves do not only broaden the width of the interface, but they can also destroy the orientational order
      in highly swollen lamellar phases. Those phases occur in mixtures of diblock-copolymers and homopolymers. The addition
      of homopolymers swells the distance between the lamellae, and the self-consistent field theory predicts that this
      distance diverges at Lifshitz points. However, general considerations show that mean-field approximations
      are bound to break down in the vicinity of Lifshitz points~\cite{HOLYST}. (The upper critical dimension is $d_u = 8$).
      This can be quantified by a Ginzburg criterion. Fluctuations are important if
      \begin{equation}
      t \equiv \frac{\chi N - \chi_t N}{\chi_t N} \ll \N^{-2/5}
      \label{eqn:lif}
      \end{equation}
      The Ginzburg criterion has a simple heuristic interpretation: Fluctuations of the interface position destroy the
      long-ranged lamellar order. Those interface fluctuations are not limited by the increase of the surface area ($\gamma=0$),
      but rather by the bending stiffness $\kappa$. The correlation length $\xi_n$ of the direction of a single, stiff interface 
      is given by $\xi_n \sim \exp(2\pi \kappa/k_BT)$~\cite{TAUPIN}. If the stiffness $\kappa$ is of order unity, there is no long-range order
      of the normal direction of the interface, and the lamellar order will be replaced by a microemulsion-like structure. In the 
      vicinity of a Lifshitz point self-consistent field theory predicts the interface stiffness to vanish like 
      $\kappa \sim k_BT \sqrt{\N}t^{5/4}$ which is compatible with the Ginzburg criterion (\ref{eqn:lif})~\cite{GERHARD}.

      Similar arguments hold for a tricritical Lifshitz point,
      where a Lifshitz point happens to merge with a regular tricritical point. Here, fluctuations are important if
      \begin{equation}
      t \equiv \frac{\chi N - \chi_t N}{\chi_t N} \ll \N^{-2/3}
      \end{equation}
      Again this behavior is compatible with the behavior of the bending stiffness, $\kappa \sim k_BT \sqrt{\N}t^{3/4}$, 
      as obtained from self-consistent field calculations~\cite{GERHARD}.
\\

\item[{\bf 5) Phase diagram of random copolymers}]~\\
      Macrophase separation can also occur in random copolymers~\cite{FML}, which consist of a random sequence of $Q$ blocks, each of which comprises
      either $M$ segments of type $A$ or $M$ segments of type $B$. Macrophase separation occurs when $\chi M$ is of order unity, \ie,
      independent from the number of blocks, and the $A$-rich and $B$-rich phases differ in their composition only by an amount of order 
      $1/\sqrt{Q}$. The strength of fluctuation effects can be quantified by the Ginzburg number~\cite{JEROME}
      \begin{equation}
      {\rm Gi} \sim \frac{Q^2}{\bar{\cal M}} \qquad \mbox{with} \qquad \bar {\cal M}=\left(\rho_m R^3_M/M \right)^2
      \end{equation}
      where $\rho_m = N \poldens$ denotes the monomer number density.
      While Gi decreases in the case of homopolymer mixtures like $1/\N$, it increases quadratically in the number of blocks for random copolymers.
      In other words, if we increase the number of blocks per molecule, fluctuation effects become stronger and stronger. Indeed, Monte Carlo simulations
      indicate the presence of a disordered, microemulsion-like structure in much of the parameter region where mean-field theory predicts macrophase separation.

\end{description}
In all but the last example, fluctuation effects in three dimensions are controlled by the chain 
length $N$, in the sense that if $N \sim \N \to \infty$,
fluctuations effects can be made arbitrary small. This provides a formal limit in which self-consistent field theory becomes
accurate. We emphasize, however, that in many practical examples fluctuations are very important. For instance,
the broadening of the interface width by capillary waves is typically a sizeable fraction of the intrinsic width calculated
by self-consistent field theory. 

In particle-based models, the mean-field approximation also neglects short-ranged correlations in the polymeric fluid,
\eg, the fluid-like packing of the particles or the correlations due to the self-avoidance of the polymers on short
length scales. There is no small parameter which controls the magnitude of these correlation effects, but they are
incorporated into the coarse-grained parameters $R_e$ and $\chi N$.

\subsection{Gaussian fluctuations and Random Phase Approximation}
\label{sec:RPA}

As long as fluctuations are weak, they can be treated within the Gaussian 
approximation. One famous example of such a treatment is the random phase approximation 
(RPA), which describes Gaussian fluctuations in homogeneous phases. 
The RPA has been extended to inhomogeneous saddle points by Shi, Noolandi, 
and coworkers~\cite{shi,laradji}.
In this section, we shall re-derive this generalized RPA theory within the 
formalism developed in the previous sections. 

Our starting point is the 
Hamiltonian $\G[U,W]$ of Eq. (\ref{eqn:Gt1}). We begin with expanding
it up to second order about the saddle point $\bar{\G} = \G[U^*,W^*]$ 
of the SCF theory. To this end, we define the averaged single chain 
correlation functions~\cite{review,shi,laradji}
\begin{equation}
K_{\alpha \beta} (\rr,\rr') 
= - \frac{\delta \phi_{\alpha}^*}{\delta W_{\beta}}
= \frac{1}{\poldens} \sum_J n_J \frac{\delta^2 
\ln \V Q_J[W_A,W_B]}{\delta W_{\alpha}(\rr) \:\delta W_{\beta}(\rr')}
\end{equation}
with $\alpha, \beta = A$ or $B$, where $\phi_{\alpha}^*$ is defined
as in Eq. (\ref{eqn:phistar}), and all derivatives are evaluated
at the saddle point values of the fields, $W_A^* = (i U^* + W^*)/2$
and $W_B^* = (i U^* - W^*)/2)$. As usual the sum $J$ runs over the 
different types of polymers in the mixture, and $n_J$ denotes the total 
number of polymers of type $J$. In the case of a homopolymer mixture, 
the mixed correlation functions $K_{AB}(\rr,\rr')$, $K_{BA}(\rr,\rr')$ 
are zero.

With these definitions, the quadratic expansion of $\G[U,W]$
(\ref{eqn:Gt1}) can be written as
\begin{eqnarray}
\frac{\G[U,W] - \bar \G }{\sqrt{\N}k_BT (\V/\red)}& = &
\frac{1}{\V} \: \frac{1}{4 \chi N} \int \dr \: \delta W(\rr)^2 
\nonumber\\
\label{eqn:gtgauss1}
&& - \: \frac{1}{2 \V} \sum_{\alpha \beta}
\int \dr \: d^3 \rr' K_{\alpha \beta}(\rr,\rr') \: 
\delta W_{\alpha}(\rr) \: \delta W_{\beta}(\rr')
\end{eqnarray}
\begin{displaymath}
\mbox{with} \qquad 
\delta W_A = (i \delta U + \delta W)/2, 
\qquad
\delta W_B = (i \delta U - \delta W)/2.
\end{displaymath}
Since $\bar \G$ is an extremum, the linear terms in $\delta U$ 
and $\delta W$ vanish. Next we expand this expression in $\delta U$
and $\delta W$. To simplify the expressions, we follow Laradji \etal,
define~\cite{review,laradji} 
\begin{eqnarray}
\Sigma    &=& K_{AA} + K_{AB}+ K_{BA}+ K_{BB}\\
C         &=& K_{AA} - K_{AB}- K_{BA}+ K_{BB}\\
\Delta    &=& K_{AA} + K_{AB}- K_{BA}- K_{BB}\\
\Delta^+  &=& K_{AA} - K_{AB}+ K_{BA}- K_{BB},
\end{eqnarray}
and adopt the matrix notation
\begin{displaymath}
\delta A \cdot K \cdot \delta B  \equiv 
\int \dr \: {\rm d}^d \rr'  K(\rr,\rr') \: \delta A(\rr) \: \delta B(\rr').
\end{displaymath}
with the unity operator ${\bf 1} = \delta(\rr - \rr')$. 
Note that the operators $\Sigma$ and $C$ are symmetric, 
and $\Delta$ and $\Delta^+$ are transposed to each other.
In the new notation, Eq. (\ref{eqn:gtgauss1}) reads
\begin{eqnarray}
\label{eqn:g2}
\lefteqn{
\frac{\G[U,W] - \bar \G }{\sqrt{\N}k_BT (\V/\red)} = 
\frac{1}{\V} \: \frac{1}{4 \chi N} \delta W \cdot {\bf 1} \cdot \delta W 
} \qquad \\
&&
+ \:\frac{1}{8 \V} \left( 
  \delta U \cdot \Sigma \cdot \delta U
- \delta W \cdot C \cdot \delta W
+ i \delta W \cdot \Delta \cdot \delta U
+ i \delta U \cdot \Delta^+ \cdot \delta W 
\right).
\nonumber
\end{eqnarray}
and can, after some algebra, be cast in the form
\begin{eqnarray}
\lefteqn{
\frac{\G[U,W] - \bar \G }{\sqrt{\N}k_BT (\V/\red)} = 
} \qquad & \\
&&
\frac{1}{8 \V}
\left\{ 
(\delta U - \delta U^*) \cdot \Sigma \cdot (\delta U - \delta U^*)
+ \frac{2}{\chi N} \delta W \cdot [{\bf 1} - \frac{\chi N}{2}\tilde{C}] \cdot \delta W
\right\}
\nonumber
\end{eqnarray}
\begin{equation}
\label{eqn:ctilde}
\mbox{with} \quad \tilde{C} = C - \Delta  \cdot \Sigma^{-1} \cdot \Delta^+
\qquad \mbox{and} \qquad
\delta U^* = - i \Sigma^{-1} \cdot \Delta^+ \cdot \delta W.
\end{equation}
The field $U^*+ \delta U^*$ corresponds to the saddle point
of $\G[U,W]$ at fixed $W = W^* + \delta W$.

Both the fluctuations $\langle \delta W^2 \rangle$ and
$\langle (\delta U - \delta U^*) \rangle$ decrease like
$1/\sqrt{\N}$. However, the fluctuations in both fields
are qualitatively different. The operator $\Sigma$ is positive,
\ie, the eigenvalues of $\Sigma$ are positive, therefore 
fluctuations in $U$ remain small and the Gaussian approximation 
is justified in the limit of large $\sqrt{\N}$.
The operator ${\bf 1} - \chi N /2 \tilde{C}$, on the other hand, 
may have negative eigenvalues at large $\chi$. In that case the saddle 
point under consideration is unstable and the true solution
of the SCF theory corresponds to a different structure (phase).
A saddle point which turns from being stable to unstable defines
a spinodal or a continuous phase transition. In addition, the 
operator ${\bf 1} - \chi N /2 \tilde{C}$ also has zero eigenvalues 
in the presence of continuous symmetries. In all 
of these cases, the expansion for small $\delta W$ becomes
inaccurate and fluctuations in $\delta W$ may give rise 
to qualitative deviations from the mean-field theory.

Within the Gaussian approximation, the average 
$\langle W(\rr) W(\rr') \rangle$ can readily be calculated:
\begin{equation}
\langle W(\rr) W(\rr') \rangle
= W^*(\rr) W^*(\rr') + \frac{2 \chi N }{\sqrt{\N}} \red \:
\Big[{\bf 1} - \frac{\chi N}{2} \tilde C \Big]^{-1}_{\rr,\rr'}
\end{equation}
Using the relation between fluctuations of $W$ and the composition,
Eq.~(\ref{eqn:Wfluc}), we obtain
\begin{eqnarray}
\lefteqn{
\langle [\hat \phi_A({\bf r})-\hat \phi_B({\bf r})] [\hat \phi_A({\bf r'})-\hat \phi_B({\bf r'})]\rangle 
- \langle \hat \phi_A(\rr)-\hat \phi_B(\rr) \rangle
  \langle \hat \phi_A(\rr')-\hat \phi_B(\rr') \rangle
} \qquad \qquad \nonumber\\
&\equiv&
\frac{2 \red }{\sqrt{\N} \chi N}
\left[  \Big[ {\bf 1} - \frac{\chi N}{2} \tilde{C} \Big]^{-1} - {\bf 1} \right]_{\rr,\rr'}
 \nonumber \\
&=&
\frac{2 \red }{\sqrt{\N} \chi N}
  \left[ \sum_{k=1}^{\infty} \Big[ \frac{\chi N}{2} \tilde{C} \Big]^k \right]_{\rr,\rr'}
=
\frac{1}{\sqrt{\N}}
\left[\tilde{C} \Big[{\bf 1} - \frac{\chi N}{2} \tilde{C} \Big] \right]_{\rr,\rr'}^{-1}
\nonumber \\
&=&
\frac{1}{\sqrt{\N}}
\Big[ \tilde{C}^{-1} - \frac{\chi N}{2} {\bf 1} \Big]^{-1}_{\rr,\rr'}.
\label{eqn:rpafluct}
\end{eqnarray}

This is the general expression for Gaussian composition fluctuations in 
incompressible polymer blends derived from \EP~theory.
The original derivation of Shi, Noolandi 
and coworkers~\cite{shi,laradji} uses as a starting point the density 
functional (\ref{eqn:dscf}) and gives the identical result.
A generalization for compressible blends can be found in Ref.~\cite{review}.

The special situation of a homogeneous saddle point, corresponding to a 
homogeneous disordered phase, is particularly interesting. In that
case, explicit analytical relations between the single chain partition 
function and the fields can be obtained, and one recovers the well-known
random phase approximation (RPA). To illustrate this approach, we shall
now derive the RPA structure factor for the case of a symmetrical
binary homopolymer blend. 

Since the reference saddle point is spatially homogeneous, all correlation
functions $K_{\alpha \beta}(\rr,\rr')$ only depend on ($\rr - \rr'$) and
it is convenient to perform the calculations in Fourier space. 
We use the following conventions for the Fourier expansion:
\begin{equation}
\hat{\phi}({\mathbf{r}})=\bar{\phi}+\sum_{{\mathbf{q}}\neq 0}\phi_{\mathbf{q}}
e^{i\mathbf{qr}}\qquad\qquad 
\phi_{\mathbf{q}}=\frac{1}{\V}\int \dr \; \hat{\phi}({\mathbf{r}})
e^{-i\mathbf{qr}}.
\end{equation}
In homopolymer melts, the mixed single chain correlation functions
$K_{AB}$, $K_{BA}$ vanish, thus one has $C = \Sigma = K_{AA} + K_{BB}$ 
and $\Delta = \Delta^+ = K_{AA} - K_{BB}$. An explicit expression for
the functional $\phi_A^*({\bf r})=\bar \phi_A \langle \hat \phi_A({\bf r}) \rangle$ 
in the limit of small fields $W_A$ can be derived:
\begin{eqnarray}
\phi_A^*({\bf r}) &=& \bar \phi_A \left(1-\sum_{{\bf q} \neq 0}g_A({\bf q}) W_{A{\bf q}} e^{i\mathbf{qr}} \right) \nonumber \\
                  &=& \bar \phi_A \left(1- \frac{1}{\V} \int \dr' \sum_{{\bf q} \neq 0}g_A({\bf q}) W_A({\bf r}') e^{i{\bf q}({\bf r}-{\bf r}')} \right) 
\end{eqnarray}
A similar expressions holds for $\phi_B^*({\bf r})$. 
Here $g({\bf q})=(2/x^2)(\exp[-x]-1+x)$ with $x \equiv R_e^2{\bf q}^2/6$ denotes
the Debye function.
Hence, the Fourier transform of $K_{AA}$ can be identified with the single chain 
structure factor:
\begin{eqnarray}
K_{AA}({\bf r},{\bf r}') &=& - \frac{\delta \phi_A^*({\bf r})}{\delta W_A({\bf r}')}= \bar \phi_A \frac{1}{\V} \sum_{{\bf q} \neq 0}g_A({\bf q}) e^{i{\bf q}({\bf r}-{\bf r}')} \nonumber \\
K_{AA}(\qq) &=& \bar{\phi}_A g_A({\bf q}), 
\end{eqnarray}
and a similar expression holds $K_{BB}$.
Using the definition of $\tilde{C}$, Eq.~(\ref{eqn:ctilde}), one obtains
\begin{displaymath}
\tilde{C}^{-1} = \frac{1}{4} \Big[ \frac{1}{K_{AA}} + \frac{1}{K_{BB}} \Big]
= \frac{1}{4} \Big[ \frac{1}{\bar \phi_A g_A} + \frac{1}{\bar \phi_B g_B} \Big].
\end{displaymath}

Inserting this expression into Eq.~(\ref{eqn:rpafluct}),
 we can also obtain an explicit expression for the \EP~Hamiltonian ${\cal H}[W]\equiv {\cal G}[U^*,W]$
of binary blends within RPA:
\begin{eqnarray}
\frac{{\cal H}_{\rm RPA}[W]-\bar{\cal H}}{\sqrt{\N}k_BT (\V/\red)}
& = &+\frac{1}{8}\sum_{{\mathbf{q}}\neq 0} \left( \frac{2}{\chi N}- \frac{4 \bar{\phi}_Ag_A({\mathbf{q}})\bar{\phi}_Bg_B({\mathbf{q}})}
{\bar{\phi}_Ag_A({\mathbf{q}})+\bar{\phi}_Bg_B({\mathbf{q}})}\right)|W_{\mathbf{q}}|^2 
\label{eqn:HRPA}
\end{eqnarray}

Combined with Eq.~(\ref{eqn:Wfluc}), this yields the well-known RPA expression
for the collective structure factor in binary polymer blends
\begin{equation}
\label{eqn:phi_rpa}
\frac{4}{\sqrt{\N} (\V/\red) \langle|\phi_{A{\bf q}}-\phi_{B{\bf q}}|^2\rangle}
= \frac{1}{\bar{\phi}_A g_A({\bf q})}+\frac{1}{\bar{\phi}_B g_B({\bf q})} -2\chi N.
\end{equation}
This calculation also justifies the use of Eq.~(\ref{eqn:Wfluc}) to calculate the fluctuations
in the \EP~theory~\cite{ELLEN1}. If we used the literal fluctuations in the \EP~theory according 
of Eq.~(\ref{eqn:d1_EPD}), we would not recover the RPA expression but rather~\cite{ELLEN1}
\begin{eqnarray}
\lefteqn{
\langle|\hat \phi_{{A\mathbf{q}}}-\hat \phi_{B{\bf q}}|^2\rangle_{\rm EP} 
}\quad \nonumber\\
&=&
\langle|\phi^*_{{A\mathbf{q}}}-\phi^*_{B{\bf q}}|^2\rangle 
- \frac{\red}{\sqrt{\bar N} V^2} \int \dr\;{\rm d}^3{\bf r}'\; e^{i {\bf q}({\bf r}-{\bf r}')}
\left\langle \frac{\delta \phi^*_A({\bf r})}{\delta W_A({\bf r}')} + \frac{\delta \phi^*_B({\bf r})}{\delta W_B({\bf r}')}\right\rangle  \nonumber \\
&=& \frac{8\chi N g^2({\bf q})\bar\phi_A^2\bar\phi_B^2}{\sqrt{\N} (V/\red)(1-2\chi N\bar \phi_A \bar \phi_B g({\bf q}))} + \frac{g({\bf q})}{\sqrt{\N} (V/\red)} \nonumber \\
&=&  \langle|\hat \phi_{{A\mathbf{q}}}-\hat \phi_{B{\bf q}}|^2\rangle + \frac{(1-4\bar \phi_A \bar \phi_B)g({\bf q})}{\sqrt{\N} (V/\red)}.
\end{eqnarray}
Both contributions are of the same order.

This example illustrates how the RPA can be used to derive explicit, analytical
expressions for Hamiltonians structure factors. The generalization to blends
that also contain copolymers is straightforward.

\subsection{Relation to Ginzburg-Landau models}
In order to make the connection to a Landau-Ginzburg theory for binary blends,
we study the behavior of
the structure factor at small wavevectors $q$ for a symmetric mixture. Using 
$g(q) \approx 1 + \frac{(qR_e)^2}{18} + \cdots$
we obtain:
\begin{equation}
\frac{4}{\sqrt{\N}(V/\red) \langle|\phi_{A{\bf q}}-\phi_{B{\bf q}}|^2\rangle}= 
\frac{1}{\bar{\phi}_A}+\frac{1}{\bar{\phi}_B} -2\chi N + \frac{q^2R_e^2}{18 \bar \phi_A} + \frac{q^2R_e^2}{18 \bar \phi_B}
\end{equation}
If we assume composition fluctuations to be Gaussian, we can write down a free energy functional compatible with
equation (\ref{eqn:phi_rpa}).

\begin{eqnarray}
\lefteqn{
\frac{{\cal F}_{\rm RPA}[\phi_{A{\bf q}}]}{\sqrt{\N}k_BT(V/\red)}
=  \frac{1}{\sqrt{\N}(V/\red)} \frac{(2\phi_{A{\bf q}}-1)^2}{2\langle|\phi_{A{\bf q}}-\phi_{B{\bf q}}|^2\rangle}
} \nonumber \\
&=&
   \frac{1}{2} \left(
   \frac{1}{\bar{\phi}_A}+\frac{1}{1-\bar{\phi}_A} -2\chi N
   \right) |\phi_{A{0}}-\bar \phi_A|^2
   \\
   && + \frac{1}{2}\sum_{{\bf q} \neq 0}  \left(
   \frac{1}{\bar{\phi}_A}+\frac{1}{1-\bar{\phi}_A} -2\chi N + \frac{q^2R_e^2}{18 \bar \phi_A} + \frac{q^2R_e^2}{18 (1-\bar \phi_A)}
   \right) |\phi_{A{\bf q}}|^2 \nonumber
\end{eqnarray}
Note that this free energy functional is Gaussian in the Fourier coefficients of the composition,
and, hence, the critical behavior still is of mean-field type.
Transforming back from Fourier expansion for the spatial dependence to real space,
we obtain for the free energy functional:

\begin{eqnarray}
   \frac{{\cal F}_{\rm RPA}[\phi_{A}({\bf r })]}{\sqrt{\N} k_BT (\V/\red)} &=&
   \frac{1}{\V} \int \dr \;  \left\{ \frac{1}{2}\left( \frac{1}{\bar{\phi}_A}+\frac{1}{1-\bar{\phi}_A}
   -2\chi N\right) \left(\phi_A({\bf r})-\frac{1}{2}\right)^2 \right. \nonumber \\
   && \hspace*{1.5cm}\left.  + \frac{R_e^2}{36 \bar \phi_A(1-\bar \phi_A)}  (\nabla \phi_A)^2 \right\}
   \nonumber \\
   &=& \frac{1}{\V} \int \dr \;  \left\{ \frac{1}{2} \frac{{\rm d}^2 f_{\rm FH}}{{\rm d}\bar \phi_A^2}
      \left(\phi_A({\bf r})-\frac{1}{2}\right)^2  + \frac{R_e^2 (\nabla \phi_A)^2}{36 \bar \phi_A(1-\bar \phi_A)} \right\}
\nonumber\\
\end{eqnarray}
Expanding the free energy of the homogeneous system (Flory-Huggins free energy of mixing), we can restore higher order terms
and obtain the Landau-de Gennes free energy functional for a symmetric binary polymer blend:

\begin{eqnarray}
\frac{{\cal F}_{\rm GL}[\phi_{A}({\bf r })]}{\sqrt{\N} k_BT (\V/\red)} &=&
\frac{1}{\V} \int \dr \;  \left\{ f_{\rm FH}(\phi_A) + \frac{R_e^2}{36 \bar \phi_A(1-\bar \phi_A)}  (\nabla \phi_A)^2 \right\} \nonumber \\
&=& \frac{1}{\V} \int \dr \;  \left\{  (2-\chi N) \left(\phi_A({\bf r})-\frac{1}{2}\right)^2 + \frac{4}{3} \left(\phi_A({\bf r})-\frac{1}{2}\right)^4 \right. \nonumber \\
&& \hspace*{1.5cm} \left. + \frac{R_e^2}{36 \bar \phi_A(1-\bar \phi_A)}  (\nabla \phi_A)^2 \right\} \label{eqn:GL}
\end{eqnarray}
A field-theory based on this simple expansion will already yield the 
correct three-dimensional Ising critical behavior. Note that the
non-trivial critical behavior is related to higher order terms in $W$ or $\phi_A-\phi_B$. Fluctuations in the
incompressibility field $U$ or the total density $\phi_A+\phi_B$ are not important.

\subsection{Field-theoretic polymer simulations}
\label{sec:FTPS}

Studying fluctuations beyond the Gaussian approximation is difficult. 
Special types of fluctuations, \eg, capillary wave fluctuations
of interfaces, can sometimes be described analytically within 
suitable approximations~\cite{andreas}. The only truly general methods 
are however computer simulations. Here we shall discuss
two different approaches to simulating field theories
for polymers: Langevin simulations and Monte Carlo simulations.

\subsubsection{Langevin simulations}
\label{sec:FTMD}

As long as one is mainly interested in composition fluctuations 
(\EP~approximation, see section \ref{sec:ep}), the problem can be
treated by simulation of a real Langevin process.
The correct ensemble is reproduced by the dynamical equation
\begin{equation}
\label{eq:real_langevin}
\frac{\partial W (\rr ,t)}{\partial t} 
 = 
- \int {\rm d}^d\rr' M(\rr , \rr') \frac{\delta \HH [W]}{\delta W(\rr' ,t )}
+ \theta (\rr ,t),
\end{equation}
where $M(\rr , \rr')$ is an (arbitrary) kinetic coefficient, and 
$\theta(\rr, t)$ is a stochastic noise. The first two moments
of $\theta$ are fixed by the fluctuation-dissipation theorem
\begin{equation}
\label{eq:fluctuation_dissipation}
\langle \theta \rangle = 0 
\qquad
\langle \theta(\rr,t) \theta (\rr',t) \rangle 
 = 2 k_BT M(\rr , \rr') \: \delta (\rr - \rr') \: \delta(t-t').
\end{equation}
The choice of $M(\rr, \rr')$ determines the dynamical properties
of the system. For example, $M(\rr , \rr')=\delta(\rr - \rr')/\tau k_BT$ 
corresponds to a non-conserved field, while field conservation
can be enforced by using a kinetic coefficient of the form 
$M(\rr , \rr')=\nabla_{\rr} \Lambda(\rr - \rr') \nabla_{\rr'}$.
Different forms for the Onsager coefficient $\Lambda$ will be discussed 
in \Sec~\ref{sec:dynamics}. In each step of the Langevin simulation one
updates all field variables simultaneously~\cite{ELLEN1} and the
the self-consistent equations for the saddle point $U^*(W)$ 
have to be solved. 

The approach is commonly referred to as external potential 
dynamics (EPD). A related approach has originally been 
introduced by Maurits and Fraaije~\cite{maurits}. 
However, these authors do not determine $U^*(W)$ exactly, 
but only approximately by solving separate Langevin equations 
for real fields $W_A$ and $W_B$. This amounts to introducing
a separate Langevin equation for a real field $iU$ 
(\ie, an imaginary $U$ in our notation) in addition
to Eq.~(\ref{eq:real_langevin}).

As the fluctuation effects discussed earlier (cf.~\Sec~\ref{sec:fluc_examples})
are essentially caused by composition fluctuations, the
\EP~approximation seems reasonable. Nevertheless, one would 
also like to study the full fluctuating field theory by computer 
simulations. In attempting to do so, however, one faces a serious 
problem: Since the field $iU$ is imaginary, the Hamiltonian $\G$ is 
in general {\em complex}, and the ``weight'' factor
$\exp[- \G]$ can no longer be used to generate a probability 
function. This is an example of a sign problem, as is
well-known from other areas of physics~\cite{signproblem}, 
\eg, lattice gauge theories and correlated fermion
theories. A number of methods have been devised to handle
complex Hamiltonians or complex
actions~\cite{klauder,parisi,lin,baeurle,andre}.
Unfortunately, none of them is as universally powerful as
the methods that sample real actions
(Langevin simulations, Monte Carlo, etc. ).

Fredrickson and 
coworkers~\cite{venkat,glenn_review,glenn2,alexander} 
have recently introduced
the complex Langevin method~\cite{klauder,parisi} into the 
field of polymer science. The idea of this method is simply
to extend the real Langevin formalism, \eg, as used
in EPD, to the case of a complex action. 
The drift term $\delta \HH/\delta W$ in Eq.~\ref{eq:real_langevin} is 
replaced by complex drift terms $\delta \G[U,W]/\delta U$
and $\delta \G[U,W]/\delta W$, which govern the dynamics
of complex fields $U$ and $W$. This generates a diffusion
process in the entire complex plane for both $U$ and $W$.
One might wonder why such a process should sample line 
integrals over $U$ and $W$. To understand that, one must
recall that the integration paths of the line integrals can be 
distorted arbitrarily in the complex plane, as long as no pole is 
crossed, without changing the result. Hence a complex Langevin 
trajectory samples an ensemble of possible line integrals. Under 
certain conditions, the density distribution converges towards a 
stationary distribution which indeed reproduces the desired 
expectation values~\cite{gausterer}. Unfortunately, these
conditions are not generally satisfied, and cases have been 
reported where a complex Langevin process did not converge 
at all, or converged towards the wrong 
distribution~\cite{lin,schoenmakers}. Thus the theoretical
foundations of this methods are still under debate. 
In the context of polymer simulations, however, no problems 
with the convergence or the uniqueness of the solution were
reported.

In our case, the complex Langevin equations that simulate
the Hamiltonian (\ref{eqn:Gt1}) read
\begin{eqnarray}
\frac{\partial U (\rr ,t) }{\partial t}  
& = & 
-\frac{\delta \G [U,W]/k_B T}{\delta U(\rr ,t )}
 + \: \theta_U (\rr ,t)
 \label{cl1} \\
\frac{\partial W (\rr ,t)}{\partial t} 
& = &
- \frac{\delta \G [U,W]/k_B T}{\delta W(\rr ,t )}
+ \theta_W (\rr ,t),
\label{eq16}
\end{eqnarray}
where $\theta_U(\rr,t)$, $\theta_W(\rr,t)$ are {\em real} Gaussian white noises
which satisfy the fluctuation-dissipation theorem
\begin{equation}
\langle \theta \rangle = 0 
\qquad
\langle \theta(\rr,t) \theta (\rr',t) \rangle 
= 2 \: \delta (\rr - \rr') \: \delta(t-t').
\end{equation}
The partial derivatives are given by
\begin{eqnarray}
\frac{\delta \G [U,W]/k_B T}{\delta W(\rr,t)}
&=& \frac{\sqrt{\N}}{2 \red}
  \: \Big[ \frac{W}{\chi N} + \phi_A^*  -\phi_B^* \Big] \\
\frac{\delta \G [U,W]/k_B T}{\delta U(\rr,t)}
&=& \frac{\sqrt{\N}}{2 \red}
  \: \Big[  \phi_A^*  + \phi_B^* - 1\Big], 
\end{eqnarray}
with $\phi_A^*$ and $\phi_B^*$ defined as in Eq. (\ref{eqn:phistar}).

\begin{figure}[t]
\begin{center}
\epsfig{file=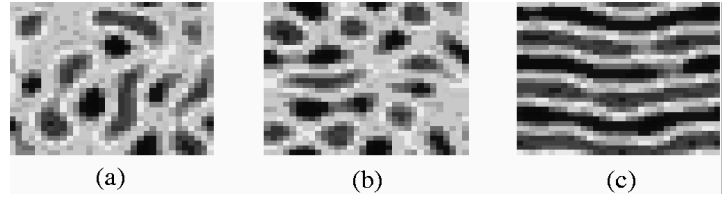,width=0.8\textwidth}
\end{center}
\caption{\label{fig:cops_venkat}
Averaged densities across the order-disorder transition in
a two-dimensional ternary system with A,B homopolymers and 
A-B copolymers (20 \% homopolymer volume fraction), 
as obtained from Complex Langevin simulation runs.
}
\end{figure}

If the noise term is turned off, the system is driven towards
the nearest saddle point. Therefore, the same set of equations
can be used to find and test mean-field solutions.
The complex Langevin method was first applied to dense
melts of copolymers~\cite{venkat}, and later to mixtures
of homopolymers and copolymers~\cite{dominik1} 
and to diluted polymers confined in a slit under good solvent
conditions~\cite{alexander}. Fig.~\ref{fig:cops_venkat}
shows examples of average ``density'' configurations 
$\langle \phi^* \rangle$ for a ternary block copolymer/
homopolymer system above and below the order/disorder
transition.

\subsubsection{Field-theoretic Monte Carlo}
\label{sec:FTMC}

As an alternative approach to sampling the fluctuating field
theory, D\"uchs \etal~\cite{dominik_diss,dominik1} have 
proposed a Monte Carlo method. Since the weight $\exp[- \G]$ 
is not positive semidefinite, the Monte Carlo algorithm
cannot be applied directly. To avoid this problem, we split
$\G$ into a real and an imaginary part $\G^R$ and 
$i \G^I$ and sample only the real contribution 
$\exp(- \G^R)$~\cite{lin}. The imaginary contribution is incorporated 
into a complex reweighting factor $\exp(-i \G^I)$. 
Statistical averages over configurations $j$ then have to 
be computed according to
\begin{equation}
\label{reweight}
\langle A \rangle = \frac{\sum_j 
\exp(- i \G^I_j) A_j}{\sum_j \exp(- i \G^I_j) },
\end{equation}
\ie, every configuration is weighted with this factor.
Furthermore, we premise that the (real) saddle point $i U^*$
contributes substantially to the integral over the imaginary field 
$iU$ and shift the integration path such that it passes through 
the saddle point. The field $iU$ is then represented as 
\begin{equation}
iU = iU^*[W] + i \omega,
\end{equation}
where $\omega$ is real. 

The Monte Carlo simulation includes two different types 
of moves: trial moves of $W$ and $\omega$. 
The moves in $\omega$ are straightforward. Implementing the
moves in $W$ is more involved. Every time that $W$ is changed, 
the new saddle point $iU^*[W]$ must be evaluated. 
In an incompressible blend, the set of self consistent equations
\begin{equation}
\phi_A^*(\rr) + \phi_B^*(\rr) =1, \qquad \forall \rr
\end{equation}
must be solved for $iU^*$. Fortunately, this does not require too 
many iterations, if $W$ does not vary very much from one 
Monte Carlo step to another.

Compared to the Complex Langevin method, the Monte Carlo method 
has the advantage of being well founded theoretically.
However, it can become very inefficient when $\G^I$ spreads
over a wide range and the reweighting factor oscillates strongly. 
In practice, it relies on the fact that the integral is indeed
dominated by one (or several) saddle points.

Can we expect this to be the case here? To estimate the 
range of the reweighting factor, we briefly re-inspect the 
Hubbard-Stratonovich transformation of the total density 
$\hat{\phi}_A + \hat{\phi}_B$ that lead to the fluctuating field $U$. 
For simplicity we consider a one-component system.
In a compressible polymer solution or blend, the contribution of 
the repulsive interaction energy to the partition function 
can be written as 
\begin{equation}
\label{comp}
\exp\Big[- \poldens N \frac{\kappa}{2} \int \dr 
(\hat{\phi}_A + \hat{\phi}_B - \phi_0)^2 \Big]
\end{equation}
(with $\phi_0 =1$ in a melt, and $\phi_0 = 0$ in a good solvent). 
The Hubbard-Stratonovich transformation of this expression is proportional 
to
\begin{equation}
\int  \DD U \exp \Big[-
\poldens \Big\{ \frac{1}{8 \kappa N}\int \dr \; U^2 
        + i \int \dr \; \frac{U}{2} 
(\hat{\phi}_A + \hat{\phi}_B-\phi_0) 
\Big\} \Big]
\end{equation}
Thus $U$ should to be distributed around the saddle 
point with a width proportional to $\kappa$. The method should work 
best for very compressible solutions with small $\kappa$.
In contrast, in a compressible blend, the contribution
(\ref{comp}) is replaced by a delta function constraint
$\delta(\hat{\phi}_A + \hat{\phi}_B-1)$. The fluctuating field 
representation of this constraint is
\begin{equation}
\int \DD U \exp [ - i \poldens 
\int \dr \; \frac{U}{2} (\hat{\phi}_A + \hat{\phi}_B-1) ].
\end{equation}
The only forces driving $U$ towards the saddle point are now those related to
the single chain fluctuations (cf. \Sec~\ref{sec:RPA}), and $U$ will be widely 
distributed on the imaginary axis. Thus it is not clear whether the Monte 
Carlo method will work for incompressible systems.

\begin{figure}[t]

\bigskip
\bigskip
\begin{center}
\epsfig{file=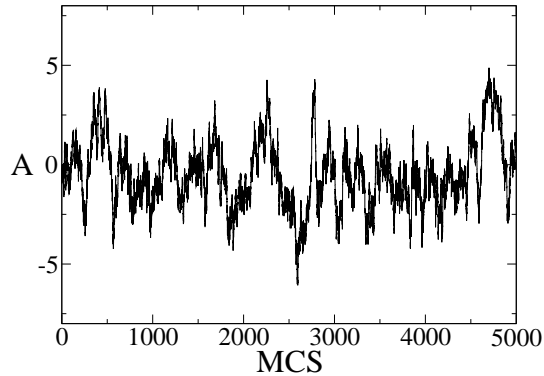,width=0.6\textwidth}
\end{center}
\caption{
\label{fig:rei}
Argument $\G^I$ of the reweighting factor in a simulation of
an incompressible blend. From Ref.~\cite{dominik1}.
}
\end{figure}

Indeed, simulations of an incompressible ternary A+B+AB
homopolymer / copolymer blend showed that the values of
the argument $\G^I$ of the reweighting factor cover a wide 
range (Fig.~\ref{fig:rei}). Hence computing statistical
averages becomes very difficult, because both the numerator
and the denominator in Eq. (\ref{reweight}) are subject
to very large relative errors.

Fortunately, a closer inspection of the data reveals that
the reweighting factor is entirely uncorrelated with the 
fluctuating field $W$. Since the latter determines all 
quantities of interest to us 
(Eq.~(\ref{eqn:Wav}) and (\ref{eqn:Wfluc})), the averages
can be computed without reweighting. 
Moreover, the time scales for variations of $W$ and
$\omega$ are decoupled. By sampling $\omega$ more often
than $W$, they can be chosen such that the time scale of 
$\omega$ is much shorter than that of $W$. 
Finally, the dynamics of the actual simulation, which is 
governed by the real factor $\exp(- \G^R)$, is virtually 
the same as that of a reference simulation with $\omega$ 
switched off~\cite{dominik1}.

Combining all these observations, we infer that the 
fluctuations in $\omega$ do not influence the composition 
structure of the blend substantially. A similar decoupling
between composition and density fluctuations is suggested
by other studies~\cite{PRISM,VILGIS,MREV}.
The composition structure 
can equivalently be studied in a Monte Carlo simulation which 
samples only $W$ and sets $\omega\equiv 0$. 
We have demonstrated this for one specific case of an 
incompressible blend and suspect that it may be a feature of 
incompressible blends in general. The observation that the 
fluctuations in $W$ and $\omega$ are not correlated with each other 
is presumably related to the fact that the (vanishing) density 
fluctuations do not influence the composition fluctuations.
If that is true, we can conclude that field-theoretic 
Monte Carlo can be used to study fluctuations in polymer 
mixtures in the limits of high and low compressibilities. 
Whether it can also be applied at intermediate compressibilities 
will have to be explored in the future.

\begin{figure}[t]
\begin{center}
\epsfig{file=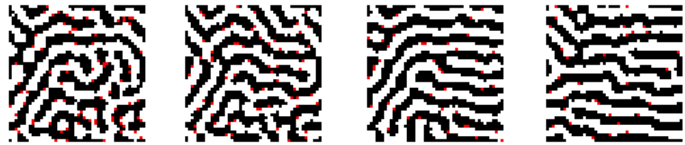,width=0.8\textwidth}
\end{center}
\caption{\label{fig:cops_mc}
Snapshots of $\phi_A^*$ for the same ternary A+B+AB
system as in Fig.~\ref{fig:cops_venkat}, at 70 \% 
homopolymer volume fraction, for different $\chi N$
above and below the order-disorder transition,
as obtained from Monte Carlo simulations with 
$\omega$ switched off (\EP~theory). From Ref.~\cite{dominik2}.
}
\end{figure}

Fig.~\ref{fig:cops_mc} shows examples of snapshots for
the ternary system in the vicinity of the order-disorder
transition.

We should note that the Monte-Carlo simulation with 
$\omega \equiv 0$ effectively samples the \EP~Hamiltonian. 
This version of field-theoretic Monte Carlo is equivalent 
to the real Langevin method (EPD), and can be used as an alternative.
Monte Carlo methods are more versatile than Langevin methods, 
because an almost unlimited number of moves can be invented 
and implemented. In our applications, the $W$ and $\omega$-moves
simply consisted of random increments of the local field values,
within ranges that were chosen such that the Metropolis
acceptance rate was roughly 35\%. On principle, much  
more sophisticated moves are conceivable, \eg, collective
moves or combined EPD/Monte Carlo moves (hybrid moves~\cite{HYBRID}). 
On the other hand, EPD is clearly 
superior to Monte Carlo when dynamical properties are studied. 
This will be discussed in the next section.

\section{Dynamics}
\label{sec:dynamics}
\subsection{Onsager coefficients and dynamic SCF theory (DSCFT) }
In order to describe the diffusive dynamics of composition fluctuations in
binary mixtures one can extend the time-dependent Ginzburg Landau methods to
the free energy functional of the SCF theory. The approach relies on two
ingredients: a free energy functional that accurately describes the chemical
potential of a spatially inhomogeneous composition distribution out of
equilibrium and an Onsager coefficient that relates the variation of the
chemical potential to the current of the composition.

We implicitly assume that the out of equilibrium configuration can be described
by the spatially varying composition, but that we do not have to specify the
details of the molecular conformations. This can be justified if the time scale
on which the composition changes is much larger than the molecular relaxation
time, such that the chain conformations are always in ``local equilibrium''
with the instantaneous composition. The time scale for the chain conformations
to equilibrate is set by $\tau=R_e^2/D$, where $D$ denotes the single chain
self-diffusion constant.  Moreover, we assume that the free energy obtained
from the SCF theory also provides an accurate description even out of
equilibrium, \ie, not close to a saddle point. 

Then, one can relate the spatial variation of the chemical potential to a current of the composition.
\begin{equation}
{\mathbf{J}}_\phi({\mathbf{r}},t)=-\int \dr^{\prime}\;\Lambda_\phi({\mathbf{r}},{\mathbf{r}}^\prime)
\nabla_{{\bf r}^\prime}\mu_\phi({\mathbf{r}}^\prime,t)
\label{eq:Onsager_intro}
\end{equation}
${\mathbf{J}}_\phi({\mathbf{r}},t)$ denotes the current density at position
${\mathbf{r}}$ at time $t$.  There is only one independent chemical potential,
$\mu_\phi$, because of the incompressibility constraint and it is given by
$\mu_\phi=\frac{\delta {\cal F}^*}{\delta\phi_A}$ with ${\cal F}^*[\Phi_A]$
from Eq.~(\ref{eqn:dscf}). 
\begin{eqnarray}
\frac{\mu_\phi({\mathbf{r}})R_e^d}{\sqrt{\N}k_BT} &=& \chi N(\phi_B({\mathbf{r}})-\phi_A({\mathbf{r}}))-(i\Omega_A^*[\phi_A({\mathbf{r}})]-i\Omega_B^*[\phi_B({\mathbf{r}})]) 
\label{eq:chem_pot} 
\\
&\stackrel{\rm RPA}{\approx} & (\chi N-2)(1-2\phi_A({\mathbf{r}})) - \frac{2R_e^2}{9} \triangle \phi_A({\mathbf{r}}) 
\end{eqnarray}
where $i\Omega_A^*$ and $i\Omega_B^*$ are real fields. In the last equation we
have used the explicit expression for the chemical potential within RPA
(cf.~Eq.~(\ref{eqn:GL})) for a symmetric binary polymer blend.

The kinetic Onsager coefficient $\Lambda_\phi({\mathbf{r}},{\mathbf{r}}^\prime)$
relates the ``chemical force'' acting on a monomer at position ${\bf r}'$ due
to the gradient of the chemical potential to the concomitant current density
at position ${\mathbf{r}}$. This describes a purely relaxational dynamics,
inertial effects are not captured.  $\Lambda_\phi$ can be modeled in different ways:
The simplest approach would be local coupling which results in the Onsager
coefficient 
\begin{equation}
\frac{\sqrt{\N}k_BT \Lambda_{\rm local}(\mathbf{r},\mathbf{r}^{\prime})}{R_e^2} 
           = \frac{D}{R_e^2} \phi_A({\mathbf{r}}^\prime,t) \phi_B({\mathbf{r}}^\prime,t) R_e^d\delta(\mathbf{r}-\mathbf{r}^{\prime})
	  \hfill \text{(local)} \label{eq:local_dyn}\\
\end{equation}
The composition dependence accounts for the fact that currents of $A$- and
$B$-densities have to exactly cancel in order to fulfill the incompressibility
constraint.  This local Onsager coefficient completely neglects the propagation
of forces along the backbone of the chain and monomers move independently.
Such a local Onsager coefficient is often used in calculations of dynamic
models based on Ginzburg-Landau type energy functionals for reasons of
simplicity~\cite{Kotnis,Chakrabarti,Glotzer}.  

Bearing in mind that the connectivity of monomeric units along the backbone of
the polymer is an essential ingredient of the single chain dynamics it is clear
that non-local coupling should lead to a better description.  In the Rouse
model forces acting on a monomer caused by the chain connectivity are also
taken into account~\cite{Rouse,Doi}. This leads to a kinetic factor that is
proportional to the intramolecular pair-correlation
function~\cite{Binder_1,deGennes_pap,Pincus}, $P({\mathbf{r}},{\mathbf{r}}^{\prime})$
\begin{equation}
\frac{\sqrt{\N}k_BT \Lambda_{\rm Rouse}(\mathbf{r},\mathbf{r}^{\prime})}{R_e^2} 
           = \frac{D}{R_e^2} \phi_A({\mathbf{r}}^\prime,t) \phi_B({\mathbf{r}}^\prime,t) P({\mathbf{r}},{\mathbf{r}}^{\prime})
	  \hfill \text{(Rouse)} \label{eq:Rouse_dyn}.
\end{equation}

The difficulty in using Rouse dynamics lies in the computational expense of
calculating the pair-correlation function in a spatially inhomogeneous
environment at each time step. If the system is still fairly homogeneous (\eg,
in the early stages of a demixing process), one can use the pair-correlation
function of a homogeneous melt, as it is given through 
RPA \cite{deGennes}.  This leads to the following Onsager
coefficient in Fourier space:

\begin{equation}
\Lambda_{\rm Rouse}(q)=\Lambda_{\rm local}(q) g(q) 
= \frac{R_e^{d+2}}{\sqrt{\N}k_BT} \frac{D}{R_e^2} \bar\phi_A\bar\phi_B\frac{2(\exp[-x]-1+x)}{x^2}
\end{equation}
$x$ is defined as $x\equiv {\mathbf{q}}^2R_e^2/6$ and $g$ is the Debye
function. Another model for non-local coupling is the reptation
model~\cite{Doi,deGennes_pap_2} which is appropriate for polymer melts with
very long chains, \ie, entangled chains.

Since the total amount of $A$-component is conserved, the continuity equation
relates the current of the composition to its time evolution:
\begin{equation}
\frac{\partial\phi_A({\mathbf{r}},t)}{\partial t}+\nabla {\mathbf{J}}_\phi({\mathbf{r}},t)=0
\label{eq:cont_eq}
\end{equation}
Eqs.~(\ref{eq:cont_eq}) and \eqref{eq:Onsager_intro} lead to the following diffusion equation: 
\begin{equation}
\frac{\partial\phi_A({\mathbf{r}},t)}{\partial t}=\nabla_{\bf r}\int \dr^{\prime}
\Lambda_\phi({\mathbf{r}},{\mathbf{r}}^\prime)\nabla_{{\bf r}^\prime}\mu_\phi({\mathbf{r}}^\prime,t)+\eta({\mathbf{r}},t)
\label{eq:diff_eq}
\end{equation}
the last term representing noise that obeys the fluctuation-dissipation theorem.
After Fourier transformation this diffusion equation adopts a very simple form:
\begin{eqnarray}
\frac{\partial\phi_A({\mathbf{q}},t)}{\partial t} &=& -\Lambda_\phi({\mathbf{q}})q^2\mu_\phi({\mathbf{q}},t)+\eta({\mathbf{q}},t) 
\label{eq:diff_eq_q} \\
  &\stackrel{\rm RPA}{\approx} & - \frac{\sqrt{\N}k_BT \Lambda_\phi (qR_e)^2}{R_e^{d+2}} \left[ -2(\chi N-2) + \frac{2}{9}(qR_e)^2 \right] \phi_A +\eta \\
  &\stackrel{\rm RPA}{\approx} & - \frac{\sqrt{\N}k_BT \Lambda_\phi (qR_e)^2}{R_e^{d+2}} \left[ \frac{1}{\bar \phi_Ag(q)} + \frac{1}{\bar \phi_Bg(q)} -2\chi N  \right] \phi_A +\eta \nonumber \\
\label{eq:DSCFT_RPA}
\end{eqnarray}
The last two equations are the explicit expressions within RPA, and $g(q)$ denotes the debye function.

We have now found all necessary equations to numerically calculate the time 
evolution of the densities in a binary polymer mixture. This leads us the 
following procedure to which we refer as the dynamic self consistent field theory 
(DSCFT) method:
\begin{enumerate}
\item[(1)] Calculate the real fields $i\Omega_A^*$ and $i\Omega_B^*$ that ``create'' the density distribution according to Eq.~(\ref{eqn:omega}) using
           the Newton-Broyden scheme.
\item[(2)] Calculate the chemical potential $\mu_\phi$ according to Eq.~(\ref{eq:chem_pot}).
\item[(3)] Propagate the density in time according to Eq.~(\ref{eq:diff_eq_q}) using a simplified Runge-Kutta method.
\item[(4)] Go back to (1).
\end{enumerate}

\subsection{External Potential Dynamics (EPD)}
\label{sec:EPD}

Instead of propagating the composition in time, we can study the time evolution
of the exchange potential $W$.  In equilibrium the density variable
$\phi_A-\phi_B$ and the field variable $W=i\Omega_A-i\Omega_B$ are related to
each other via $\phi_A-\phi_B=-W/\chi N$, see Eq.~(\ref{eqn:Wav}). We also use
this identification to relate the time evolution of the field $W$ to the time
dependence of the composition. Since the composition is a conserved quantity
and it is linearly related to the field variable, we also expect the field $W$
with which we are now describing our system to be conserved. Therefore we can
use the free energy functional ${\cal H}[W]$ from Eq.~(\ref{eqn:H}) and
describe the dynamics  of the field $W$ through the relaxational dynamics of a
model B system, referring to the classification introduced by Hohenberg and
Halperin~\cite{Hohenberg}.
\begin{equation}
\frac{\partial W({\mathbf{r}},t)}{\partial t}=\nabla_{{\mathbf{r}}}
\int \dr\; \Lambda_{\rm EPD}({\mathbf{r}},{\mathbf{r}}^{\prime})\nabla_{{\mathbf{r}}^\prime}
\mu_W({\mathbf{r}}^\prime,t)+\eta({\mathbf{r}},t)
\label{eq:diff_eq_EPD}
\end{equation}
with the chemical potential $\mu_W$ being the first derivative of the free energy ${\cal H}[W]$ with
respect to the order parameter $W$,
\begin{equation}
\frac{\mu_W({\mathbf{r}})}{k_BT}=\frac{\delta {\cal H}[W({\mathbf{r}})]}{\delta W({\mathbf{r}})}
=\frac{\sqrt{\N}k_BT}{R_e^d}\frac{W+\chi N\left[\phi_A^*({\mathbf{r}})-\phi_B^*({\mathbf{r}})\right]}{2\chi N}
\label{eq:chem_pot_EPD},
\end{equation}
$\Lambda_{\rm EPD}$ is a kinetic Onsager coefficient, and $\eta$ denotes noise that satisfies the fluctuation-dissipation theorem.
The Fourier transform of this new diffusion equation is simply:
\begin{eqnarray}
\frac{\partial W(q,t)}{\partial t} &=& -\frac{\sqrt{\N}k_BT\Lambda_{EPD}(qR_e)^2 }{R_e^{d+2}}\frac{W+\chi N(\phi_A^*-\phi_B^*)}{2\chi N} +\eta 
\label{eq:diff_eq_EPD_q}
\\
                                   &\stackrel{\rm RPA}{\approx}& -\frac{\sqrt{\N}k_BT\Lambda_{EPD}(qR_e)^2}{R_e^{d+2}}\frac{W \left(1 - 2 \chi N \bar \phi_A \bar \phi_B g(q)\right)}{2\chi N} +\eta
\label{eq:EPD_RPA}
\end{eqnarray}
$\eta(q,t)$ is white noise that obeys the fluctuation-dissipation theorem. In the last equation we have used the 
Random-Phase Approximation for the Hamiltonian of the \EP~theory (cf.~Eq.~(\ref{eqn:HRPA})).
We refer to the method which uses this diffusion equation as the external potential dynamics (EPD)~\cite{maurits}.
As shown in \Sec~\ref{sec:FTMD}, any Onsager coefficient (in junction with the fluctuation-dissipation theorem)
will reproduce the correct thermodynamic equilibrium. But how is the dynamics of the field $W$ 
related to the collective dynamics of composition fluctuations?

Comparing the diffusion equations of the dynamic SCF theory and the EP Dynamics, Eqs.~(\ref{eq:DSCFT_RPA}) and (\ref{eq:EPD_RPA}),
and using the relation $W(q,t) = - 2\chi N \phi_A(q,t)$, we obtain a relation between the Onsager coefficients within RPA:
\begin{equation}
\Lambda_{\rm EPD}(q) = \frac{2 \chi N}{\bar\phi_A \bar \phi_B g(q)} \Lambda_\phi(q)
\end{equation}
In particular, the non-local Onsager coefficient $\Lambda_{\rm Rouse}$ that mimics Rouse-like dynamics (cf.~Eq.~(\ref{eq:Rouse_dyn})
corresponds to a local Onsager coefficient in the EP Dynamics
\begin{equation}
\frac{\sqrt{\N} k_BT \Lambda_{\rm EPD-Rouse}(q)}{R_e^{d}} = 2 \chi N D
\label{eq:Onsager_EPD}
\end{equation}

Generally, one can approximately relate the time evolution of the field $w$ to the
dynamic SCF theory~\cite{maurits}. The saddle point approximation in the
external fields leads to a bijective relation between the external fields
$i\Omega_A^*$, $i\Omega_B^*$ and $\phi_A$, $\phi_B$. In the vicinity of the
saddle point,  we can therefore choose with which of the two sets of variables
we would like to describe the system. Using the approximation, 
\begin{equation}
\nabla_{{\mathbf{r}}}P({\mathbf{r,r'}})\cong-\nabla_{\mathbf{r'}}P(\mathbf{r,r'})
\label{eq:approx_EPD}
\end{equation}
one can derive Eq.~(\ref{eq:Onsager_EPD}) without invoking the random phase
approximation.  The approximation (\ref{eq:approx_EPD}) is obviously exactly
valid for a homogeneous system, because the pair-correlation function only
depends on the distance $|{\mathbf{r}}-{\mathbf{r}}^{\prime}|$ between two
points. It is expected to be reasonably good for weakly perturbed chain
conformations.  

To study the time evolution in the EPD we use the following scheme~\cite{ELLEN1}:
\begin{enumerate}
\item[(0)] Find initial fields, $W=i\Omega_A^*-i\Omega_B^*$ and $U=i\Omega_A^*+i\Omega_B^*$, that create the initial density distribution
\item[(1)] Calculate the chemical potential $\mu_W$ according to Eq.~(\ref{eq:chem_pot_EPD}).
\item[(2)] Propagate $W$ according to Eq.~(\ref{eq:diff_eq_EPD_q}) using a simplified Runge-Kutta method.
\item[(3)] Adjust $U$ to make sure the incompressibility constraint $\phi_A^*+\phi_B^*=1$ 
      is fulfilled again using the Newton-Broyden method.
\item[(4)] Go back to (1).
\end{enumerate}

The EPD method has two main advantages compared to DSCFT: First of all it 
incorporates non-local coupling corresponding to the Rouse dynamics via a local Onsager coefficient.
Secondly it proves to be computationally faster by up to one order of magnitude.
There are two main reasons for this huge speed up:
In EPD the number of equations that have to be solved via the Newton-Broyden method
to fulfill incompressibility
is just the number of Fourier functions used. The number of equations in DSCFT that have
to be solved
to find the new fields after integrating the densities is twice as large. On the 
other hand, comparing the diffusion equation (\ref{eq:diff_eq}) used in the DSCFT method with equation 
(\ref{eq:diff_eq_EPD}) in EPD, it is easily seen, that the right hand side of the latter is 
a simple multiplication with the squared wave vector of the relevant mode, whereas
the right hand side of equation (\ref{eq:diff_eq}) is a complicated multiplication of three
spatially dependent variables.

\section{Extensions}
The dynamic self-consistent field theory has been widely used in form of MESODYN~\cite{MESODYN}.
This scheme has been extended to study the effect of shear on phase separation
or microstructure formation, and to investigate the morphologies of block copolymers
in thin films. In many practical applications, however, rather severe numerical 
approximations (\eg, very large discretization in space or 
contour length) have been invoked, that make a quantitative comparison to the 
the original model of the
SCF theory difficult, and only the qualitative behavior could be captured.

More recently, stress and strain have been incorporated as conjugated pair of slow dynamic variables 
to extend the model of the SCF theory. This allows to capture some effects of viscoelasticity~\cite{glenn2}.
Similar to the evaluation of the single chain partition function by enumeration of
explicit chain conformations, one can simulate an ensemble of mutually non-interacting 
chains exposed to the effective, self-consistent fields, $U$ and $W$, in order to obtain
the densities~\cite{BESOLD}. Possibilities to extend this scheme to incorporate non-equilibrium chain
conformations have been explored~\cite{VENKAT}.

\section{Applications}

\subsection{Homopolymer-copolymer mixtures}
\begin{figure}[t]

\bigskip
\bigskip
\begin{center}
\epsfig{file=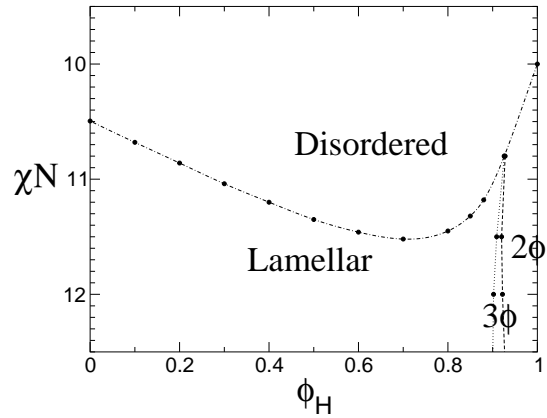,width=0.6\textwidth}
\end{center}
\caption{\label{fig:phase_mf}
Mean-field phase diagram for a ternary symmetrical
A+B+AB blend as a function of the incompatibility parameter
$\chi N$ and the homopolymer volume fraction $\Phi_H$.
The chain lengths of the copolymers is five time that of 
homopolymers. 2$\phi$ denotes a region of two-phase coexistence 
between an A-rich and a B-rich phase, 3 $\phi$ one of three-phase 
coexistence between an A-rich, a B-rich, and a lamellar phase. 
From Ref.~\cite{dominik1}.
}
\end{figure}

As discussed in \Sec~\ref{sec:fluc_examples}, one prominent example
of a situation where the SCF theory fails on a qualitative
level is the microemulsion channel in ternary mixtures
of A and B homopolymers and AB diblock copolymers.
Fig.~\ref{fig:phase_mf} shows an example of a mean-field phase
diagram for such a system. Four different phases are
found: A disordered phase, an ordered (lamellar)
phase (see Fig.~\ref{fig:cops_mc}, right snapshot), an
A-rich and a B-rich phase. The SCF theory 
predicts the existence of a point where all three
phases meet and the distance of the lamellar sheets 
approaches infinity, an isotropic Lifshitz
point~\cite{hornreich1,hornreich2}.

It seems plausible that fluctuations affect the Lifshitz
point. If the lamellar distance is large enough that the 
interfaces between A and B sheets can bend around, the lamellae 
may rupture and form a globally disordered structure.
A Ginzburg analysis reveals that the upper critical 
dimension of isotropic Lifshitz points is as high as 8
(see also \Sec~\ref{sec:fluc_examples}). The lower critical dimension
of isotropic Lifshitz points is 
$d^* = 4$~\cite{diehl,lifshitz_critical}.
Thus fluctuations destroy the transition in two and three dimensions.

\begin{figure}[t]
\begin{center}

\bigskip
\epsfig{file=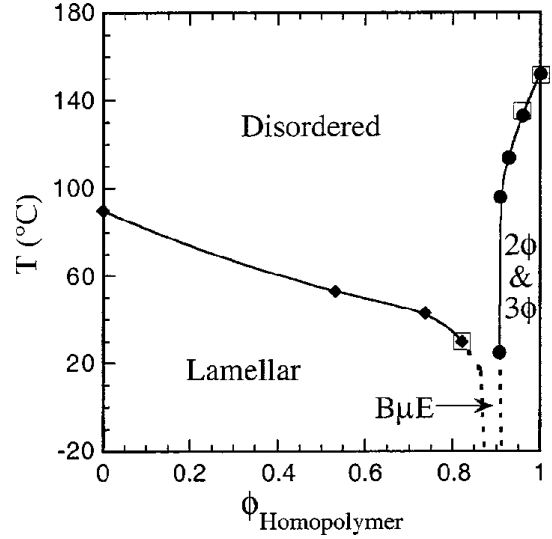,width=0.6\textwidth}
\end{center}
\caption{\label{fig:phase_exp}
Experimental phase diagram of a PDMS-PE/PDMS/PE blend
with copolymers about five times as long as the homopolymers.
From Ref.~\cite{morkved}.
Reproduced by permission of the Royal Society of Chemistry.
}
\end{figure}

\begin{figure}[b]
\begin{center}
\epsfig{file=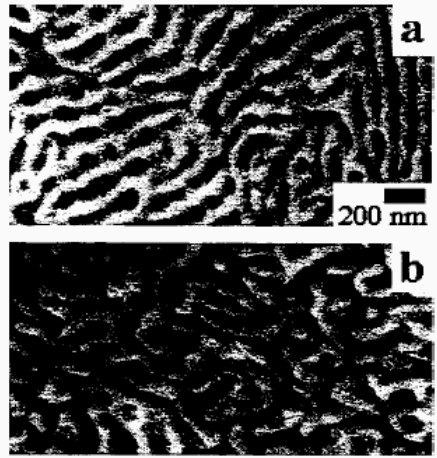,width=0.5\textwidth}
\end{center}
\caption{\label{fig:tem}
Transmission electron micrographs from symmetric PE/PEP/PE-PEP
blends: (a) fluctuating lamellae (b) microemulsion phase.
From Ref.~\cite{morkved}.
Reproduced by permission of the Royal Society of Chemistry.
}
\end{figure}

Indeed, the experimentally observed phase behavior differs
substantially from the mean-field phase diagram. An example
is shown in Fig.~\ref{fig:phase_exp}. The Lifshitz point
is destroyed, the three phase coexistence region between
the lamellar phase and the demixed A-rich and B-rich
phases is removed and gives way to a macroscopically mixed
phase. A transmission electron micrograph of this phase is shown 
and compared with the structure of the lamellar phase
in Fig.~\ref{fig:tem}.


\begin{figure}[t]
\begin{center}
\epsfig{file=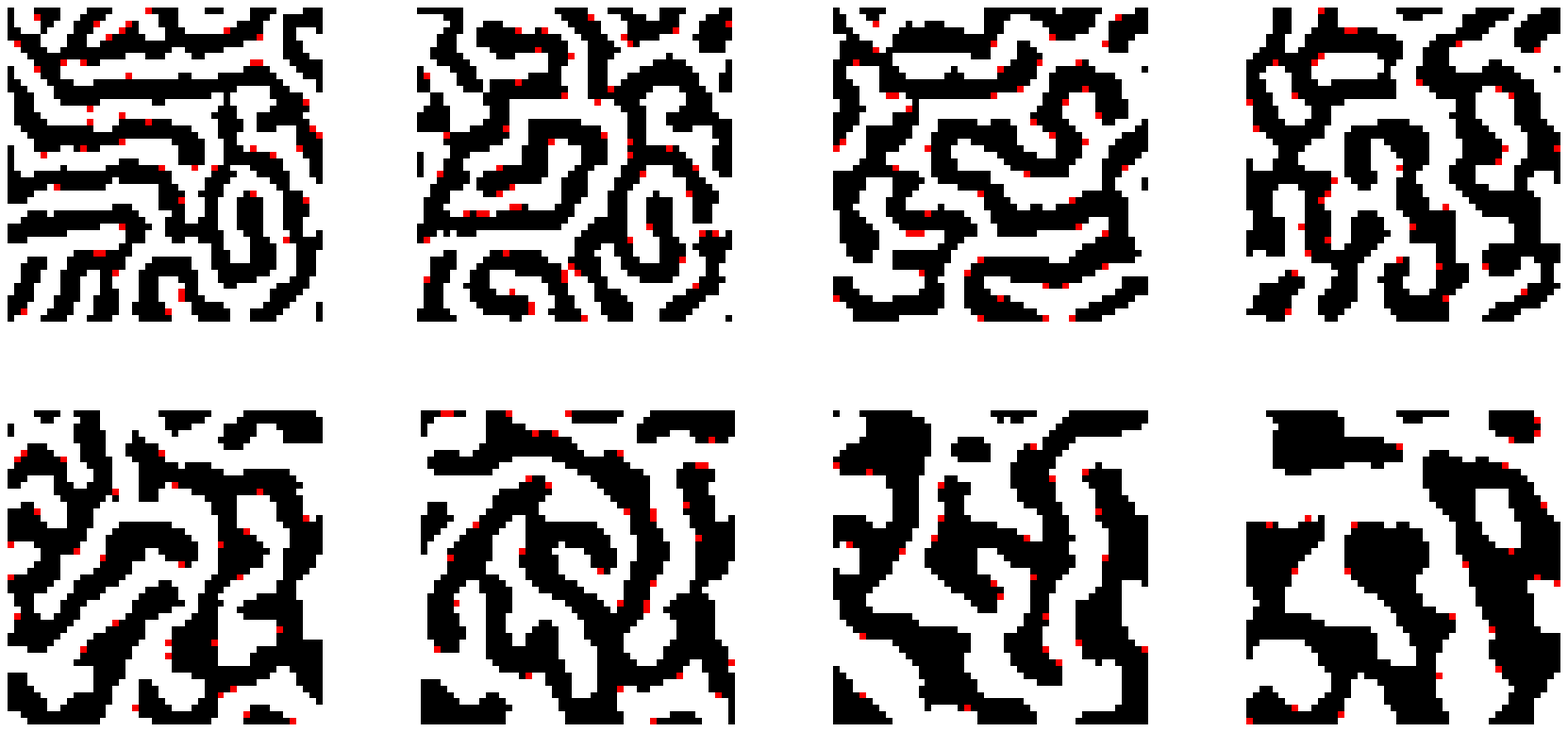,width=0.8\textwidth}
\end{center}
\caption{\label{fig:snap2}
Snapshots of $\phi_A^*$ for the ternary A+B+AB system of 
Fig.~\ref{fig:phase_mf} at $\chi N = 12.5$ for homopolymer
volume fraction $\phi_H =$ 0.74,0.78,0.8, 0.82 (first row
from left), and 0.84, 0.86, and 0.9 (second row from left).
}
\end{figure}

In order to study this effect, D\"uchs \etal~have performed 
field-theoretic Monte Carlo simulations of the system of 
Fig.~\ref{fig:phase_mf}~\cite{dominik1,dominik2,dominik3}, 
in two dimensions. (For the reasons explained in Sec.~\ref{sec:FTMC},
most of these simulations were carried out in the EP approximation).
Some characteristic snapshots were already
shown in Fig.~\ref{fig:cops_mc}. Here we show another
series of snapshot at $\chi N = 12.5$ for increasing
homopolymer concentrations (Fig.~\ref{fig:snap2}).
For all these points, the self-consistent field theory
would predict an ordered lamellar phase. In the 
simulation, only configurations with lower homopolymer
concentrations exhibit (defective) lamellar order.
At higher homopolymer concentration, the order breaks
up and a microemulsion emerges. 

\begin{figure}[tb]

\bigskip
\bigskip
\begin{center}
\epsfig{file=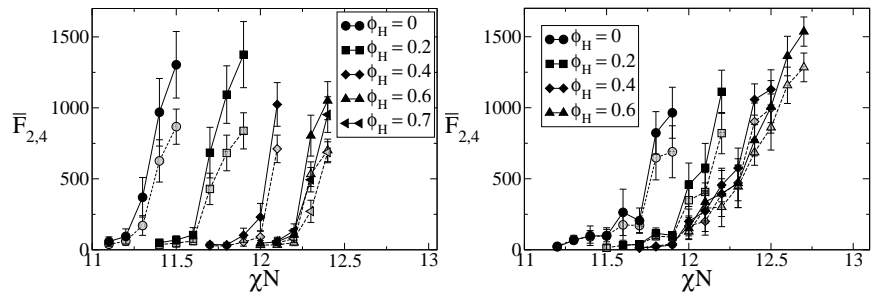,width=0.48\textwidth}
\epsfig{file=f2f4_vg.eps,width=0.48\textwidth}
\end{center}
\caption{\label{fig:f24}
Anisotropy parameters $\bar{F}_2$ (solid) and $\bar{F}_4$ (dashed)
vs. $\chi N$ for different homopolymer volume fractions
$\Phi_H$ at the order/disorder transition. 
Left: From Monte Carlo Simulations (\EP~theory)
Right: From Complex Langevin simulations.
After Ref.~\cite{dominik1}
}
\end{figure}

The phase transition can be identified by defining 
appropriate order parameters. We use anisotropy parameters,
which are defined in terms of the Fourier transform 
$F({\bf q})$ of images as those shown in Fig.~\ref{fig:snap2}:
\begin{equation}
\label{fn}
F_n(q) \equiv \frac{1}{2\pi} \Big| \int_0^{2\pi} \!\!\!
d \phi |F({\bf q})|^2 e^{i n\phi} \Big|.
\end{equation}
$F_n(q)$ is zero for isotropic (disordered) configurations 
and nonzero for anisotropic (lamellar) configurations.
The anisotropy information can be condensed into
a single dimensionless parameter
\begin{equation}
\bar{F}_n := \frac{\int \! d q F_n(q)}{\sigma(q)|_{F_n}},
\end{equation}
with the normalization
\begin{equation}
\sigma(q)|_{F_n} \equiv
\Big[ \frac{\int \! d q \: q^2 F_n(q)}{\int \! d q \: F_n(q)}
- \Big(\frac{\int \! d q \: q \: F_n(q)}
{\int \! d q \: F_n(q)}\Big)^2
\Big]^{1/2}.
\end{equation}
Examples of anisotropy parameters are shown in Fig.~\ref{fig:f24}(left).
They are clearly suited to characterize the phase transition between 
the disordered phase at low $\chi N$ and the lamellar phase at high 
$\chi N$. For comparison, the same system was also examined
with complex Langevin simulations. The results are shown in
Fig.~\ref{fig:f24}(right). The transition points are the same.

Unfortunately, the order of the transition could not yet
determined from these simulations. In the self-consistent field theory,
the transition is continuous. According to theoretical 
predictions~\cite{brazovskii,glenn_fluc}, fluctuations shift it 
to lower $\chi N$ and change the order of the transition.
Our simulation data are consistent both with the assumption that the
transition is continuous or weakly first order. In particular,
we have established that the miscibility gap between the ordered
and the disordered phase is very small (data not shown). 

\begin{figure}[thb]
\begin{center}

\bigskip
\bigskip
\epsfig{file=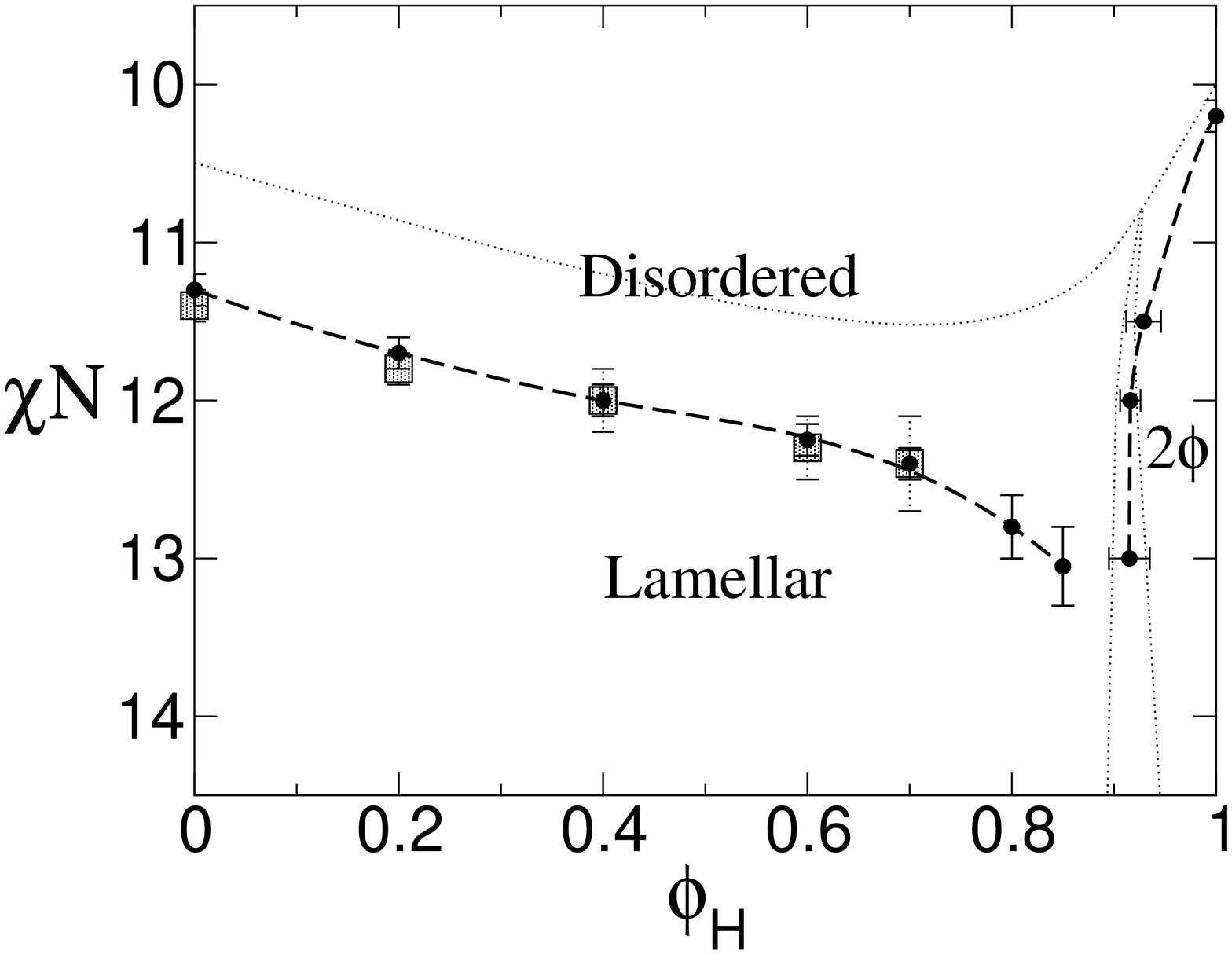,width=0.65\textwidth}
\end{center}
\caption{\label{fig:phase_final}
Phase diagram from simulations for the same system as above. 
The circles show the results from Monte Carlo simulations,
the squares those from the complex Langevin simulations.
The dotted lines correspond to the mean-field prediction
from Fig.~\ref{fig:phase_mf}. From Ref.~\cite{dominik1}.
}
\end{figure}

Besides the order/disorder transition, we also need to determine
the demixing transition. This was done in the grand canonical
ensemble by monitoring the difference between A and B monomer
densities. It exhibits a jump at the phase transition.
All phase transition points are summarized in the phase
diagram of Fig.~\ref{fig:phase_final}. It is in good qualitative 
agreement with the experimental phase diagram of Fig.~\ref{fig:phase_exp}.
In particular, we reproduce the cusp-like region of the 
microemulsion.

\begin{figure}[thb]
\begin{center}
\bigskip
\epsfig{file=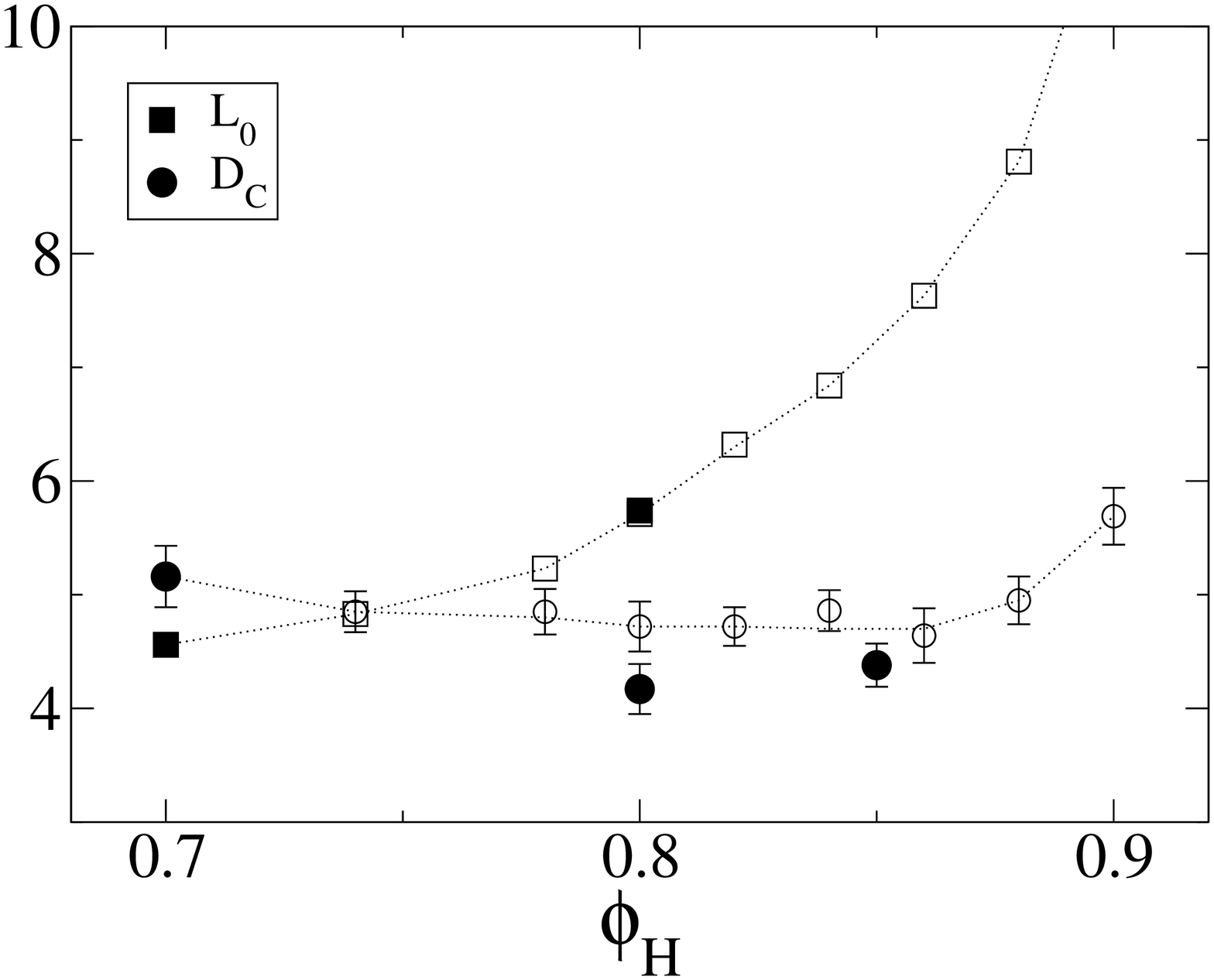,width=0.6\textwidth}
\end{center}
\caption{\label{fig:length_scales}
Characteristic length scales at $\chi N = 12.5$ as a function
of homopolymer concentration $\phi_H$.
$D_C$ (circles) shows the mean (absolute) curvature radius,
and $L_0$ the preferential length scale of layer distances,
obtained from the maximum of $F_0(q)$ (Eq. (\ref{fn})). 
From Ref.~\cite{dominik1}
}
\end{figure}

The simulations allow to analyze the phase transition
and the microemulsion phase in more detail. 
Fig.~\ref{fig:length_scales} compares different characteristic length 
scales of the system at the order/disorder transition.
One length scale, $L_0$, is obtained from the maximum of $F_0(q)$ 
(Eq. (\ref{fn})) and gives the typical distance between layers.
The other is the mean absolute curvature radius $D_C$, and
characterizes the length scale for layer bending.
Whereas the typical layer distance increases with the
homopolymer concentration, the curvature radius remains 
roughly constant. The microemulsion phase transition takes
place at the value of $\phi_H$ where the two lines cross.
Thus we can identify the mechanism by which fluctuations 
generate the microemulsion: The lamellae break up when their 
width grows larger than the length scale which characteristizes 
the boundary fluctuations.

To summarize, the example of homopolymer/copolymer mixtures
demonstrates nicely how field-theoretic simulations can be
used to study non-trivial fluctuation effects in polymer
blends within the Gaussian chain model. The main advantage
of these simulations is that they can be combined in a natural way
with standard self-consistent field calculations. 
As mentioned earlier, the self-consistent field theory is
one of the most powerful method for the theoretical description
of polymer blends, and it is often accurate on a quantitative
level. In many regions of the parameter space, fluctuations
are irrelevant for large chain lengths (large $\N$) and 
simulations are not necessary. Field-theoretic simulations
are well suited to complement self-consistent field theories
in those parameter regions where fluctuation effects
become important.

\subsection{Spinodal decomposition}
To illustrate the application of the method to study the dynamics of collective
composition fluctuations, we now consider the spontaneous phase separation
(\ie, spinodal decomposition) that ensues after a quench into the miscibility
gap at the critical composition, $\bar \phi_A = \bar \phi_B = 1/2$, of the
symmetric blend. Different time regimes can be distinguished: During the early
stages, composition fluctuations are amplified. Fourier modes with a wavevector
$q$ smaller than a critical value $q_c$ exponentially grow and modes with
different wavevectors are decoupled.  As the amplitude of the modes continues
to grow, the composition in the domains begins to saturate and different modes
begin to interact. At even later stages of phase ordering, hydrodynamics
dominates the coarsening.

In the following, we restrict our attention to the early stages of spinodal
decomposition.  In the analysis of experiments one often uses the Landau-de
Gennes functional (\ref{eqn:GL}). The Cahn-Hilliard-Cook theory~\cite{CHC} for
the early stages of phase separation. This treatment predicts that Fourier
modes of the composition independently evolve and increase exponentially in time
with a wavevector-dependent rate, $\phi_A(q,t)^2 \sim \exp [ R({\bf q}) t ]$.
Therefore, it is beneficial to expand the spatial dependence of the composition
in our dynamic SCF or \EP~calculations in a Fourier basis of plane waves. As
the linearized theory suggests a decoupling of the Fourier modes at early
stages, we can describe our system by a rather small number of Fourier modes.

Using Eq.~(\ref{eq:DSCFT_RPA}) one reads off the rate $R(q)$:
\begin{equation}
R({\bf q}) = 4 \frac{\sqrt{\N}k_BT\Lambda_\phi (qR_e^2)}{R_e^{d+2}} \left[ \chi N -2 - \frac{(qR_e)^2}{9}\right]
\end{equation}
and obtains for local dynamics and Rouse-like dynamics
\begin{eqnarray}
\frac{R(q)R_e^2}{D} &=& (qR_e)^2 (\chi N -2) \left[ 1 - \left(\frac{q}{q_c}\right)^2\right]  \label{eqn:Rloc}\\
\frac{R(q)R_e^2}{D} &=& (qR_e)^2 (\chi N -2) g(q) \left[ 1 - \left(\frac{q}{q_c}\right)^2 \right] ,
\end{eqnarray}
respectively,
where the critical wavevector is given by $q_c R = \sqrt{9(\chi N -2)}$. During
the early stage, composition modes with wavevectors $q<q_c$ will grow. In the
case of local dynamics, the fastest growth occurs at $q_c/\sqrt{2}$. In the
case of Rouse-like dynamics, the maximum of the growth rate is shifted to
smaller wavevectors.  The expressions above use the RPA which becomes accurate
at weak segregation (WSL).  Of course, the applicability of the linearized
theory of spinodal decomposition using the Landau-de Gennes free energy
functional is severely limited: In the ultimate vicinity of the critical point,
the free energy functional does not even give an accurate description of the
thermodynamic equilibrium, and the relaxation times are characterized by a
non-classical, dynamical critical exponent. 

\begin{figure}[!t]
\begin{center}
\epsfig{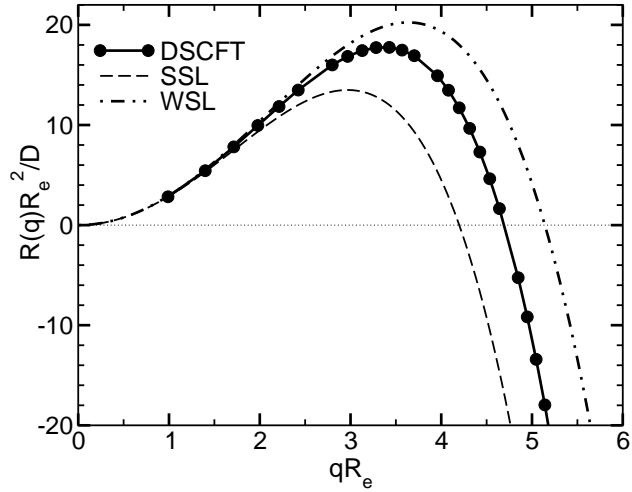}
\end{center}
\caption{\label{fig:E1}
Growth rates of density modes for a quench of a symmetric mixture to $\chi
N=5$. Below the critical wave vector $q_c$ the density modes increase
spontaneously.  Modes with larger wave vectors are damped. 
From Ref.~\cite{ELLEN1}.
} 
\end{figure}

In the opposite limit of strong segregation (SSL) $\chi N \ll 2$, the free
energy can also be expressed in terms of a square gradient functional, and one
obtains $q_c R = \sqrt{6(\chi N -2)}$.  At intermediate segregation, which is
most relevant to experiments, the square-gradient approximation breaks down and
there is no straightforward way to interpolate between the simple analytical
expression in the weak and strong segregation limit.  

The predictions of the rate $R(q)$ for the weak and the strong segregation
limit are presented in Fig.~\ref{fig:E1}.  As expected the results of the
dynamic SCF theory with a local Onsager coefficient fall between the results of the Cahn-Hilliard-Cook theory in
the two limits of weak and strong segregation.  Comparing the predictions for
the kinetics of phase separation to experiments or simulations, it often
remains unclear whether deviations are caused by approximating the free energy
functional by the Landau-de Gennes expressions or the Onsager coefficient.
Indeed, earlier Monte Carlo simulations found rather pronounced
deviations~\cite{EARL} from the prediction of the Cahn-Hilliard-Cook
theory~\cite{CHC}.

\begin{figure}[!t]
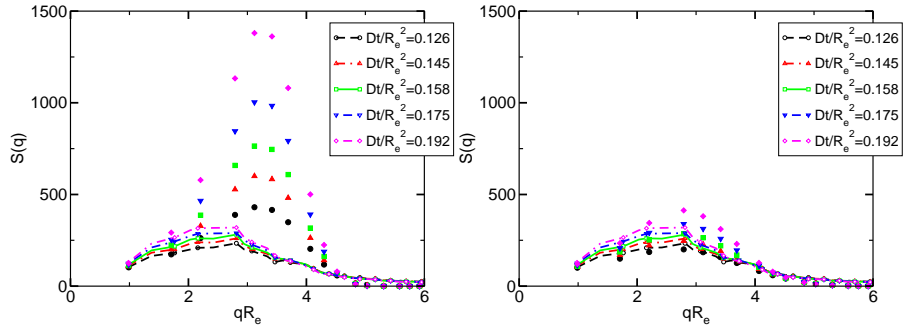

\begin{minipage}{0.5\textwidth}
\epsfig{file=Glostr_SCF_MC.eps,width=1\textwidth,clip=}\\
\end{minipage}
\begin{minipage}{0.5\textwidth}
\epsfig{file=Glostr_MC_Rouse.eps,width=1\textwidth,clip=}\\
\end{minipage}
\caption{
Global structure factor versus wave vector for different times for a quench
from the mixed state at $\chi N = 0.314$ to $\chi N=5$.  Lines represent Monte
Carlo results, filled symbols dynamic SCF theory results.  Panel (a)
compares dynamic SCF theory using a local Onsager coefficient with the Monte
Carlo simulations.  Local dynamics obviously overestimates the growth rate and
shifts the wavevector that corresponds to maximal growth rate to larger values.
Panel (b) compares dynamic SCF theory using a non-local Onsager
coefficient that mimics Rouses dynamics with Monte Carlo results showing better
agreement. From Ref.~\cite{ELLEN1}.  \label{fig:E2}
}
\end{figure}

To overcome this difficulty we use \EP~dynamics or dynamic SCF theory, which do
not invoke a gradient expansion of the free energy functional and yield an
accurate description of the free energy costs of composition fluctuations.
Quantitatively comparing dynamical calculations to computer simulations of the
bond fluctuation model~\cite{BFM}, we investigate the relation between the
dynamics of collective composition fluctuations and the underlaying dynamics of
single chains, which is encoded in the Onsager coefficient $\Lambda$. The
parameters of the bond fluctuation model can be related to the coarse-grained
parameters, $\chi N$, $R_e$ and $\N$, of the standard SCF model and the static
properties of the bond fluctuation model (\eg, the bulk phase behavior~\cite{MREV,M0} and
interface properties~\cite{W2,WET,A0,andreas}) have been quantitatively compared to SCF theory. Note
that the mapping between the particle-based simulation model and the SCF model
does not involve any adjustable parameter: the Flory-Huggins parameter $\chi N$
is identified via the energy of mixing, the length scale is set by $R_e$ and
the time scale is determined by the self-diffusion constant of a single polymer
chain in a dense melt. All these quantities are readily measurable in the
simulations. In the following example, we consider a quench to $\chi N=5$. This
is a compromise: On the one hand it is sufficiently below the critical point,
$\chi_c N=2$, such that critical composition fluctuations are not important. On
the other hand the interfaces between the coexisting phases are still wide
enough such that the polymers are still describable by the Gaussian chain model
on the length scale of the (intrinsic) interfacial width~\cite{STIFF}. The
chains in the bond fluctuation model comprise $N=64$ segments which corresponds
to an invariant degree of polymerization, $\N = 240$.  The time scale is set by
$\tau = R_e^2/D = 1.5 \cdot 10^7$ Monte Carlo steps, where each segment
attempts on average one random, local displacement during a Monte Carlo step.
In the simulations we use a cubic system of size $L= 6.35 R_e$ and average the
results over 64 independent realizations of the temperature quench.

Fig.~\ref{fig:E2} compares the time evolution of the dynamic structure factor
as calculated by dynamic SCF theory using a local Onsager coefficient (panel a) and non-local Onsager coefficient (panel b)
with the results of the Monte Carlo
simulation. As qualitatively expected from Cahn-Hilliard-Cook
theory~\cite{CHC}, there is a well-defined peak in the structure factor $S({\bf
q},t)$
\begin{equation}
S({\bf q},t) \equiv \left \langle \left | \int \dr \; \left( \hat \phi_A({\bf r},t) - \hat \phi_B({\bf r},t) \right) \exp(i {\bf qr}) \right |^2 \right \rangle
\end{equation}
that exponentially growth with time. For these calculations we have omitted the
thermal noise. Therefore, composition fluctuations with $q<q_c$ grow, while
those with $q>q_c$ are damped, and the structure factor for different times
exhibit a common intersection at $q_c$. No such well-defined critical
wavevector $q_c$ is observed in Monte Carlo simulations or experiments, where
thermal fluctuations limit the decay of fluctuations with $q>q_c$. Clearly,
there is a sizeable difference between dynamic SCF calculations using a local
Onsager coefficient and those which use a non-local $\Lambda$ corresponding to
Rouse-like dynamics.  The peak of the structure factor in the dynamic SCF
calculations with the local Onsager coefficient grows much too fast, and also
the peak in the calculations occurs at larger wavevectors compared to the Monte
Carlo simulations.  Apparently, the calculations with the non-local Onsager
coefficient agree much better with the simulation results.

\begin{figure}[!t]
\begin{center}
\epsfig{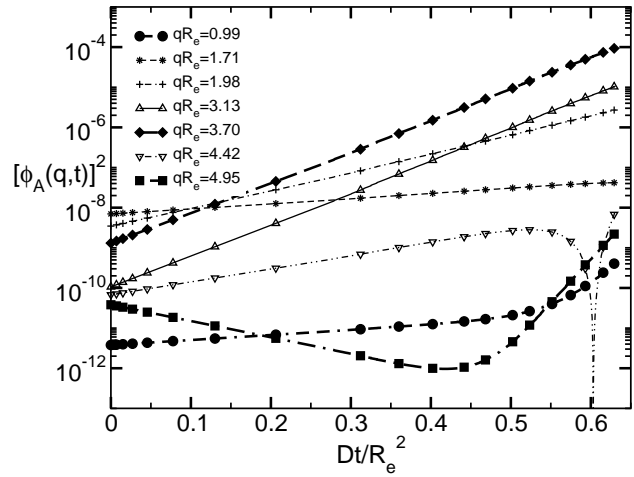}
\end{center}
\caption{\label{fig:E3} Several density modes displayed on a logarithmic scale
versus time. The results were obtained through DSCFT calculations in a three
dimensional system of length $L_x\!=\!L_y\!=\!L_z\!=\!6.35 R_e$ using
$7\!\times\!7\!\times\!7$ functions for a quench from $\chi N=0.314$ to $\chi
N=5$. The expected exponential behavior during early stages of demixing is well
reproduced. From Ref.~\cite{ELLEN1}.
} 
\end{figure}

To quantify this observation, we investigate the growth (or decay) of density modes. The data
in Fig.~\ref{fig:E3} show that the time evolution during the early stages of phase separation indeed
is exponential and one can define a growth rate. The results for a local Onsager coefficient are
presented in Fig.~\ref{fig:E1} and are bracketed by the approximations for the weak and strong
segregation limits.

\begin{figure}[!t]
\begin{center}
\epsfig{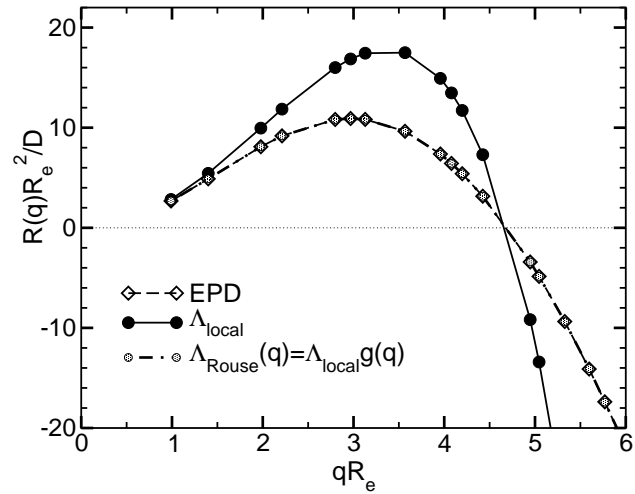}
\end{center}
\caption{\label{fig:E4} Growth rates obtained through two dimensional DSCFT calculations
using local and Rouse dynamics and EPD calculations. The DSCFT results using
the pair-correlation function of a homogeneous melt and the EPD results are in
good agreement.  From Ref.~\cite{ELLEN1}.
}
\end{figure}
In Fig.~\ref{fig:E4} we corroborate that the results of the dynamical SCF calculations using the
non-local Onsager coefficient (\ref{eq:Rouse_dyn}) agree with the EPD using the local Onsager 
coefficient (\ref{eq:Onsager_EPD}) for the field $W$. As detailed in \Sec~\ref{sec:dynamics}, however, the
EPD calculations are computationally much less demanding.

\begin{figure}[!t]
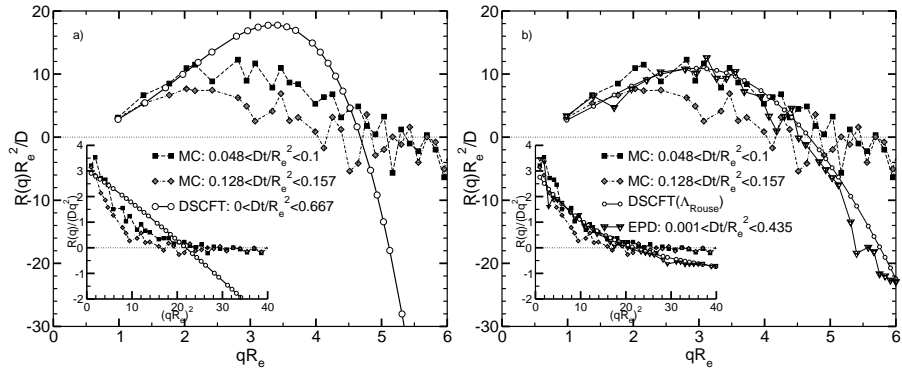

\begin{minipage}{0.5\textwidth}
\epsfig{file=Relax_Cahn_local.eps,width=\textwidth,clip=}\\
\end{minipage}
\begin{minipage}{0.5\textwidth}
\epsfig{file=Relax_Cahn_Rouse.eps,width=\textwidth,clip=}\\
\end{minipage}
\caption{\label{fig:E5}
Corresponding relaxation rates to
Fig.~\ref{fig:E2}. Panel a) compares the Monte Carlo relaxation 
rates with DSCFT calculations with local dynamics. Panel b) compares Monte Carlo results with EPD and DSCFT calculations with Rouse dynamics. For earlier 
times good agreement in the growth region is found but Monte Carlo simulations
show an earlier change in the exponential behaviour. From Ref.~\cite{ELLEN1}.} 
\end{figure}

The growth rates for the dynamic SCF theory with local Onsager coefficient are
displayed in Fig.~\ref{fig:E5} a) and compared to the results of the Monte
Carlo simulations. The maximal growth rate occurs to too large wavevectors and
the growth rate for $q<q_c$ is too large. The Cahn plot, $R(q)/D q^2$ {\em
vs.}~ $q^2$ presented in the inset, shows a linear behavior
(cf.~Eq.~(\ref{eqn:Rloc}) in contrast to the simulations. 

Much of the discrepancy between calculations and Monte Carlo simulations is
removed when we employ a non-local Onsager coefficient or EPD. For $q<q_c$ we
obtain almost quantitative agreement for the growth rate at early times. In the
Monte Carlo simulations, however, deviations from the exponential growth is
observed earlier than in the calculations and there is no decay of fluctuations
for $q>q_c$ in the simulations. 

\begin{figure}[!t]
\begin{center}
\epsfig{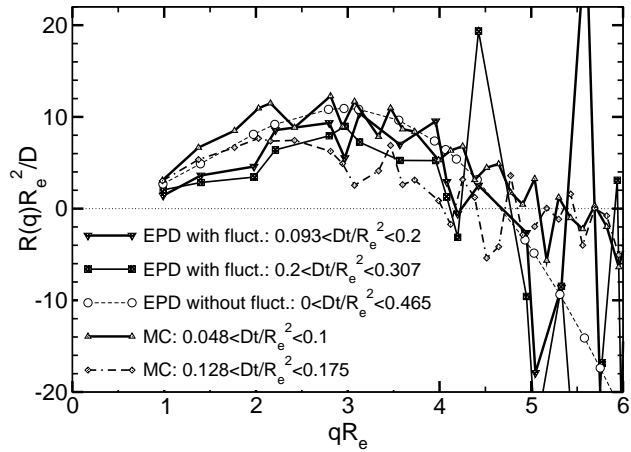}
\end{center}
\caption{\label{fig:E6}Relaxation rates obtained 
through Monte Carlo simulations and EPD calculations with random fluctuations
for different time intervals. Fluctuations lead in both methods to an earlier
change in the exponential behaviour of the increasing modes. From Ref.~\cite{ELLEN1}.}
\end{figure} 

Both effects can be captured if we include fluctuations in the EPD
calculations.  The results of EPD simulations including fluctuations are
presented in Fig.~\ref{fig:E6}.  For $q<q_c$ and early times the calculations
with and without fluctuations mutually agree, and resemble the simulation
results. At later times, the growth rate decreases in the EPD and the
simulations. For $q>q_c$, fluctuations do not decay in the EPD but rather adopt
their finite equilibrium strength. The growth rates extracted from the EPD for
$q>q_c$ rather mirror the fact that some modes grow and other shrink during the
time window of the simulations, although the results have been averaged over 64
independent, two-dimensional EPD calculations.

The influence of the chain connectivity on the dynamics of the composition
fluctuations does not only influence two-point correlation functions like the
global structure factor, but it is also visible in the time evolution of
composition profiles in the vicinity of a surface~\cite{ELLEN2}. A description
of the kinetics of phase separation is important because many multi-component
polymer systems do not reach equilibrium on large length scales and the
morphology of interfaces or self-assembled structures depend on processing
conditions.

The example highlights that both an accurate free energy functional as well as
an Onsager coefficient that describes the dynamics of the molecules are
important for a quantitative description. Of course, this study has
concentrated on the simplest possible scenario. The assumption of the chain
conformations being in equilibrium with the instantaneous composition field and
the neglect of hydrodynamics have to be address to explore the behavior of more
complex systems.

\section{Summary and outlook}
We have briefly reviewed methods which extend the self-consistent mean-field
theory in order to investigate the statics and dynamics of collective composition
fluctuations in polymer blends. Within the standard model of the
self-consistent field theory, the blend is described as an ensemble of Gaussian
threads of extension $R_e$.  There are two types of interactions: zero-ranged
repulsions between threads of different species with strength $\chi N$ and an
incompressibility constraint for the local density.

In the \EP~theory one regards fluctuation of the
composition but still invokes a mean-field approximation for fluctuations of the
total density. This approach is accurate for dense mixtures of long molecules,
because composition fluctuations decouple from fluctuations of the density.
The energy per monomer due to composition fluctuations is typically on the
order of $\chi k_BT \sim k_BT/N$, and it is therefore much smaller than the
energy of repulsive interactions (on the order $1 k_BT$) in the polymer fluid
that give rise to the incompressibility constraint. If there were a coupling
between composition and density fluctuations, a better description of the
repulsive hard-core interactions in the  compressible mixture would be 
required in the first place.

Both the statics and the collective dynamics of composition fluctuations can be
described by these methods, and one can expect these schemes to capture the
essential features of fluctuation effects of the field theoretical model for
dense polymer blends. The pronounced effects of composition fluctuations have
been illustrated by studying the formation of a microemulsion~\cite{dominik1}. 
Other situations where composition fluctuations are very important and where we 
expect that these methods can make straightforward contributions to our
understanding are, \eg, critical points of the demixing in a polymer blend, 
where one observes a crossover from mean
field to Ising critical behavior~\cite{SCHWAHN,HPD}, or random copolymers, where a
fluctuation-induced microemulsion is observed~\cite{JEROME} instead of macrophase separation
which is predicted by mean-field theory\cite{FML}.

The application of the dynamic SCF theory~\cite{MESODYN} or EPD~\cite{maurits,ELLEN1,ELLEN2} to the collective dynamics
of concentration fluctuations and the relation between the dynamics of
collective concentration fluctuations and the single chain dynamics is an
additional, practically important aspect. We have merely illustrated the
simplest possible case -- the early stages of spontaneous phase separation
within a purely diffusive dynamics. In applications hydrodynamic effects~\cite{HYD,PANDERSON}, shear
and viscoelasticity~\cite{TANAKA} might become important.  Even deceptively simple situations
-- like nucleation phenomena in binary polymer blends -- still pose challenging
questions~\cite{BALSARA}. Also the assumption of local equilibrium for the chain
conformations, which allows us to use the SCF free energy functional, has to be
questioned critically. Methods have been devised to incorporate some of these
complications~\cite{glenn2,VENKAT,TANAKA,PANDERSON}, but the development in this area is 
still in its early stages.

Field theoretical simulations~\cite{venkat,glenn_review,dominik1} avoid any saddle point approximation and provide
a formally exact solution of the standard model of the self-consistent field
theory.  To this end one has to deal with a complex free energy functional as a
function of the composition and density. This significantly increases the
computational complexity. Moreover, for certain parameter regions, it is very
difficult to obtain reliable results due to the sign problem that a complex
weight imparts onto thermodynamical averages~\cite{dominik1}. We have illustrated that for a
dense binary blend the results of the field theoretical simulations and the \EP~theory 
agree quantitatively, \ie, density and composition fluctuation
decouple~\cite{dominik1}. In other systems, however, density fluctuations are of crucial
importance. Let us mention two examples: (i) Field theoretical simulation
methods can be used to study semi-dilute solutions~\cite{glenn_review,alexander}, \ie, 
a compressible
ensemble of Gaussian threads with repulsive interactions. In dilute solutions
the chains adopt a self-avoiding walk behavior, while in semi-dilute solutions
the excluded volume interactions are screened on large length scale~\cite{deGennes}. The very
same model has been extensively studied by series expansions~\cite{SER} and
renormalization group theory~\cite{REN} and those results provide an excellent testing bed
to gauge the efficency of the field theoretical simulation method. (ii) It is a challenge
to apply field theoretical simulations to particle-based models. This might provide valuable
insight into the relation between coarse-grained
parameters, $R_e$ and $\chi N$, of the standard model of the self-consistent
field theory and the parameters (temperature, pressure/density and chain
length) of a particle-based model.

\subsection*{Acknowledgment}
It is a great pleasure to thank K. Binder, D. D{\"u}chs, V. Ganesan, G.~H. Fredrickson, 
E. Reister, and M. Schick for their enjoyable and fruitful collaboration on this topic.
Financial support was provided by the DFG under grants Mu 1674/1, Bi 314/17, and Schm 985/7
and the calculations were performed at the NIC in J{\"u}lich and the HLR Stuttgart.


\begin{thebibliography}{999}
\bibitem{APPLICATION} R.~W. Cahn, P. Haasen and E.~J. Kramer, {\em Materials Science and Technology, A Comprehensive Treatment}, Vol 12, VCH, Weinheim (1993); F. Garbassi, M. Morra and E. Occhiello, {\em Polymer Surface: From Physics to Technology}, Wiley, Chichester (2000).

\bibitem{RUSSELL} K.~R. Shull, A.~M. Mayes and T.~P.  Russell, Macromolecules {\bf 26}, 3929 (1993); A.~M. Mayes, T.~P. Russell, S.~K. Satija, and C.~F. Majkrzak, Macromolecules {\bf 25}, 6523 (1992); T.~P. Russell, S.~H. Anastasiadis, A. Menelle, G.~P. Flecher and S.~K. Satija, Macromolecules {\bf 24}, 1575 (1991); P.~F. Green and T.~P. Russell, Macromolecules {\bf 24}, 2931 (1991).

\bibitem{KRAMER}  K.~H. Dai, L.~J. Norton and E.~J. Kramer, Macromolecules {\bf 27}, 1949 (1994); K.~H. Dai and E.~J. Kramer, Polymer {\bf 35}, 157 (1994); K.~H. Dai, E.~J. Kramer and K.~R. Shull, Macromolecules {\bf 25}, 220 (1992); R.~A.~L. Jones, E.~J. Kramer, M.~H.  Rafailovich and S.~A. Schwarz, Phys.Rev.Lett {\bf 62}, 280 (1989).

\bibitem{BUD} A. Budkowski, J. Klein, and L. Fetters, Macromolecules {\bf 28}, 8571 (1995).

\bibitem{HASHI} H. Tanaka, H. Hasegawa, and T. Hashimoto, Macromolecules {\bf 24}, 240 (1991).

\bibitem{jansen}
  B.~J.~P. Jansen, S. Rastogi, H.~E.~H.~ Meijer, P.~J. Lemstra,
  Macromolecules {\bf 34}, 3998 (2001).

\bibitem{tucker}
  C.~L. Tucker, P. Moldenaers, Ann. Rev. Fluid Mechanics {\bf 34}, 177 (2002).

\bibitem{EDWARDS} S.~F. Edwards, Proc. Phys. Soc. {\bf 85}, 613 (1965).

\bibitem{helfand} E. Helfand, Y. Tagami, J. Polym. Sci. {\bf B 9}, 741 (1971); J. Chem. Phys. {\bf 56}, 3592 (1971);  {\bf 57}, 1812 (1972); E. Helfand, A.~M. Sapse, J. Chem. Phys. {\bf 62}, 1327 (1975); E. Helfand, J. Chem. Phys. {\bf 62}, 999 (1975).

\bibitem{hong} K.~M. Hong, J. Noolandi, Macromolecules {\bf 14}, 727 (1981); {\bf 14}, 737 (1981).

\bibitem{scheutjens} J.~M.~H.~M. Scheutjens, G.~J. Fleer, J. Chem. Phys. {\bf 83}, 1619 (1979).

\bibitem{matsen}
  M.~W. Matsen and M. Schick, Phys. Rev. Lett. {\bf 72}, 2660 (1994);
  M.~W. Matsen, Phys. Rev. Lett. {\bf 74}, 4225 (1995);
               Macromolecules {\bf 28}, 5765 (1995);
              J. Physics: Cond. Matt. {\bf 14}, R21 (2001).


\bibitem{review} F. Schmid, J. Phys.: Cond. Matter {\bf 10}, 8105 (1998).

\bibitem{FH}  P.~J. Flory, J. Chem. Phys. {\bf  9}, 660 (1941); H.~L. Huggins, J. Chem. Phys. {\bf  9}, 440 (1941).

\bibitem{COMMENT} In $d=2$ dimensions $\N$ is independet from the number of segments $N$ per molecule and mean field theory is inaccurate, cf.~A. Cavallo, M. M{\"u}ller und K. Binder Europhys. Lett. {\bf 61}, 214 (2003).


\bibitem{deGennes} P.~G. de~Gennes.  {\em Scaling Concepts in Polymer Physics}, Cornell University Press, Ithaca (1979).

\bibitem{WORM}  O. Kratky and G. Porod, Rec. Trav. Chim. {\bf 68}, 1106 (1949); N. Saito, K. Takahashi, Y. Yunoki, J. Phys. Soc. Jpn. {\bf 22}, 219 (1967).

\bibitem{SZLEIFER} I. Szleifer, Curr. Opin. Colloid Interface Sci. {\bf 2}, 416 (1997); I. Szleifer, M.~A. Carignano, Adv. Chem. Phys. {\bf 1996}, {\bf 94}, 742 (1996).

\bibitem{MSROD} M. M{\"u}ller and M. Schick, Macromolecules {\bf 29}, 8900 (1996). 

\bibitem{LV} M. M\"uller and L.~G. MacDowell, Macromolecules {\bf 33}, 3902 (2000); M. M\"uller, L.~G. MacDowell, and A. Yethiraj, J. Chem. Phys. {\bf 118}, 2929 (2003). 

\bibitem{MCONF} M. M{\"u}ller, Macromolecules {\bf 31}, 9044 (1998).

\bibitem{MatsenRev} M.~W. Matsen, J. Phys.: Cond. Matter {\bf 14}, 21 (2002).

\bibitem{CMIX} J.~S. Rowlinson and F.~L. Swinton, {\em Liquids and liquid mixtures}, Butterworths, London (1982); P. Van Konynenburg and R.~L. Scott, Philos. Trans. Soc. London Series {\bf A298}, 495 (1980).

\bibitem{BH} A. Barker and D. Henderson, J. Chem. Phys. {\bf 47}, 4714 (1967).

\bibitem{HMC} J.~P. Hansen and I.~R. McDonald, {\em Theory of simple liquids}, Academic Press (1986).

\bibitem{PH} R.~A. Orwoll and P.~A. Arnold, {\em Polymer-Solvent Interaction Parameter $\chi$}, in Physical Properties of Polymers, Handbook,
J.E. Mark (ed.), Chapter 14 , AIP, Woodbury, New York (1996).

\bibitem{ELLEN_DIS} E. Reister, dissertation, Johannes Gutenberg-Universit\"at, Mainz (2001).

\bibitem{ELLEN1} E. Reister, M. M{\"u}ller and K. Binder, Phys. Rev. {\bf E 64}, 041804 (2001).

\bibitem{COMMENTW} We note that on the right hand side of
Eq.~(\ref{eqn:commentw}) the second term is smaller than the first term by a
factor of $\red/\V$ and can therefore be neglected in large systems. $\red
\delta \phi_\alpha^*({\bf r})/\delta W_\beta({\bf r}')$ is independent of the
system size.

\bibitem{maurits} N.~M. Maurits and J.~G.~E.~M. Fraaije, J. Chem. Phys. {\bf 107}, 5879 (1997).

\bibitem{numerical_recipes} W. H. Press, B. P. Flannery, S. A. Teukolsky, and W. T. Vetterling, {\em Numerical Recipes} 1986, (Cambridge Univ. Press, Cambridge, 1986).

\bibitem{QNEWTON} T. Yamamoto, J. Comp. App. Math. {\bf 124}, 1 (2000).  X. Chen, J. Comp. App. Math. {\bf 80}, 105 (1997); J.M. Martinez, J. Comp. App. Math. {\bf 124}, 45 (2000).

\bibitem{scft} F. Schmid, M. M\"uller, Macromolecules {\bf 28}, 8639 (1995).

\bibitem{anderson} D. G. Anderson, J. Assoc. Comput. Mach. 12, 547 (1965).

\bibitem{eyert} V. Eyert, J. Comp. Phys. {\bf 124}, 271 (1996). 

\bibitem{russell} R.~B. Thompson, K.~\O. Rasmussen and T. Lookman, J. Chem. Phys. {\bf 120}, 31 (2004).

\bibitem{drolet} F. Drolet and G.~H. Frederickson, Phys. Rev. Lett. {\bf 83}, 4317 (1999).

\bibitem{fdm} A.~R. Mitchell and D.~F. Griffiths, {\em The Finite Difference Method in Partial Differential Equations} (Wiley, Chichester, 1980).

\bibitem{feit} M.~D. Feit, J.~A. Fleck, Jr. and A. Steiger, J. Comp. Phys. {\bf 47}, 412 (1982).

\bibitem{rasmussen} K.~\O. Rasmussen and G. Kalosakas, J. Polym. Sci. {\bf B 40}, 777 (2002).

\bibitem{dominik_diss} D. D\"uchs, Dissertation Universit\"at Bielefeld, (2003).

\bibitem{W1} D.~C. Morse and G.~H. Fredrickson, Phys. Rev. Lett. {\bf 73}, 3235 (1994).

\bibitem{W2} F. Schmid and M. M\"uller, Macromolecules {\bf 28}, 8639 (1995).

\bibitem{MREV}M. M{\"u}ller, Macromol. Theory Simul. {\bf 8}, 343 (1999).

\bibitem{WET}M. M{\"u}ller und K. Binder, Macromolecules {\bf 31}, 8323 (1998).

\bibitem{GINZ} V.~L. Ginzburg, Sov. Phys. Solid State  {\bf 1}, 1824 (1960); P.~G. de Gennes,  J. Phys. Lett. (Paris) {\bf 38}, L-441 (1977); J.~F. Joanny, J. Phys. A {\bf 11}, L-117 (1978); K. Binder, Phys. Rev. A  {\bf 29}, 341 (1984).

\bibitem{MON} M.~E. Fisher, Rev. Mod. Phys. {\bf 46}, 587 (1974).

\bibitem{SCHWAHN}  M.~D. Gehlen, J.~H. Rosedale, F.~S. Bates, G.~D. Wignall, K. Almdal, Phys. Rev. Lett. {\bf 68}, 2452 (1992); D. Schwahn, G. Meier, K. Mortensen, and S. Janssen, J. Phys. II (France) {\bf 4}, 837 (1994); H. Frielinghaus, D. Schwahn, L. Willner, and T. Springer, Physica {\bf B 241}, 1022 (1998);

\bibitem{HPD} H.~P. Deutsch and K. Binder, Macromolecules {\bf 25}, 6214 (1992); J. Phys. II (France) {\bf 3} 1049 (1993).

\bibitem{M0}M. M{\"u}ller und K. Binder, Macromolecules {\bf 28}, 1825 (1995).

\bibitem{brazovskii} S.~A. Brazovskii, Sov. Phys. JETP {\bf 41}, 85 (1975).

\bibitem{glenn_fluc} G.~H. Fredrickson and E.~Helfand, J. Chem. Phys. {\bf 87}, 697 (1987).

\bibitem{CAP} F.~P. Buff, R.~A. Lovett and F.~H. Stillinger, Phys. Rev. Lett. {\bf 15}, 621 (1965).

\bibitem{A0} A. Werner, F. Schmid, M. M\"uller, K. Binder, J. Chem. Phys. {\bf 107}, 8175 (1997). 

\bibitem{andreas} A. Werner, F. Schmid, M. M\"uller, K. Binder, Phys. Rev. {\bf E 59}, 728 (1999).

\bibitem{albano} M. M\"uller, E.~V. Albano, and K. Binder, Phys. Rev. {\bf E 62}, 5281 (2000). 

\bibitem{MMCOP}M. M{\"u}ller und M. Schick, J. Chem. Phys. {\bf 105}, 8885 (1996).

\bibitem{HOLYST}  R. Holyst and M. Schick, J.Chem.Phys. {\bf 96}, 7728 (1992).

\bibitem{TAUPIN} C. Taupin and P.~G. de~Gennes, J. Phys. Chem. {\bf 86}, 2294
(1982).

\bibitem{GERHARD}M. M{\"u}ller und G. Gompper, Phys.Rev. {\bf E 66}, 041805 (2002). 


\bibitem{FML} E.~I. Shakhnovich and A.~M. Gutin, J.Phys.France {\bf 50}, 1843 (1989);
              G.~H. Fredrickson, S.~T.  Milner and L. Leibler Macromolecules {\bf 25}, 6341 (1992);
              G.~H. Fredrickson and S.~T.  Milner, Phys. Rev. Lett. {\bf 67}, 835 (1991);
	      A.~V. Subbotin and A.~N. Semenov, Eur.Phys.J {\bf 7}, 49 (2002).

\bibitem{JEROME}J. Houdayer und M. M{\"u}ller, Europhys.Lett. {\bf 58}, 660 (2002); J. Houdayer und M. M{\"u}ller, Macromolecules (in press).

\bibitem{shi} A.~C. Shi, J. Noolandi, and R.~C. Desai, Macromolecules {\bf 29}, 6487 (1996).

\bibitem{laradji} M. Laradji, A.~C. Shi, J. Noolandi, and R.~C. Desai, Phys. Rev. Lett. {\bf 78}, 2577 (1997); Macromolecules {\bf 30}, 3242 (1997).

\bibitem{signproblem} I. Montvay and G. M\"unster, {\em Quantum fields on the lattice}, Cambridge Univ. Press, Cambridge, (1994).

\bibitem{klauder} J.~R. Klauder, J. Phys. A {\bf 16}, L317 (1983).

\bibitem{parisi} G. Parisi, Phys. Lett. {\bf 131B}, 393 (1983).

\bibitem{lin} H.~Q. Lin and J.~E. Hirsch, Phys. Rev. B {\bf 34}, 1964 (1986).

\bibitem{baeurle} S.~A. Baeurle, Phys. Rev. Lett. {\bf 89}, 080602 (2002).

\bibitem{andre} A.~G. Moreira, S.~A. Baeurle, G.~H. Fredrickson, Phys. Rev. Lett. {\bf 91}, 150201 (2003).

\bibitem{venkat} V. Ganesan and G.~H. Fredrickson, Europhys. Lett. {\bf 55}, 814 (2001).

\bibitem{glenn_review} G.~H. Fredrickson, V. Ganesan, F. Drolet, Macromolecules {\bf 35}, 16 (2002).

\bibitem{glenn2} G.~H. Fredrickson, J. Chem. Phys. {\bf 117}, 6810 (2002).

\bibitem{alexander} A. Alexander-Katz, A.~G. Moreira, G.~H. Fredrickson, J. Chem. Phys. {\bf 118}, 9030 (2003).

\bibitem{gausterer} H. Gausterer, Nucl. Phys. {\bf 642}, 239 (1998).

\bibitem{schoenmakers} W.~J. Schoenmakers, Phys. Rev. D {\bf 36}, 1859 (1987).

\bibitem{dominik1} D. D\"uchs, V. Ganesan, G.~H. Fredrickson, F. Schmid, Macromolecules {\bf 36}, 9237 (2003).

\bibitem{PRISM} A. Yethiraj and K.~S. Schweizer, J. Chem. Phys. {\bf 98}, 9080 (1993); K.~S. Schweizer and A. Yethiraj, J. Chem.Phys. {\bf 98}, 9053 (1993).

\bibitem{VILGIS} R. Holyst and T.~A. Vilgis, J. Chem. Phys. {\bf 99}, 4835 (1993); Phys. Rev. {\bf E 50}, 2087 (1994).

\bibitem{dominik2} D. D\"uchs, F. Schmid, NIC Series Vol. 20, p. 343 (2004).

\bibitem{HYBRID}  S. Duane, A.~D. Kennedy, B.~J. Penleton, and D. Roweth, Physs. Lett. {\bf B 195}, 216 (1987); B. Mehlig, D.~W. Heermann and B.~M. Forrest, Phys. Rev. {\bf B 45}, 679 (1992); B.~M. Forrest, U.~W. Suter, J. Chem. Phys. {\bf 101}, 2616 (1994).

\bibitem{Kotnis} M. Kotnis and M. Muthukumar, Macromolecules {\bf 25}, 1716 (1992).

\bibitem{Chakrabarti} A. Chakrabarti, R. Toral, J.~D. Gunton, and M. Muthukumar,  Phys. Rev. Lett. {\bf 63}, 2072 (1989).

\bibitem{Glotzer} C. Castellano and S.~C. Glotzer, J. Chem. Phys. {\bf 103}, 9363 (1995).

\bibitem{Rouse} P.~E. Rouse,  J. Chem. Phys. {\bf 21}, 1272 (1953).

\bibitem{Doi} M. Doi and S.~F. Edwards, {\em The Theory of Polymer Dynamics}. Oxford University Press (1994).

\bibitem{Binder_1} K. Binder,  J. Chem. Phys. {\bf 79}, 6387 (1983).

\bibitem{deGennes_pap} P.~G. de~Gennes,  J. Chem. Phys. {\bf 72}, 4756 (1980).

\bibitem{Pincus} P. Pincus, J. Chem. Phys. {\bf 75}, 1996 (1981).

\bibitem{deGennes_pap_2} P.~G. de~Gennes, J. Chem. Phys. {\bf 55}, 572 (1971).

\bibitem{Hohenberg} P.~C. Hohenberg and B.~I. Halperin, Rev. Mod. Phys. {\bf 49}, 435 (1977).

\bibitem{MESODYN}  P.~Altevogt, O.~A.Evers, J.~G.~E.~M. Fraaije, N.~M. Maurits, B.~A.~C. van Vlimmeren, J. Mol. Struc. Theochem. {\bf 463}, 139 (1999);
                   J.~G.~E.~M. Fraaije, J. Chem. Phys. {\bf 99}, 9202 (1993).

\bibitem{BESOLD} K.~G. Soga, M.J. Zuckermann, and H. Guo, Macromolecules 29, 1998 (1996), G. Besold \etal J. Polym. Sci {\bf 38}, 1053 (2000).

\bibitem{VENKAT} V. Ganesan and V. Pryamitsyn, J. Chem. Phys. {\bf 118}, 4345 (2003).

\bibitem{hornreich1} R.~M. Hornreich, M. Luban and S. Shtrikman, Phys. Rev. Lett. {\bf 35}, 1678 (1975).

\bibitem{hornreich2} R.~M. Hornreich J. Magn. Magn. Mat.{\bf 15}, 387 (1980).

\bibitem{diehl} H.~W. Diehl, acta physica slovaca {\bf 52}, 271 (2002).

\bibitem{lifshitz_critical} According to Ref.~\cite{diehl}, the lower critical dimension of an $m$-axial Lifshitz point is given by $d^* = 2 + m/2$.  In our case, we have an isotropic Lifshitz point with $m=d$, thus $d^*=4$.

\bibitem{morkved} T.~L. Morkved, B.R.~ Chapman, F.~S. Bates, T.~P. Lodge, P. Stepanek, K. Almdal, Faraday Discuss. {\bf 11}, 335 (1999).

\bibitem{dominik3} D. D\"uchs, F. Schmid, submitted to J. Chem. Phys. (2004).

\bibitem{CHC} J.~W. Cahn and J.~E. Hilliard,  J. Chem. Phys. {\bf 28}, 258 (1958); {\em ibid} {\bf 31}, 668 (1959); H.~E. Cook, Acta metall. {\bf 18}, 297 (1970).

\bibitem{EARL} A.~Sariban and K.~Binder,  Polym. Comm. {\bf 30}, 205 (1989); Macromolecules {\bf 24}, 578 (1991); A.~Baumg{\"a}rtner and D.W. Heermann, Polymer {\bf 27}, 1777 (1986).

\bibitem{BFM} I. Carmesin and K. Kremer, Macromolecules {\bf 21},  2819 (1988).  H.~P. Wittmann and K. Kremer Comp. Phys. Com. {\bf 61}, 309 (1990).  H.~P. Deutsch and K. Binder,  J. Chem. Phys. {\bf 94},  2294 (1991).

\bibitem{STIFF}M. M{\"u}ller und A. Werner,  J. Chem. Phys. {\bf 107}, 10764 (1997).

\bibitem{ELLEN2}E. Reister und M. M{\"u}ller, J.Chem.Phys. {\bf 118}, 8476 (2003).

\bibitem{HYD} N.~M. Maurits, A.~V. Zvelindovsky, and J.~G.~E.~M. Fraaije, J. Chem. Phys. {\bf 108}, 9150 (1998);
              T. Koga, K. Kawasaki, M. Takenaka, and T. Hashimoto,  Physica {\bf A198}, 473 (1993).

\bibitem{PANDERSON} B.~J. Keesta, P.~C.~J. van Puyvelde, P.~D. Anderson and H.~E.~H. Heijer, Phys. Fluids {\bf 15}, 2567 (2003).

\bibitem{TANAKA} H.~M. Tanaka, J. Phys.: Cond. Mat {\bf 12}, R207 (2000); Prog. Theo. Phys. {\bf 101}, 863 (1999).

\bibitem{BALSARA} A.~A. Lefebvre, J.~H. Lee, N.~P. Balsara, C. Vaidyanathan, J. Chem. Phys. {\bf 117}, 9063 and 9074 (2002).

\bibitem{SER} M. Muthukumar and B.~G. Nickel, J. Chem. Phys. {\bf 86}, 460 (1987).

\bibitem{REN}  J. des Cloizeaux and G. Jannink, {\em Polymers in Solution: Their Modeling and Structure}, Oxford Science Publications, Oxford (1990); K.~F. Freed, {\em Renormalization Group Theory of Macromolecules}, Wiley, New York (1987); J.~C. Le Guillou and J. Zinn-Justin, J. Phys. (France) {\bf  50}, 1365 (1989); L. Sch{\"a}fer, {\em Excluded Volume Effects in Polymer Solutions}, Springer, Berlin (1999).


\end{thebibliography}
\end{document}